\documentclass[twocolumn,secnumarabic,superscriptaddress,amssymb, nobibnotes, aps, prl]{revtex4-2} %fleqn: make all eqns left alighed  ,showpacs
\usepackage{epsfig,amssymb,amsmath,graphicx,subfigure,hyperref  }  % pdfpages and graphicx are added   hyperref(set hyperlink for content)
\usepackage{mathrsfs}
\usepackage{color}
\usepackage{multirow}
\usepackage{tabularx}
\usepackage[normalem]{ulem}
\usepackage{physics}

\usepackage{cancel}
\renewcommand{\vec}[1]{\boldsymbol{#1}}
\usepackage{xcolor}

\begin{document}
\title{Emergent Haldane Model and Photon-Valley Locking in Chiral Cavities}

\author{Liu Yang} 
\affiliation{Tsung-Dao Lee Institute, Shanghai Jiao Tong University, Shanghai 200240, China}
\affiliation{School of Physics and Astronomy, Shanghai Jiao Tong University, Shanghai 200240, China}
%\email{}
\author{Qing-Dong Jiang}
\email{qingdong.jiang@sjtu.edu.cn}
\affiliation{Tsung-Dao Lee Institute, Shanghai Jiao Tong University, Shanghai 200240, China}
\affiliation{School of Physics and Astronomy, Shanghai Jiao Tong University, Shanghai 200240, China}
\affiliation{Shanghai Branch, Hefei National Laboratory, Shanghai 201315, China}

%\author{}
%\affiliation{} %

\begin{abstract}
The realization of Haldane's topological graphene model in practical materials has presented significant challenges. Here, we propose achieving this model by embedding graphene in chiral cavities, using the asymptotically decoupled framework detailed in Ref.~[Phys. Rev. Lett. {\bf 126}, 153603 (2021)]. Additionally, we introduce an equilibrium strategy for achieving valley polarization in this system with $C_2$-symmetry breaking. Through numerical methods, we quantify the locking of photon numbers 
with Bloch electrons and calculate the topology-induced imbalance of valley photons. Furthermore, we elucidate that topological phase transition is characterized by the sign change of photon numbers during interband excitation.  These findings underscore the remarkable potential of utilizing cavity quantum fluctuations to engineer electronic and photonic properties specific to valleys and topologies, particularly within the realm of strong light-matter coupling.
\end{abstract}

\maketitle

\textit{Introduction}.—Haldane's topological graphene model stands as the pioneering model capable of inducing the quantum Hall effect without the reliance of Landau levels~\cite{Mlattice1976,HaldaneModel,shouchengReview2011,RMP_QAHE2023}. Formulated on a honeycomb lattice, this model reveals the behavior of non-interacting electrons with staggered flux, which averages to zero in each unit cell. Despite the absence of total magnetic flux, a topological bands with nonzero Chern numbers can still emerge due to an intrinsic time-reversal symmetry breaking~\cite{TKNN,QWZmodel2006}. It demonstrates that, in the realization of topological phases, time-reversal symmetry breaking is more fundamental than external magnetic fields.
The Haldane model plays an essential role in the theoretical understanding of topological phases of matter~\cite{RMP2010qian,shouchengReview2011,RMP_QAHE2023}. Although its experimental realization has been achieved in artificial lattices like ultra-cold atom and superconducting circuit systems~\cite{Haldane_experiment2014,roushan2014observation}, implementing it in real materials is a formidable task due to the tiny scale of material lattice.

To circumvent the challenge of realizing a unit-cell sized staggered flux configuration required by Haldane's original model, physicists have proposed alternative methods to achieve the quantum anomalous Hall effect, incorporating ingredients such as spin-orbital coupling, magnetism, disorder, spontaneous time-reversal symmetry breaking, or laser radiation~\cite{liu2008quantum,yu2010quantized,qiao2010quantum,jiang2012quantum,PhysRevLett.109.116803,zhang2012electrically,xu2015intrinsic,tse2011quantum,xiao2011interface,cook2014double,RealizingHaldane2017,nandkishore2010quantum,mciver2020light}. In sharp contrast, our work aims at realizing the original Haldane model in pure graphene, where the emergence of staggered flux is provided by the unique environment of a vacuum chiral cavity.

In recent years, both experimental and theoretical investigations have shown the effective manipulation of material properties by utilizing vacuum cavities~\cite{appugliese2022breakdown,jarc2023cavity,chiral_cavity2021,RevQEDgas2021,review_cavity2022}. The underlying idea is that when a material is placed inside a cavity, the vacuum cavity's electromagnetic fluctuations become sufficiently strong to modify the material's physical properties~\cite{amelio2021optical,kiffner2019manipulating}. Compared to the conventional method of engineering material properties through laser irradiation~\cite{wang2013observation}, 
employing cavity vacuum to, for example, design band structures, conductance, and superconductivity offers the distinct advantage of avoiding heat generation and inexplicable nonequilibrium phenomena~\cite{cavitygraphene2011,cavityHall2021,Ashida2023,jiang2023engineering,cavityConductance2021,eckhardt2022quantum,cavityTransition2023,cavitySC2018,cavityMediated2019,cavitySC2019,cavitySC2020,cavitySC2022,Rokaj2023cavity}.

We unequivocally prove the emergence of the Haldane model from single-layer graphene within a chiral cavity~\cite{chiralmirrors2015,Baranov2020,Baranov2020nature,chiral_cavity2021,chiralQED2022}. The vacuum chiral cavity provides the source of time-reversal symmetry breaking without introducing magnetic fields. Physicists have recognized the significant potential of utilizing electromagnetic fluctuations in proximity to symmetry-broken materials or within symmetry-broken cavities to tailor the properties of materials or molecules \cite{jiang2019atmospherics,butcher2012casimir,jiang2019axial,Baranov2020,ke2023vacuum}. For example, studies have shown that electromagnetic fluctuations in chiral cavities is capable of engineering the topology of materials~\cite{chiral_cavity2021,chiralQED2022,espinosa2014semiconductor,CavityChern2019,jiang2023engineering}. Furthermore, references such as \cite{CavityChern2019,Ashida2023} have illustrated how quantum fluctuations within a chiral cavity can yield an integer Chern number in single-layer graphene. 
While scholars in Ref.~\cite{Ashida2023} speculate on the potential connection between this emerging phenomenon and the Haldane model, a definitive equivalence has yet to be established.

In the first part of this Letter, we derive the emergent stagger-flux gauge field in graphene induced by cavity electromagnetic fluctuations by employing the effective Hamiltonian in the asymptotically decoupled (AD) frame~\cite{CavityQED2021,Ashida2023}. With additional sublattice-symmetry breaking, we then propose an equilibrium route for generating valley polarization.
In the second part, we investigate graphene's reciprocal influence on cavity photons. We derive a neat formula to quantify the average photon number per unit cell associated with a Bloch electron. Through numerical analysis, we investigate the valley-dependent photonic distribution in various topological phases, revealing the potential to engineer cavity photonics through the topology of the light-matter interacting states.

\textit{Effective Hamiltonian in the AD frame}.—The advancement of strong light-matter coupling within cavities~\cite{ultrastrong2009,ultrastrong2012,ultrastrong2016,ultrastrongNP2017,ultrastrong2017,ultrastrong2019,ultrastrongRMP2019,Baranov2020,ultrastrong2020,Baranov2020nature} has led to the development of efficient theoretical frameworks across all interaction strengths~\cite{CavityQED2021,Ashida2023}.  Following the derivation outlined in Ref.~\cite{Ashida2023}, we introduce the effective Hamiltonian in the AD frame.  Considering non-interacting electrons in the material embedded in a single-mode chiral cavity, the Hamiltonian for an electron coupled with chiral photons in the Coulomb gauge is given by:
\begin{align} 
\hat{H}^C & =\frac{(\hat{\boldsymbol{p}}+e\hat{\mathbf{A}})^2}{2 m}+V(\boldsymbol{r})+\hbar \omega_c\left(\hat{a}^{\dagger} \hat{a}+\frac{1}{2}\right),\\
\hat{\mathbf{A}}&=A_0\left(\boldsymbol{\epsilon} \hat{a}+\boldsymbol{\epsilon}^* \hat{a}^{\dagger}\right)\label{eq:A_chiral}
\end{align}
Here, $e$ ($>0$) and $m$ represent the electric charge and mass, respectively. $\omega_c$ is the frequency of the cavity photons, and $\hat{\mathbf{A}}$ denotes the quantized vector potential of the photonic field. $a^\dagger$ ($a$) corresponds to the creation (annihilation) operator associated with the circular polarization $\boldsymbol{\epsilon}=(1,-i)^T/\sqrt{2}$. Additionally, $\boldsymbol{G}$ signifies a reciprocal lattice vector of the material, with a periodic potential $V(\boldsymbol{r})=\sum_{\boldsymbol{G}} V_{\boldsymbol{G}} e^{i \boldsymbol{G r}}$. The coupling strength of the light-matter interaction is captured by a dimensionless factor $g/\omega_c=eA_0(m\hbar\omega_c)^{-1/2}$. For $\hat{H}^C$, we apply a unitary transformation~\cite{LeeTrans1953,CavityQED2021,Ashida2023}:
\begin{align}
    \hat{U}=\exp\left[-\mathrm{i}\xi\frac{\hat{\boldsymbol{p}}}{\hbar}\cdot\hat{\boldsymbol{\pi}}\right],\label{eq:ADunitary}
\end{align}
and project the transformed Hamiltonian onto the photon vacuum as an approximation
\begin{align}
    \hat{H}^U&=\langle 0_\text{photon}|\hat{U}^\dagger \hat{H}^C\hat{U}|0_\text{photon}\rangle\nonumber\\&=\frac{\hat{p}^2}{2m_{\text{eff}}}+\widetilde{V}(\boldsymbol{r}+\frac{\xi^2}{2\hbar}\hat{\boldsymbol{p}}\times \boldsymbol{e}_z),\label{eq:H_U}
    \end{align}
where $m_{\text{eff}}=m(1+g^2/\omega_c^2)$, $\xi=\sqrt{\frac{\hbar}{m\omega_c}}g/\omega_c[1+(g/\omega_c)^2]^{-1}$, $\hat{\boldsymbol{\pi}}=\mathrm{i}\left(\boldsymbol{\epsilon} \hat{a}-\boldsymbol{\epsilon}^* \hat{a}^{\dagger}\right)$ and $\widetilde{V}(\boldsymbol{r})=\sum_{\boldsymbol{G}}V_{\boldsymbol{G}}e^{-\xi^2\boldsymbol{G}^2/4+\mathrm{i}\boldsymbol{G}\cdot\boldsymbol{r}}+\hbar\omega_c(1+g^2/\omega_c^2)/2$. The advantage to use this effective Hamiltonian $\hat{H}^U$ is that it does not contain the degree of freedom of photons. As $\xi$ is usually small~\cite{CavityQED2021,Ashida2023}, we can expand the Hamiltonian to the first-order of $\xi$ and obtain the effective Hamiltonian as
\begin{align}
     \hat{H}^U\approx\frac{[\hat{\boldsymbol{p}}+e\boldsymbol{A}(\boldsymbol{r})]^2}{2m_{\text{eff}}}+V_{\text{eff}}(\boldsymbol{r}).\label{eq:Heff}
\end{align}
Here, we define a gauge potential
\begin{align}
    \boldsymbol{A}(\boldsymbol{r})=\frac{\beta}{2e\omega_c}\boldsymbol{e}_z\times\nabla \widetilde{V},\label{eq:Aeff}
\end{align}
and an effective periodic scalar potential $V_{\text{eff}}(\boldsymbol{r})=\widetilde{V}(\boldsymbol{r})-\beta^3|\nabla\widetilde{V}(\boldsymbol{r})|^2/(8m g^2)$, where $\beta=(g/\omega_c)^2[1+(g/\omega_c)]^{-2}$.  
We observe an emergent vector potential $\boldsymbol{A}(\boldsymbol{r})$ associated with the periodic scalar potential function. It induces an effective magnetic field with the same crystalline periodicity, given by
\begin{align}
\boldsymbol{B}(\boldsymbol{r})=\nabla\times\boldsymbol{A}(\boldsymbol{r})=\frac{\beta}{2e\omega_c}\nabla^2\widetilde{V}\boldsymbol{e}_z.\label{eq:mageff}
\end{align}
Our work uses a two-dimensional potential to mimic graphene~\cite{TBgraphene2002,Ashida2023}, termed the continuum graphene model. The potential is described by
\begin{equation}
 V(\boldsymbol{r})=\sum_{i=1}^6\left(V_0+\frac{\Delta}{9}e^{-\mathrm{i}\boldsymbol{b}_i\cdot\boldsymbol{\delta}}\right)\exp(\mathrm{i}\boldsymbol{b}_i\cdot\boldsymbol{r}),\label{eq:potential}
\end{equation}
where $\boldsymbol{\delta}=(1,0)$, the reciprocal lattice vector $\boldsymbol{b}_i=4\pi(3a_0)^{-1}(\cos{2\pi i/3},\sin{2\pi i/3})$. Throughout our work, we set $a_0=1$, $\hbar=1$, and $e=1$. For later convenience, we fix the material parameters $V_0=m=1.5$ and the cavity parameter $\omega_c=2$. We show the potential function $V(\boldsymbol{r})$ and the effective potential $V_{\text{eff}}(\boldsymbol{r})$ at the coupling strength $g/\omega_c=1$ in the supplementary material (SM)~\cite{SM}. 

\begin{figure}
    \centering
    \includegraphics[width=0.48\textwidth]{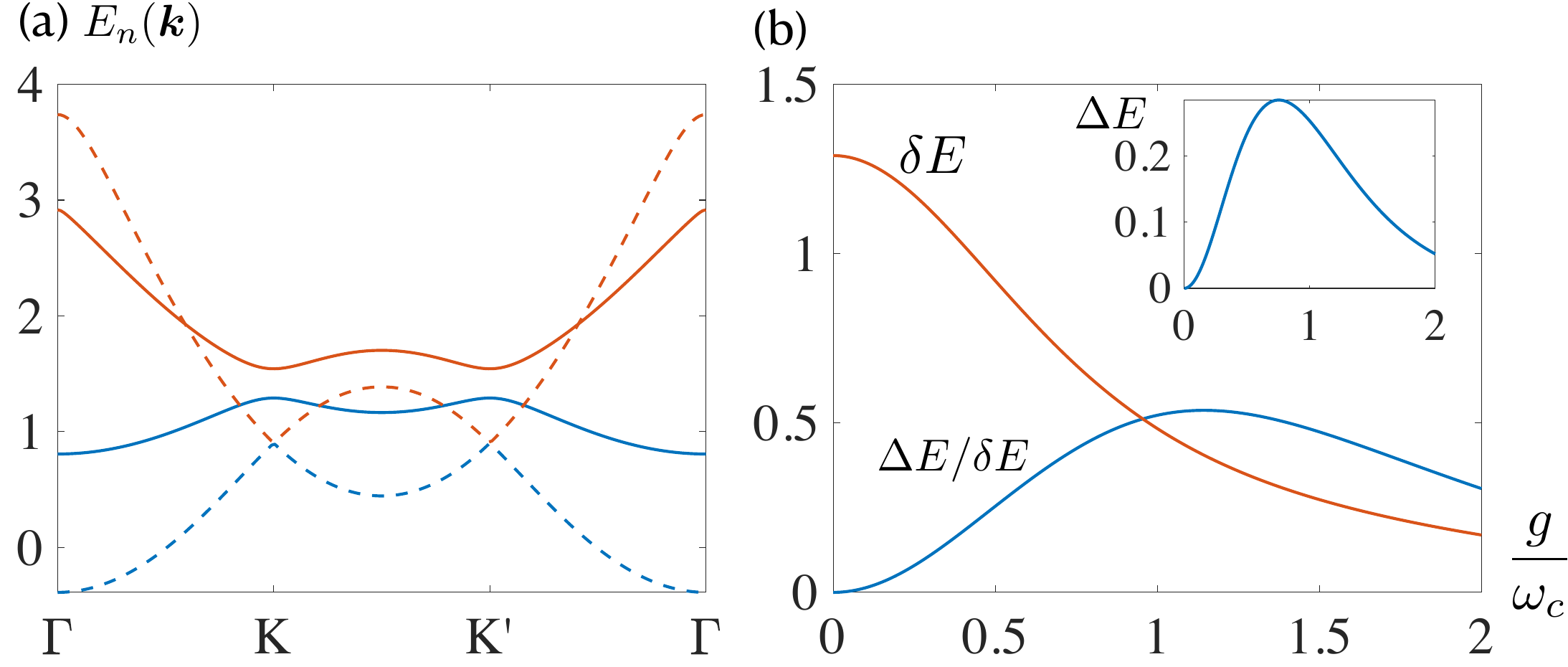}
    \caption{(a): The two lowest-energy bands of the bare graphene (dashed lines, $g/\omega_c=0$) and graphene in a chiral cavity (solid lines, $g/\omega_c=1$). We see that the Dirac points at valleys are opened for the cavity graphene. For numerical calculation, we have chosen $a_0=1$, $\hbar=1$, $e=1$, $\omega_c=2$, and the material parameters $V_0=m=1.5$ and $\Delta=0$. (b) Energy gap ($\Delta E$) of the lowest two bands and the lowest bandwidth ($\delta E$) for the cavity graphene as a function of the light-matter coupling strength $g/\omega_c$. }\label{fig:twobands}
\end{figure}

By plane-wave expansion technique~\cite{SM}, we compute the material spectra modified by the chiral cavity. The energy distributions are plotted along the trajectory $\Gamma=\boldsymbol{0}\to K=2(\boldsymbol{b}_1+\boldsymbol{b}_2)/3\to K'=(\boldsymbol{b}_1+\boldsymbol{b}_2)/3\to\Gamma$ in Figs.~\ref{fig:twobands} (a). Dashed (solid) lines represent the bare (cavity quantum-fluctuations-modified) graphene. We observe two Dirac points in the lowest two bands of the bare graphene, attributed to the nearest-neighbor hopping in honeycomb lattice. Additionally, the sublattice symmetry is slightly broken around the $\Gamma$ point because of the non-vanishing next-nearest hopping. In contrast, in the chiral cavity, the energy gap is opened around the $K$ and $K'$ valleys due to the time-reversal symmetry breaking. In Fig.~\ref{fig:twobands} (b), we show the first energy gap $\Delta E$ and width of the lowest band $\delta E$ as a function of $g/\omega_c$ ranging from 0 to 2. We observe that the bandwidth decreases as the coupling strength increases, and the first energy gap initially reaches a maximum value before decreasing.

\begin{figure}
    \centering
    \includegraphics[width=0.48\textwidth]{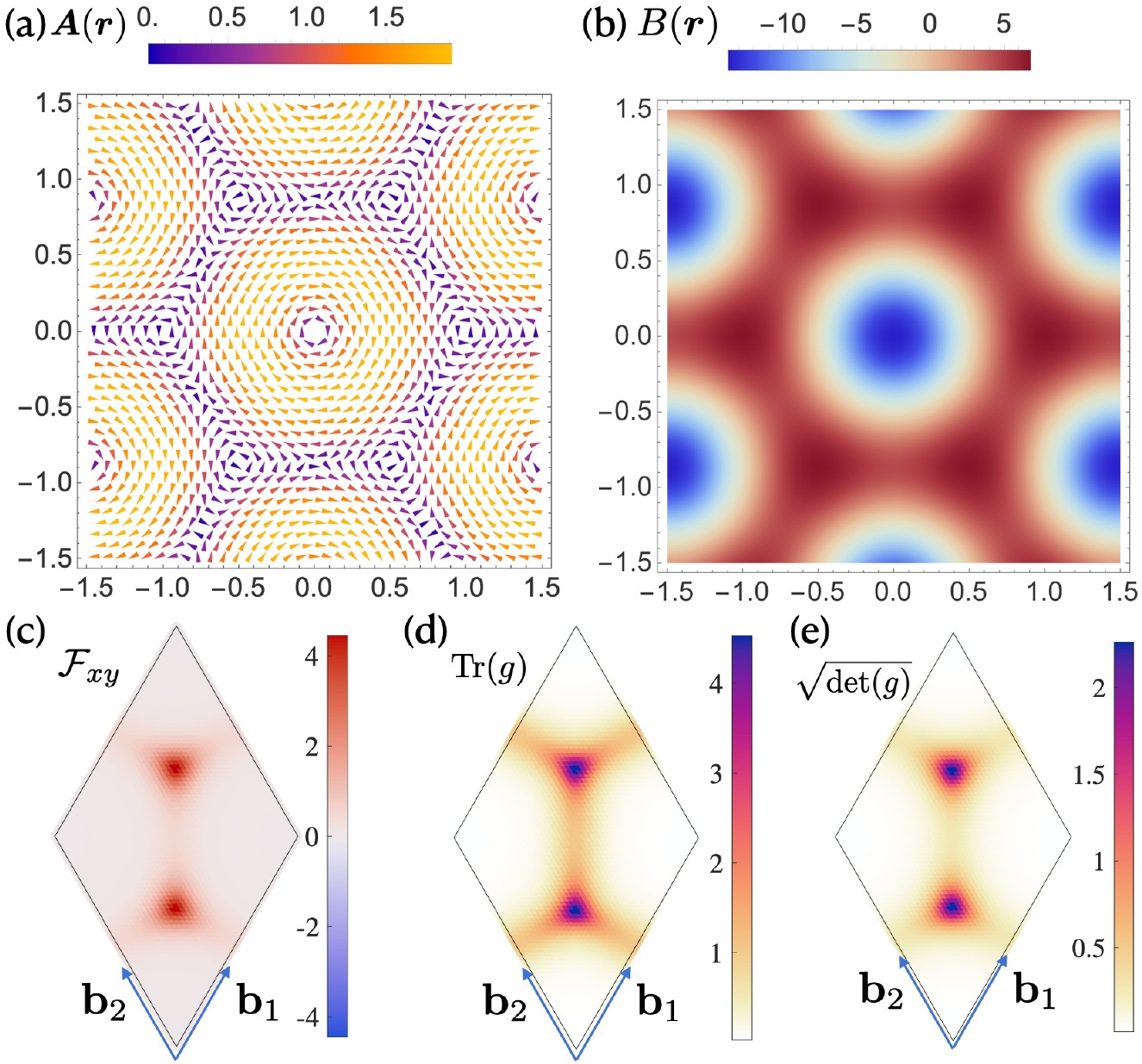}
    \caption{(a)-(b) Emergent vector potential (see Eq.~\ref{eq:Aeff}) and magnetic field (see Eq.~\ref{eq:mageff}) in the effective Hamiltonian Eq.~\ref{eq:Heff} of the cavity graphene. The magnetic field around the sub-lattices A and B (see the locations in Fig.~\ref{fig:twobands}) is positive, while it is negative around the center of one hexagon unit cell. (c)-(e): Quantum geometry of the lowest energy band of the emergent Haldane model. (c), (d) and (e) correspond to distributions of the Berry curvature, trace of the quantum metric, and volume form of the quantum metric over the Brillouin zone, respectively.}\label{fig:Haldane_cavity}
\end{figure}

\textit{Emergent Haldane model}.—We now prove that graphene in a chiral cavity is equivalent to the Haldane model (Details of the Haldane's model are reviewed in SM~\cite{SM}). The emergence of a gauge potential $\boldsymbol{A}(\boldsymbol{r})$ in the AD frame leads to a periodic magnetic field $B(\boldsymbol{r})$, thereby inducing a phase in the next-nearest hopping. Figs.~\ref{fig:Haldane_cavity} (a) and (b) show the emergent gauge potential and magnetic field, respectively. The emergent magnetic field exhibits negative around the center of a unit cell and positive around the lattice points. Due to periodicity, the magnetic flux in a whole unit cell is zero. A recent proposal for realizing the Haldane model in artificial graphene has also presented the same pattern of an external magnetic potential~\cite{HaldaneModel2018}. Therefore, the emergent magnetic field Eq.~(\ref{eq:mageff}) in the AD frame manifests the magnetic field pattern required by the Haldane model. 

Principally, the continuum graphene model can be described by a two-band tight-binding model with the nearest and second neighbor hopping amplitudes. We denote the hopping amplitude as $t_{ij}$, where $i$ and $j$ labels the orbitals at the places $\boldsymbol{r}_i$ and $\boldsymbol{r}_j$. Then, a straightforward way to incorporate gauge potential $\boldsymbol{A}(\boldsymbol{r})$ is by using the Peierls substitution, replacing $t_{ij}$ by $t_{ij}\,\exp(-\mathrm{i}\varphi_{ij})$~\cite{Peierlsub}, where the phase factor $\varphi_{ij}$ is defined as
\begin{align}
    \varphi_{ij}=-\frac{e}{\hbar}\int_{\boldsymbol{r}_i}^{\boldsymbol{r}_j}\boldsymbol{A}(\boldsymbol{r})\cdot d\boldsymbol{r}.
\end{align}
By using the potential Eq.~(\ref{eq:potential}), we find that $\varphi_{ij}=0$ for the nearest neighbor hopping and $\varphi_{ij}=\varphi=2\pi\beta V_0\exp(-4\pi^2\xi^2/9)/(\hbar\omega_c)$ for the next-nearest hopping, assuming the same parameters as those under Eq.~(\ref{eq:potential}), $\varphi=3\pi/4\exp(-\pi^2/27)$ at $g/\omega_c=1$. Adding such a phase to the second-neighbor hopping, we therefore prove that the minimum two-band model of the effective Hamiltonian Eq.~(\ref{eq:Heff}) is a Haldane model~\cite{SM,HaldaneModel}. 

To further analyze the emergent model in the chiral cavity, we study its topology and quantum geometry. The quantum geometric tensor, also known as the Fubini-Study metric, describes the associated topological and geometrical properties of the band structures~\cite{RMP_Resta1994,Resta2006,Resta2011,bandgeometry2014,Bandgeometry2017} (reviewed in SM~\cite{SM}). For a single-occupied band, it is defined as
\begin{align}
    Q_{\mu\nu}(\vec{k})=\langle\partial_{k_\mu}u_{n}(\vec{k})|\partial_{k_\nu}u_{n}(\vec{k})\rangle-\mathcal{A}_\mu(\vec{k})\mathcal{A}_\nu(\vec{k}),
\end{align}
where $|u_n(\vec{k})\rangle$ is the periodic part of the Bloch state and the Abelian Berry connection $\mathcal{A}_\mu(\vec{k})$ is defined as $\mathrm{i}\langle u_n(\vec{k})|\partial_{k_\mu}u_n(\vec{k})\rangle$. Specifically, the real part of the quantum geometric tensor $g_{\mu\nu}$ is termed the quantum metric. For the emergent Haldane model in the chiral cavity, we show the Berry curvature, trace, and volume form of the quantum metric over the lowest band for the cavity graphene in Figs.~\ref{fig:Haldane_cavity} (c)-(e) (with same parameters as used in Fig.~\ref{fig:twobands}). We observe that the Berry curvature over the BZ is positive, indicating a non-trivial Chern number $+1$. To confirm the exact value of the Chern number, we calculate the phase flow of the Wilson loops~\cite{SM}. Also, we have checked the inequality relation $\text{Tr}(g)\geq 2\sqrt{\text{det}(g)}\geq|\mathcal{F}_{xy}|$ between the trace of the quantum metric and the volume form $\sqrt{\text{det}(g)}$~\cite{bandgeometry2014,topobound2020,relation2021,hu2023anomalous}. This lower bound has recently been used to explain the superfluid phase stiffness of the twisted bilayer graphene~\cite{topobound2020,fracTBG2020,hu2023anomalous}. Unlike the Landau level, which has constant Berry curvature and quantum geometric metric over BZ, the Haldane model has a non-uniform distribution of the geometric tensor in the momentum space\cite{bandgeometry2014,relation2021}. For the quantum geometry of the tight-binding Haldane model, one can refer to SM~\cite{SM}.

\begin{figure}
    \centering
\includegraphics[width=0.49\textwidth]{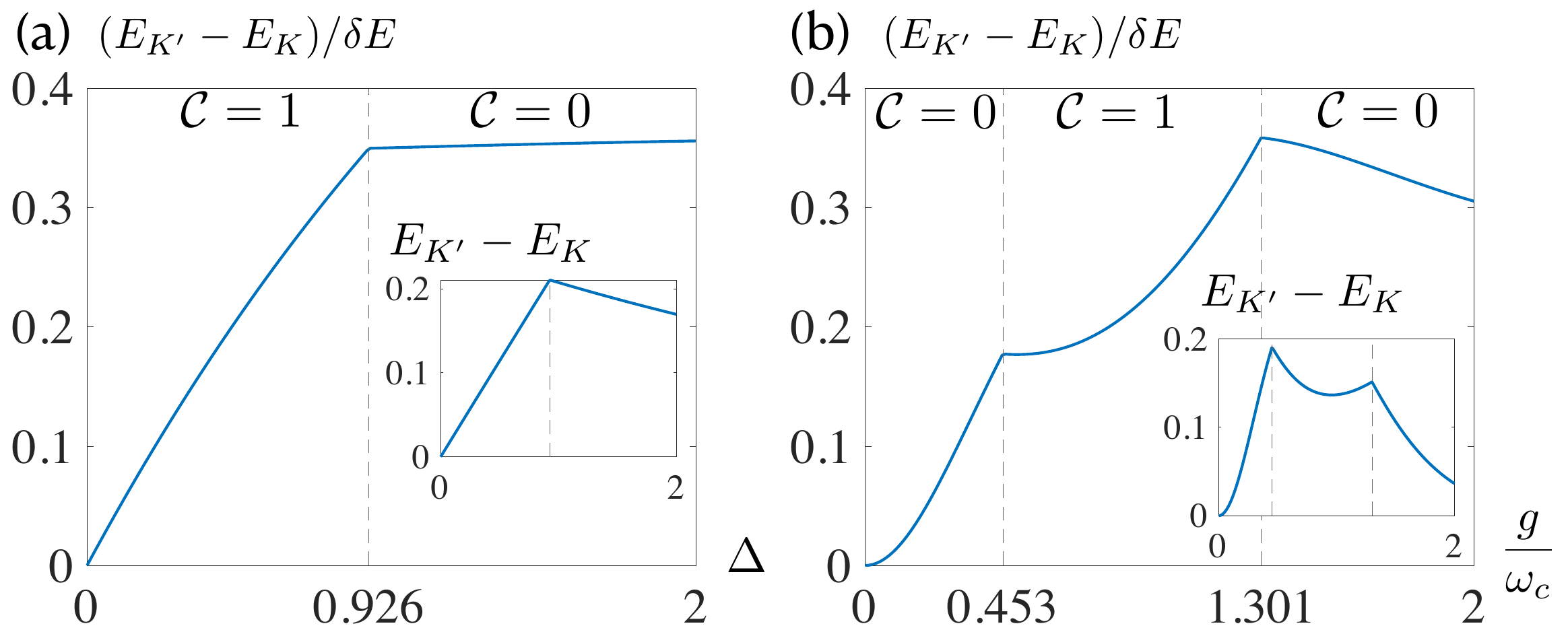}
    \caption{The ratio of valley energy split $E_K-E_{K'}$ to the lowest bandwidth $\delta E$, in the function of sub-lattice energy splitting $\Delta$ or $g/\omega_c$. In (a), $g/\omega_c=1$. In (b), $\Delta=0.6$. $\mathcal{C}$ is the Chern number of the labeled domains, and the dashed lines label the topological transition points.}
    \label{fig:valleysplit}
\end{figure}

\textit{Valley splitting of graphene in chiral cavities}.—In the previous analysis, our focus has been on cavity graphene with $C_2$ symmetry, where the energies of the two valleys around the $K$ and $K'$ points remain degenerate. However, when the sub-lattice energies split, the valley energies shift to different values. In Figs.~\ref{fig:valleysplit} (a) and (b) show the influence of onsite energy {difference $\Delta$ of A-sites and B-sites }and coupling strength $g/\omega_c$ on these valley energies. 
%We set $g/\omega_c=1$ and $\Delta=0.6$ in (a) and (b), respectively. 
In Fig. 3(a), we observe that the valley splitting ratio (defined as the ratio of valley energy split $E_{K'}-E_K$ to the bandwidth $\delta E$) increases continuously with respect to {the onsite energy splitting} $\Delta$ in the topological phase ($\Delta=0.926$ marks the topological phase transition point), and it remains at a plateau in the topologically trivial phase. For a specific system with a fixed $\Delta$, the valley splitting can be engineered by adjusting the coupling strength $g/\omega_c$.
In Fig. 3(b), we notice two topological phase transitions occurring as the coupling strength $g/\omega_c$ surpasses 0.453 and 1.301. Before the re-entrant transition at $g/\omega_c=1.301$, as observed in Ref.~\cite{Ashida2023}, the valley splitting ratio undergoes a continuous rise, only to subsequently decrease after entering the second topologically trivial phase. It is worth noting that the valley splitting ratio always experiences abrupt changes in its derivatives associated with the topological transition.

When the Fermi energy is between $E_K$ and $E_{K'}$, the number of electrons in the two valleys will differ when the light-matter interacting state holds. Thus, using the chiral cavity to generate the valley splitting can be a promising approach to generate valley polarization at the equilibrium. Compared with the classical optical pumping method in the mono-layer MoS$_2$~\cite{valley_symmetry2008,Valley_selection2012,Valleyoptical2012,Valley_helicity2012,ValleytronicsView2016,lightHall2010,lightHall2019,lightHall2020}, which is a non-equilibrium process, using quantum equilibrium state avoids heating issues~\cite{decoherence2011,decoherence2015,decoherence2016,decoherence2018,decoherence2023,Rokaj2023cavity}. 

Reversing the chirality of cavity photons should cause the valley energy differences to change smoothly between their value and inverse. To observe this transition, we introduce a hybrid polarization $\vec{\epsilon}\cos{\theta}+\vec{\epsilon}^\ast\sin{\theta}$ in a generalized cavity and analyze the valley splitting as a function of $\theta$ in SM~\cite{SM}. 

\textit{The photon number locking in the chiral cavity}.—Due to strong light-matter interaction, the ground state of the hybridized system deviates from the vacuum state of photons, leading to a stable confinement of a finite number of photons within the cavity. Upon transforming the state to the AD frame, we obtain the photon number $n_p$ (per unit cell) of the lowest-energy band bounded with a Bloch electron at momentum $\boldsymbol{k}$:
\begin{align}
    n_p(\boldsymbol{k})&=\langle \hat{a}^\dagger\hat{a}\rangle_C=\frac{\beta T_{\vec{k}}}{\hbar\omega_c},\label{eq:photon_n}
\end{align}
where the parameter $\beta$ has been defined under Eq.~(\ref{eq:Aeff}) and $T_{\vec{k}}$ is the average kinetic energy $\langle\hat{p}^2/(2m_\text{eff})\rangle_U$ of the Bloch state at momentum $\boldsymbol{k}$ in the AD frame. In deriving this result, we assumed that the photon part of the state in the AD frame is approximated as the zero-photon state~\cite{CavityQED2021,Ashida2023}. Eq.~(\ref{eq:photon_n}) shows that the average photon number in the chiral cavity is determined by the electron's kinetic energy rather than the state's total energy.

\begin{figure}
    \centering
\includegraphics[width=0.48\textwidth]{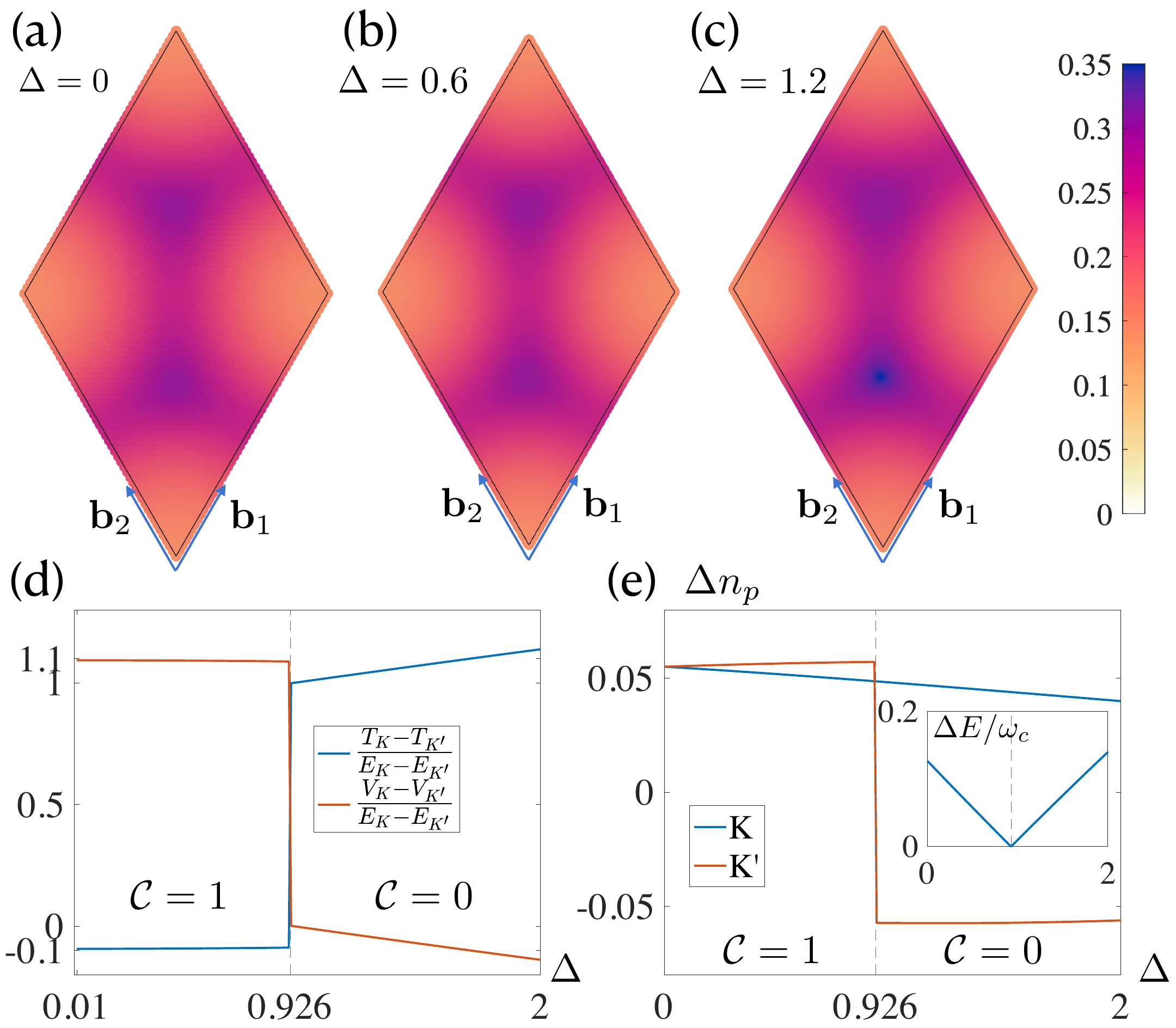}
    \caption{(a)-(c) The coupled photon number $n_p(\vec{k})$ by a Bloch electron at momentum $\boldsymbol{k}$ in the lowest-energy band of the cavity graphene with different sub-lattice energy splits $\Delta$ (labeled upon the figures). (d) The proportions of the kinetic and potential energy splits (in the AD frame) to the total energy difference between $K$ and $K'$ points in a function of $\Delta$ (from 0 to 2). The kinetic energy part dominates the total valley energy split in the region $\Delta<0.926$ and $\mathcal{C}=1$. (e) The change of the coupled photon numbers $\Delta n_p$ at $K$ and $K'$ points when the light-matter interacting state is excited from the lowest band to the upper band in a function of $\Delta$ at $g/\omega_c=1$. We see that the sign of the Chern number determines the sign of $\Delta n_p$. The inset figure shows the energy gap is much smaller than $\omega_c$.}
    \label{fig:photon}
\end{figure}

In Figs.~\ref{fig:photon} (a)-(c), we exhibit the locked photon number density $n_p(\boldsymbol{k})$ over the BZ of three systems with $\Delta=0,0.6,1.2$. Here, we focus on the strong coupling regime and fix $g/\omega_c$ to be 1, and we observe that the electrons at $K$ and $K'$ points can couple more photons than other states. 

While the valley energy splitting occurs in both (b) and (c), the imbalance in valley photon numbers (i.e., $n_p(K)-n_p(K')$) is more noticeable in (c) than in (b). This discrepancy is attributed to the system's topology: cases (a) and (b) exhibit a +1 Chern number, whereas case (c) has a zero Chern number. In Fig.~\ref{fig:photon} (d), we plot the kinetic and potential energy split of the lowest-band valleys in the AD frame as a function of {the sub-lattice energy splitting
} $\Delta$. Notably, when the Chern number $\mathcal{C}=1$, the energy difference between $K$ and $K'$ points is primarily due to the potential energy $V_{\vec{k}}=\langle\widetilde{V}(\boldsymbol{r}+\frac{\xi^2}{2\hbar}\hat{\boldsymbol{p}}\times \boldsymbol{e}_z)\rangle_U$ {of the Bloch state at $\boldsymbol{k}$}(where $\widetilde{V}(\boldsymbol{r})$ is defined under Eq.~(\ref{eq:H_U})), while the kinetic energy split dominates the total valley energy split after the transition point. Thus, the distinct kinetic energy splittings in different topological phases explain the varying photon number imbalances at $K$ and $K'$.

We now focus on the photonic features associated the two valleys induced by interband excitation. We define $\Delta n_p(\vec{k})$ as the change in photon numbers during the transition from the lowest to the upper band at momentum $\vec{k}$.  In Fig.~\ref{fig:photon} (e), we display $\Delta n_p(K)$ and $\Delta n_p(K')$ as functions of sub-lattice split $\Delta$ at fixed light-matter coupling $g/\omega_c=1$. Here, we consider the {off-resonant} case where $\Delta E\ll\omega_c$~\cite{Ashida2023,yaowang2023} (see the inset figure in (e)), meaning that the interband excitation costs less energy than photon excitation in the AD frame. Fig.~\ref{fig:photon} (e) shows $\Delta n_p(K')$ maintains its sign across the entire phase diagram whereas $\Delta n_p(K')$ flips sign at topological transition point~\cite{HaldaneModel,SM}. 
%Recalling that the Berry curvature of $K'$ point reverses the sign at the transition point~\cite{HaldaneModel,SM}, we see that it also determines the sign of $\Delta n_p$. 
This indicates the cavity vacuum annihilates (produces) photons during the interband excitation at $K'$ in the topological (trivial) phase. Further analysis on the photon number change as a function of the coupling strength $g/\omega_c$ can be referred to SM~\cite{SM}.

\textit{Conclusion}.—We have derived an emergent Haldane model from a continuum graphene model in a chiral cavity and analyzed the photon-valley locking by the lowest-energy light-matter interacting band. Particularly, we have unambiguously explained how the non-trivial topology of cavity graphene emerges from an emergent periodic magnetic field. In our investigations of valley energy splitting in models with finite sublattice energy difference, we propose an equilibrium pathway to achieve valley polarization. Furthermore, we have examined the reciprocal influence of graphene on cavity photons, revealing distinct characteristics of photon-valley locking in both topological and trivial phases.
In addition, we have obtained the valley-specified photon number change during interband excitation.
Unlike previous studies focusing solely on total energy bands, our work highlights the significance of kinetic bands in manipulating cavity light-matter hybridized quantum states related to geometric and topological properties.

\textit{Acknowledgements}.—We are grateful for useful discussions with Hans Hansson. The authors are sponsored by National Natural Science Foundation of China (NSFC) under Grant No. 23Z031504628, Jiaoda2030 Program Grant No.WH510363001, TDLI starting up grant, and Innovation Program for Quantum Science and Technology Grant No.2021ZD0301900.

\bibliography{ref.bib}

\begin{thebibliography}{90}%
\makeatletter
\providecommand \@ifxundefined [1]{%
 \@ifx{#1\undefined}
}%
\providecommand \@ifnum [1]{%
 \ifnum #1\expandafter \@firstoftwo
 \else \expandafter \@secondoftwo
 \fi
}%
\providecommand \@ifx [1]{%
 \ifx #1\expandafter \@firstoftwo
 \else \expandafter \@secondoftwo
 \fi
}%
\providecommand \natexlab [1]{#1}%
\providecommand \enquote  [1]{``#1''}%
\providecommand \bibnamefont  [1]{#1}%
\providecommand \bibfnamefont [1]{#1}%
\providecommand \citenamefont [1]{#1}%
\providecommand \href@noop [0]{\@secondoftwo}%
\providecommand \href [0]{\begingroup \@sanitize@url \@href}%
\providecommand \@href[1]{\@@startlink{#1}\@@href}%
\providecommand \@@href[1]{\endgroup#1\@@endlink}%
\providecommand \@sanitize@url [0]{\catcode `\\12\catcode `\$12\catcode
  `\&12\catcode `\#12\catcode `\^12\catcode `\_12\catcode `\%12\relax}%
\providecommand \@@startlink[1]{}%
\providecommand \@@endlink[0]{}%
\providecommand \url  [0]{\begingroup\@sanitize@url \@url }%
\providecommand \@url [1]{\endgroup\@href {#1}{\urlprefix }}%
\providecommand \urlprefix  [0]{URL }%
\providecommand \Eprint [0]{\href }%
\providecommand \doibase [0]{https://doi.org/}%
\providecommand \selectlanguage [0]{\@gobble}%
\providecommand \bibinfo  [0]{\@secondoftwo}%
\providecommand \bibfield  [0]{\@secondoftwo}%
\providecommand \translation [1]{[#1]}%
\providecommand \BibitemOpen [0]{}%
\providecommand \bibitemStop [0]{}%
\providecommand \bibitemNoStop [0]{.\EOS\space}%
\providecommand \EOS [0]{\spacefactor3000\relax}%
\providecommand \BibitemShut  [1]{\csname bibitem#1\endcsname}%
\let\auto@bib@innerbib\@empty
%</preamble>
\bibitem [{\citenamefont {Hofstadter}(1976)}]{Mlattice1976}%
  \BibitemOpen
  \bibfield  {author} {\bibinfo {author} {\bibfnamefont {D.~R.}\ \bibnamefont
  {Hofstadter}},\ }\bibfield  {title} {\bibinfo {title} {Energy levels and wave
  functions of bloch electrons in rational and irrational magnetic fields},\
  }\href {https://doi.org/10.1103/PhysRevB.14.2239} {\bibfield  {journal}
  {\bibinfo  {journal} {Phys. Rev. B}\ }\textbf {\bibinfo {volume} {14}},\
  \bibinfo {pages} {2239} (\bibinfo {year} {1976})}\BibitemShut {NoStop}%
\bibitem [{\citenamefont {Haldane}(1988)}]{HaldaneModel}%
  \BibitemOpen
  \bibfield  {author} {\bibinfo {author} {\bibfnamefont {F.~D.~M.}\
  \bibnamefont {Haldane}},\ }\bibfield  {title} {\bibinfo {title} {Model for a
  quantum hall effect without landau levels: Condensed-matter realization of
  the "parity anomaly"},\ }\href {https://doi.org/10.1103/PhysRevLett.61.2015}
  {\bibfield  {journal} {\bibinfo  {journal} {Phys. Rev. Lett.}\ }\textbf
  {\bibinfo {volume} {61}},\ \bibinfo {pages} {2015} (\bibinfo {year}
  {1988})}\BibitemShut {NoStop}%
\bibitem [{\citenamefont {Qi}\ and\ \citenamefont
  {Zhang}(2011)}]{shouchengReview2011}%
  \BibitemOpen
  \bibfield  {author} {\bibinfo {author} {\bibfnamefont {X.-L.}\ \bibnamefont
  {Qi}}\ and\ \bibinfo {author} {\bibfnamefont {S.-C.}\ \bibnamefont {Zhang}},\
  }\bibfield  {title} {\bibinfo {title} {Topological insulators and
  superconductors},\ }\href {https://doi.org/10.1103/RevModPhys.83.1057}
  {\bibfield  {journal} {\bibinfo  {journal} {Rev. Mod. Phys.}\ }\textbf
  {\bibinfo {volume} {83}},\ \bibinfo {pages} {1057} (\bibinfo {year}
  {2011})}\BibitemShut {NoStop}%
\bibitem [{\citenamefont {Chang}\ \emph {et~al.}(2023)\citenamefont {Chang},
  \citenamefont {Liu},\ and\ \citenamefont {MacDonald}}]{RMP_QAHE2023}%
  \BibitemOpen
  \bibfield  {author} {\bibinfo {author} {\bibfnamefont {C.-Z.}\ \bibnamefont
  {Chang}}, \bibinfo {author} {\bibfnamefont {C.-X.}\ \bibnamefont {Liu}},\
  and\ \bibinfo {author} {\bibfnamefont {A.~H.}\ \bibnamefont {MacDonald}},\
  }\bibfield  {title} {\bibinfo {title} {Colloquium: Quantum anomalous hall
  effect},\ }\href {https://doi.org/10.1103/RevModPhys.95.011002} {\bibfield
  {journal} {\bibinfo  {journal} {Rev. Mod. Phys.}\ }\textbf {\bibinfo {volume}
  {95}},\ \bibinfo {pages} {011002} (\bibinfo {year} {2023})}\BibitemShut
  {NoStop}%
\bibitem [{\citenamefont {Thouless}\ \emph {et~al.}(1982)\citenamefont
  {Thouless}, \citenamefont {Kohmoto}, \citenamefont {Nightingale},\ and\
  \citenamefont {den Nijs}}]{TKNN}%
  \BibitemOpen
  \bibfield  {author} {\bibinfo {author} {\bibfnamefont {D.~J.}\ \bibnamefont
  {Thouless}}, \bibinfo {author} {\bibfnamefont {M.}~\bibnamefont {Kohmoto}},
  \bibinfo {author} {\bibfnamefont {M.~P.}\ \bibnamefont {Nightingale}},\ and\
  \bibinfo {author} {\bibfnamefont {M.}~\bibnamefont {den Nijs}},\ }\bibfield
  {title} {\bibinfo {title} {Quantized hall conductance in a two-dimensional
  periodic potential},\ }\href@noop {} {\bibfield  {journal} {\bibinfo
  {journal} {Phys. Rev. Lett.}\ }\textbf {\bibinfo {volume} {49}},\ \bibinfo
  {pages} {405} (\bibinfo {year} {1982})}\BibitemShut {NoStop}%
\bibitem [{\citenamefont {Qi}\ \emph {et~al.}(2006)\citenamefont {Qi},
  \citenamefont {Wu},\ and\ \citenamefont {Zhang}}]{QWZmodel2006}%
  \BibitemOpen
  \bibfield  {author} {\bibinfo {author} {\bibfnamefont {X.-L.}\ \bibnamefont
  {Qi}}, \bibinfo {author} {\bibfnamefont {Y.-S.}\ \bibnamefont {Wu}},\ and\
  \bibinfo {author} {\bibfnamefont {S.-C.}\ \bibnamefont {Zhang}},\ }\bibfield
  {title} {\bibinfo {title} {Topological quantization of the spin hall effect
  in two-dimensional paramagnetic semiconductors},\ }\href
  {https://doi.org/10.1103/PhysRevB.74.085308} {\bibfield  {journal} {\bibinfo
  {journal} {Phys. Rev. B}\ }\textbf {\bibinfo {volume} {74}},\ \bibinfo
  {pages} {085308} (\bibinfo {year} {2006})}\BibitemShut {NoStop}%
\bibitem [{\citenamefont {Xiao}\ \emph {et~al.}(2010)\citenamefont {Xiao},
  \citenamefont {Chang},\ and\ \citenamefont {Niu}}]{RMP2010qian}%
  \BibitemOpen
  \bibfield  {author} {\bibinfo {author} {\bibfnamefont {D.}~\bibnamefont
  {Xiao}}, \bibinfo {author} {\bibfnamefont {M.-C.}\ \bibnamefont {Chang}},\
  and\ \bibinfo {author} {\bibfnamefont {Q.}~\bibnamefont {Niu}},\ }\bibfield
  {title} {\bibinfo {title} {Berry phase effects on electronic properties},\
  }\href {https://doi.org/10.1103/RevModPhys.82.1959} {\bibfield  {journal}
  {\bibinfo  {journal} {Rev. Mod. Phys.}\ }\textbf {\bibinfo {volume} {82}},\
  \bibinfo {pages} {1959} (\bibinfo {year} {2010})}\BibitemShut {NoStop}%
\bibitem [{\citenamefont {Jotzu}\ \emph {et~al.}(2014)\citenamefont {Jotzu},
  \citenamefont {Messer}, \citenamefont {Desbuquois}, \citenamefont {Lebrat},
  \citenamefont {Uehlinger}, \citenamefont {Greif},\ and\ \citenamefont
  {Esslinger}}]{Haldane_experiment2014}%
  \BibitemOpen
  \bibfield  {author} {\bibinfo {author} {\bibfnamefont {G.}~\bibnamefont
  {Jotzu}}, \bibinfo {author} {\bibfnamefont {M.}~\bibnamefont {Messer}},
  \bibinfo {author} {\bibfnamefont {R.}~\bibnamefont {Desbuquois}}, \bibinfo
  {author} {\bibfnamefont {M.}~\bibnamefont {Lebrat}}, \bibinfo {author}
  {\bibfnamefont {T.}~\bibnamefont {Uehlinger}}, \bibinfo {author}
  {\bibfnamefont {D.}~\bibnamefont {Greif}},\ and\ \bibinfo {author}
  {\bibfnamefont {T.}~\bibnamefont {Esslinger}},\ }\bibfield  {title} {\bibinfo
  {title} {Experimental realization of the topological haldane model with
  ultracold fermions},\ }\href {https://doi.org/10.1038/nature13915} {\bibfield
   {journal} {\bibinfo  {journal} {Nature}\ }\textbf {\bibinfo {volume}
  {515}},\ \bibinfo {pages} {237} (\bibinfo {year} {2014})}\BibitemShut
  {NoStop}%
\bibitem [{\citenamefont {Roushan}\ \emph {et~al.}(2014)\citenamefont
  {Roushan}, \citenamefont {Neill}, \citenamefont {Chen}, \citenamefont
  {Kolodrubetz}, \citenamefont {Quintana}, \citenamefont {Leung}, \citenamefont
  {Fang}, \citenamefont {Barends}, \citenamefont {Campbell}, \citenamefont
  {Chen} \emph {et~al.}}]{roushan2014observation}%
  \BibitemOpen
  \bibfield  {author} {\bibinfo {author} {\bibfnamefont {P.}~\bibnamefont
  {Roushan}}, \bibinfo {author} {\bibfnamefont {C.}~\bibnamefont {Neill}},
  \bibinfo {author} {\bibfnamefont {Y.}~\bibnamefont {Chen}}, \bibinfo {author}
  {\bibfnamefont {M.}~\bibnamefont {Kolodrubetz}}, \bibinfo {author}
  {\bibfnamefont {C.}~\bibnamefont {Quintana}}, \bibinfo {author}
  {\bibfnamefont {N.}~\bibnamefont {Leung}}, \bibinfo {author} {\bibfnamefont
  {M.}~\bibnamefont {Fang}}, \bibinfo {author} {\bibfnamefont {R.}~\bibnamefont
  {Barends}}, \bibinfo {author} {\bibfnamefont {B.}~\bibnamefont {Campbell}},
  \bibinfo {author} {\bibfnamefont {Z.}~\bibnamefont {Chen}}, \emph {et~al.},\
  }\bibfield  {title} {\bibinfo {title} {Observation of topological transitions
  in interacting quantum circuits},\ }\href@noop {} {\bibfield  {journal}
  {\bibinfo  {journal} {Nature}\ }\textbf {\bibinfo {volume} {515}},\ \bibinfo
  {pages} {241} (\bibinfo {year} {2014})}\BibitemShut {NoStop}%
\bibitem [{\citenamefont {Liu}\ \emph {et~al.}(2008)\citenamefont {Liu},
  \citenamefont {Qi}, \citenamefont {Dai}, \citenamefont {Fang},\ and\
  \citenamefont {Zhang}}]{liu2008quantum}%
  \BibitemOpen
  \bibfield  {author} {\bibinfo {author} {\bibfnamefont {C.-X.}\ \bibnamefont
  {Liu}}, \bibinfo {author} {\bibfnamefont {X.-L.}\ \bibnamefont {Qi}},
  \bibinfo {author} {\bibfnamefont {X.}~\bibnamefont {Dai}}, \bibinfo {author}
  {\bibfnamefont {Z.}~\bibnamefont {Fang}},\ and\ \bibinfo {author}
  {\bibfnamefont {S.-C.}\ \bibnamefont {Zhang}},\ }\bibfield  {title} {\bibinfo
  {title} {Quantum anomalous hall effect in hg 1- y mn y te quantum wells},\
  }\href@noop {} {\bibfield  {journal} {\bibinfo  {journal} {Physical review
  letters}\ }\textbf {\bibinfo {volume} {101}},\ \bibinfo {pages} {146802}
  (\bibinfo {year} {2008})}\BibitemShut {NoStop}%
\bibitem [{\citenamefont {Yu}\ \emph {et~al.}(2010)\citenamefont {Yu},
  \citenamefont {Zhang}, \citenamefont {Zhang}, \citenamefont {Zhang},
  \citenamefont {Dai},\ and\ \citenamefont {Fang}}]{yu2010quantized}%
  \BibitemOpen
  \bibfield  {author} {\bibinfo {author} {\bibfnamefont {R.}~\bibnamefont
  {Yu}}, \bibinfo {author} {\bibfnamefont {W.}~\bibnamefont {Zhang}}, \bibinfo
  {author} {\bibfnamefont {H.-J.}\ \bibnamefont {Zhang}}, \bibinfo {author}
  {\bibfnamefont {S.-C.}\ \bibnamefont {Zhang}}, \bibinfo {author}
  {\bibfnamefont {X.}~\bibnamefont {Dai}},\ and\ \bibinfo {author}
  {\bibfnamefont {Z.}~\bibnamefont {Fang}},\ }\bibfield  {title} {\bibinfo
  {title} {Quantized anomalous hall effect in magnetic topological
  insulators},\ }\href@noop {} {\bibfield  {journal} {\bibinfo  {journal}
  {science}\ }\textbf {\bibinfo {volume} {329}},\ \bibinfo {pages} {61}
  (\bibinfo {year} {2010})}\BibitemShut {NoStop}%
\bibitem [{\citenamefont {Qiao}\ \emph {et~al.}(2010)\citenamefont {Qiao},
  \citenamefont {Yang}, \citenamefont {Feng}, \citenamefont {Tse},
  \citenamefont {Ding}, \citenamefont {Yao}, \citenamefont {Wang},\ and\
  \citenamefont {Niu}}]{qiao2010quantum}%
  \BibitemOpen
  \bibfield  {author} {\bibinfo {author} {\bibfnamefont {Z.}~\bibnamefont
  {Qiao}}, \bibinfo {author} {\bibfnamefont {S.~A.}\ \bibnamefont {Yang}},
  \bibinfo {author} {\bibfnamefont {W.}~\bibnamefont {Feng}}, \bibinfo {author}
  {\bibfnamefont {W.-K.}\ \bibnamefont {Tse}}, \bibinfo {author} {\bibfnamefont
  {J.}~\bibnamefont {Ding}}, \bibinfo {author} {\bibfnamefont {Y.}~\bibnamefont
  {Yao}}, \bibinfo {author} {\bibfnamefont {J.}~\bibnamefont {Wang}},\ and\
  \bibinfo {author} {\bibfnamefont {Q.}~\bibnamefont {Niu}},\ }\bibfield
  {title} {\bibinfo {title} {Quantum anomalous hall effect in graphene from
  rashba and exchange effects},\ }\href@noop {} {\bibfield  {journal} {\bibinfo
   {journal} {Physical Review B}\ }\textbf {\bibinfo {volume} {82}},\ \bibinfo
  {pages} {161414} (\bibinfo {year} {2010})}\BibitemShut {NoStop}%
\bibitem [{\citenamefont {Jiang}\ \emph
  {et~al.}(2012{\natexlab{a}})\citenamefont {Jiang}, \citenamefont {Qiao},
  \citenamefont {Liu},\ and\ \citenamefont {Niu}}]{jiang2012quantum}%
  \BibitemOpen
  \bibfield  {author} {\bibinfo {author} {\bibfnamefont {H.}~\bibnamefont
  {Jiang}}, \bibinfo {author} {\bibfnamefont {Z.}~\bibnamefont {Qiao}},
  \bibinfo {author} {\bibfnamefont {H.}~\bibnamefont {Liu}},\ and\ \bibinfo
  {author} {\bibfnamefont {Q.}~\bibnamefont {Niu}},\ }\bibfield  {title}
  {\bibinfo {title} {Quantum anomalous hall effect with tunable chern number in
  magnetic topological insulator film},\ }\href@noop {} {\bibfield  {journal}
  {\bibinfo  {journal} {Physical Review B}\ }\textbf {\bibinfo {volume} {85}},\
  \bibinfo {pages} {045445} (\bibinfo {year} {2012}{\natexlab{a}})}\BibitemShut
  {NoStop}%
\bibitem [{\citenamefont {Jiang}\ \emph
  {et~al.}(2012{\natexlab{b}})\citenamefont {Jiang}, \citenamefont {Qiao},
  \citenamefont {Liu}, \citenamefont {Shi},\ and\ \citenamefont
  {Niu}}]{PhysRevLett.109.116803}%
  \BibitemOpen
  \bibfield  {author} {\bibinfo {author} {\bibfnamefont {H.}~\bibnamefont
  {Jiang}}, \bibinfo {author} {\bibfnamefont {Z.}~\bibnamefont {Qiao}},
  \bibinfo {author} {\bibfnamefont {H.}~\bibnamefont {Liu}}, \bibinfo {author}
  {\bibfnamefont {J.}~\bibnamefont {Shi}},\ and\ \bibinfo {author}
  {\bibfnamefont {Q.}~\bibnamefont {Niu}},\ }\bibfield  {title} {\bibinfo
  {title} {Stabilizing topological phases in graphene via random adsorption},\
  }\href {https://doi.org/10.1103/PhysRevLett.109.116803} {\bibfield  {journal}
  {\bibinfo  {journal} {Phys. Rev. Lett.}\ }\textbf {\bibinfo {volume} {109}},\
  \bibinfo {pages} {116803} (\bibinfo {year} {2012}{\natexlab{b}})}\BibitemShut
  {NoStop}%
\bibitem [{\citenamefont {Zhang}\ \emph {et~al.}(2012)\citenamefont {Zhang},
  \citenamefont {Lazo}, \citenamefont {Bl{\"u}gel}, \citenamefont {Heinze},\
  and\ \citenamefont {Mokrousov}}]{zhang2012electrically}%
  \BibitemOpen
  \bibfield  {author} {\bibinfo {author} {\bibfnamefont {H.}~\bibnamefont
  {Zhang}}, \bibinfo {author} {\bibfnamefont {C.}~\bibnamefont {Lazo}},
  \bibinfo {author} {\bibfnamefont {S.}~\bibnamefont {Bl{\"u}gel}}, \bibinfo
  {author} {\bibfnamefont {S.}~\bibnamefont {Heinze}},\ and\ \bibinfo {author}
  {\bibfnamefont {Y.}~\bibnamefont {Mokrousov}},\ }\bibfield  {title} {\bibinfo
  {title} {Electrically tunable quantum anomalous hall effect in graphene
  decorated by 5 d transition-metal adatoms},\ }\href@noop {} {\bibfield
  {journal} {\bibinfo  {journal} {Physical review letters}\ }\textbf {\bibinfo
  {volume} {108}},\ \bibinfo {pages} {056802} (\bibinfo {year}
  {2012})}\BibitemShut {NoStop}%
\bibitem [{\citenamefont {Xu}\ \emph {et~al.}(2015)\citenamefont {Xu},
  \citenamefont {Lian},\ and\ \citenamefont {Zhang}}]{xu2015intrinsic}%
  \BibitemOpen
  \bibfield  {author} {\bibinfo {author} {\bibfnamefont {G.}~\bibnamefont
  {Xu}}, \bibinfo {author} {\bibfnamefont {B.}~\bibnamefont {Lian}},\ and\
  \bibinfo {author} {\bibfnamefont {S.-C.}\ \bibnamefont {Zhang}},\ }\bibfield
  {title} {\bibinfo {title} {Intrinsic quantum anomalous hall effect in the
  kagome lattice cs 2 limn 3 f 12},\ }\href@noop {} {\bibfield  {journal}
  {\bibinfo  {journal} {Physical review letters}\ }\textbf {\bibinfo {volume}
  {115}},\ \bibinfo {pages} {186802} (\bibinfo {year} {2015})}\BibitemShut
  {NoStop}%
\bibitem [{\citenamefont {Tse}\ \emph {et~al.}(2011)\citenamefont {Tse},
  \citenamefont {Qiao}, \citenamefont {Yao}, \citenamefont {MacDonald},\ and\
  \citenamefont {Niu}}]{tse2011quantum}%
  \BibitemOpen
  \bibfield  {author} {\bibinfo {author} {\bibfnamefont {W.-K.}\ \bibnamefont
  {Tse}}, \bibinfo {author} {\bibfnamefont {Z.}~\bibnamefont {Qiao}}, \bibinfo
  {author} {\bibfnamefont {Y.}~\bibnamefont {Yao}}, \bibinfo {author}
  {\bibfnamefont {A.}~\bibnamefont {MacDonald}},\ and\ \bibinfo {author}
  {\bibfnamefont {Q.}~\bibnamefont {Niu}},\ }\bibfield  {title} {\bibinfo
  {title} {Quantum anomalous hall effect in single-layer and bilayer
  graphene},\ }\href@noop {} {\bibfield  {journal} {\bibinfo  {journal}
  {Physical Review B}\ }\textbf {\bibinfo {volume} {83}},\ \bibinfo {pages}
  {155447} (\bibinfo {year} {2011})}\BibitemShut {NoStop}%
\bibitem [{\citenamefont {Xiao}\ \emph {et~al.}(2011)\citenamefont {Xiao},
  \citenamefont {Zhu}, \citenamefont {Ran}, \citenamefont {Nagaosa},\ and\
  \citenamefont {Okamoto}}]{xiao2011interface}%
  \BibitemOpen
  \bibfield  {author} {\bibinfo {author} {\bibfnamefont {D.}~\bibnamefont
  {Xiao}}, \bibinfo {author} {\bibfnamefont {W.}~\bibnamefont {Zhu}}, \bibinfo
  {author} {\bibfnamefont {Y.}~\bibnamefont {Ran}}, \bibinfo {author}
  {\bibfnamefont {N.}~\bibnamefont {Nagaosa}},\ and\ \bibinfo {author}
  {\bibfnamefont {S.}~\bibnamefont {Okamoto}},\ }\bibfield  {title} {\bibinfo
  {title} {Interface engineering of quantum hall effects in digital transition
  metal oxide heterostructures},\ }\href@noop {} {\bibfield  {journal}
  {\bibinfo  {journal} {Nature communications}\ }\textbf {\bibinfo {volume}
  {2}},\ \bibinfo {pages} {596} (\bibinfo {year} {2011})}\BibitemShut {NoStop}%
\bibitem [{\citenamefont {Cook}\ and\ \citenamefont
  {Paramekanti}(2014)}]{cook2014double}%
  \BibitemOpen
  \bibfield  {author} {\bibinfo {author} {\bibfnamefont {A.~M.}\ \bibnamefont
  {Cook}}\ and\ \bibinfo {author} {\bibfnamefont {A.}~\bibnamefont
  {Paramekanti}},\ }\bibfield  {title} {\bibinfo {title} {Double perovskite
  heterostructures: Magnetism, chern bands, and chern insulators},\ }\href@noop
  {} {\bibfield  {journal} {\bibinfo  {journal} {Physical Review Letters}\
  }\textbf {\bibinfo {volume} {113}},\ \bibinfo {pages} {077203} (\bibinfo
  {year} {2014})}\BibitemShut {NoStop}%
\bibitem [{\citenamefont {Kim}\ and\ \citenamefont
  {Kee}(2017)}]{RealizingHaldane2017}%
  \BibitemOpen
  \bibfield  {author} {\bibinfo {author} {\bibfnamefont {H.-S.}\ \bibnamefont
  {Kim}}\ and\ \bibinfo {author} {\bibfnamefont {H.-Y.}\ \bibnamefont {Kee}},\
  }\bibfield  {title} {\bibinfo {title} {Realizing haldane model in fe-based
  honeycomb ferromagnetic insulators},\ }\href
  {https://doi.org/10.1038/s41535-017-0021-z} {\bibfield  {journal} {\bibinfo
  {journal} {npj Quantum Materials}\ }\textbf {\bibinfo {volume} {2}},\
  \bibinfo {pages} {20} (\bibinfo {year} {2017})}\BibitemShut {NoStop}%
\bibitem [{\citenamefont {Nandkishore}\ and\ \citenamefont
  {Levitov}(2010)}]{nandkishore2010quantum}%
  \BibitemOpen
  \bibfield  {author} {\bibinfo {author} {\bibfnamefont {R.}~\bibnamefont
  {Nandkishore}}\ and\ \bibinfo {author} {\bibfnamefont {L.}~\bibnamefont
  {Levitov}},\ }\bibfield  {title} {\bibinfo {title} {Quantum anomalous hall
  state in bilayer graphene},\ }\href@noop {} {\bibfield  {journal} {\bibinfo
  {journal} {Physical Review B}\ }\textbf {\bibinfo {volume} {82}},\ \bibinfo
  {pages} {115124} (\bibinfo {year} {2010})}\BibitemShut {NoStop}%
\bibitem [{\citenamefont {McIver}\ \emph
  {et~al.}(2020{\natexlab{a}})\citenamefont {McIver}, \citenamefont {Schulte},
  \citenamefont {Stein}, \citenamefont {Matsuyama}, \citenamefont {Jotzu},
  \citenamefont {Meier},\ and\ \citenamefont {Cavalleri}}]{mciver2020light}%
  \BibitemOpen
  \bibfield  {author} {\bibinfo {author} {\bibfnamefont {J.~W.}\ \bibnamefont
  {McIver}}, \bibinfo {author} {\bibfnamefont {B.}~\bibnamefont {Schulte}},
  \bibinfo {author} {\bibfnamefont {F.-U.}\ \bibnamefont {Stein}}, \bibinfo
  {author} {\bibfnamefont {T.}~\bibnamefont {Matsuyama}}, \bibinfo {author}
  {\bibfnamefont {G.}~\bibnamefont {Jotzu}}, \bibinfo {author} {\bibfnamefont
  {G.}~\bibnamefont {Meier}},\ and\ \bibinfo {author} {\bibfnamefont
  {A.}~\bibnamefont {Cavalleri}},\ }\bibfield  {title} {\bibinfo {title}
  {Light-induced anomalous hall effect in graphene},\ }\href@noop {} {\bibfield
   {journal} {\bibinfo  {journal} {Nature physics}\ }\textbf {\bibinfo {volume}
  {16}},\ \bibinfo {pages} {38} (\bibinfo {year}
  {2020}{\natexlab{a}})}\BibitemShut {NoStop}%
\bibitem [{\citenamefont {Appugliese}\ \emph {et~al.}(2022)\citenamefont
  {Appugliese}, \citenamefont {Enkner}, \citenamefont {Paravicini-Bagliani},
  \citenamefont {Beck}, \citenamefont {Reichl}, \citenamefont {Wegscheider},
  \citenamefont {Scalari}, \citenamefont {Ciuti},\ and\ \citenamefont
  {Faist}}]{appugliese2022breakdown}%
  \BibitemOpen
  \bibfield  {author} {\bibinfo {author} {\bibfnamefont {F.}~\bibnamefont
  {Appugliese}}, \bibinfo {author} {\bibfnamefont {J.}~\bibnamefont {Enkner}},
  \bibinfo {author} {\bibfnamefont {G.~L.}\ \bibnamefont
  {Paravicini-Bagliani}}, \bibinfo {author} {\bibfnamefont {M.}~\bibnamefont
  {Beck}}, \bibinfo {author} {\bibfnamefont {C.}~\bibnamefont {Reichl}},
  \bibinfo {author} {\bibfnamefont {W.}~\bibnamefont {Wegscheider}}, \bibinfo
  {author} {\bibfnamefont {G.}~\bibnamefont {Scalari}}, \bibinfo {author}
  {\bibfnamefont {C.}~\bibnamefont {Ciuti}},\ and\ \bibinfo {author}
  {\bibfnamefont {J.}~\bibnamefont {Faist}},\ }\bibfield  {title} {\bibinfo
  {title} {Breakdown of topological protection by cavity vacuum fields in the
  integer quantum hall effect},\ }\href@noop {} {\bibfield  {journal} {\bibinfo
   {journal} {Science}\ }\textbf {\bibinfo {volume} {375}},\ \bibinfo {pages}
  {1030} (\bibinfo {year} {2022})}\BibitemShut {NoStop}%
\bibitem [{\citenamefont {Jarc}\ \emph
  {et~al.}(2023{\natexlab{a}})\citenamefont {Jarc}, \citenamefont
  {Mathengattil}, \citenamefont {Montanaro}, \citenamefont {Giusti},
  \citenamefont {Rigoni}, \citenamefont {Sergo}, \citenamefont {Fassioli},
  \citenamefont {Winnerl}, \citenamefont {Dal~Zilio}, \citenamefont
  {Mihailovic} \emph {et~al.}}]{jarc2023cavity}%
  \BibitemOpen
  \bibfield  {author} {\bibinfo {author} {\bibfnamefont {G.}~\bibnamefont
  {Jarc}}, \bibinfo {author} {\bibfnamefont {S.~Y.}\ \bibnamefont
  {Mathengattil}}, \bibinfo {author} {\bibfnamefont {A.}~\bibnamefont
  {Montanaro}}, \bibinfo {author} {\bibfnamefont {F.}~\bibnamefont {Giusti}},
  \bibinfo {author} {\bibfnamefont {E.~M.}\ \bibnamefont {Rigoni}}, \bibinfo
  {author} {\bibfnamefont {R.}~\bibnamefont {Sergo}}, \bibinfo {author}
  {\bibfnamefont {F.}~\bibnamefont {Fassioli}}, \bibinfo {author}
  {\bibfnamefont {S.}~\bibnamefont {Winnerl}}, \bibinfo {author} {\bibfnamefont
  {S.}~\bibnamefont {Dal~Zilio}}, \bibinfo {author} {\bibfnamefont
  {D.}~\bibnamefont {Mihailovic}}, \emph {et~al.},\ }\bibfield  {title}
  {\bibinfo {title} {Cavity-mediated thermal control of metal-to-insulator
  transition in 1t-tas2},\ }\href@noop {} {\bibfield  {journal} {\bibinfo
  {journal} {Nature}\ }\textbf {\bibinfo {volume} {622}},\ \bibinfo {pages}
  {487} (\bibinfo {year} {2023}{\natexlab{a}})}\BibitemShut {NoStop}%
\bibitem [{\citenamefont {H{\"u}bener}\ \emph {et~al.}(2021)\citenamefont
  {H{\"u}bener}, \citenamefont {De~Giovannini}, \citenamefont {Sch{\"a}fer},
  \citenamefont {Andberger}, \citenamefont {Ruggenthaler}, \citenamefont
  {Faist},\ and\ \citenamefont {Rubio}}]{chiral_cavity2021}%
  \BibitemOpen
  \bibfield  {author} {\bibinfo {author} {\bibfnamefont {H.}~\bibnamefont
  {H{\"u}bener}}, \bibinfo {author} {\bibfnamefont {U.}~\bibnamefont
  {De~Giovannini}}, \bibinfo {author} {\bibfnamefont {C.}~\bibnamefont
  {Sch{\"a}fer}}, \bibinfo {author} {\bibfnamefont {J.}~\bibnamefont
  {Andberger}}, \bibinfo {author} {\bibfnamefont {M.}~\bibnamefont
  {Ruggenthaler}}, \bibinfo {author} {\bibfnamefont {J.}~\bibnamefont
  {Faist}},\ and\ \bibinfo {author} {\bibfnamefont {A.}~\bibnamefont {Rubio}},\
  }\bibfield  {title} {\bibinfo {title} {Engineering quantum materials with
  chiral optical cavities},\ }\href
  {https://doi.org/10.1038/s41563-020-00801-7} {\bibfield  {journal} {\bibinfo
  {journal} {Nature Materials}\ }\textbf {\bibinfo {volume} {20}},\ \bibinfo
  {pages} {438} (\bibinfo {year} {2021})}\BibitemShut {NoStop}%
\bibitem [{\citenamefont {Farokh~Mivehvar}\ and\ \citenamefont
  {Ritsch}(2021)}]{RevQEDgas2021}%
  \BibitemOpen
  \bibfield  {author} {\bibinfo {author} {\bibfnamefont {T.~D.}\ \bibnamefont
  {Farokh~Mivehvar}, \bibfnamefont {Francesco~Piazza}}\ and\ \bibinfo {author}
  {\bibfnamefont {H.}~\bibnamefont {Ritsch}},\ }\bibfield  {title} {\bibinfo
  {title} {Cavity qed with quantum gases: new paradigms in many-body physics},\
  }\href {https://doi.org/10.1080/00018732.2021.1969727} {\bibfield  {journal}
  {\bibinfo  {journal} {Advances in Physics}\ }\textbf {\bibinfo {volume}
  {70}},\ \bibinfo {pages} {1} (\bibinfo {year} {2021})}\BibitemShut {NoStop}%
\bibitem [{\citenamefont {Schlawin}\ \emph {et~al.}(2022)\citenamefont
  {Schlawin}, \citenamefont {Kennes},\ and\ \citenamefont
  {Sentef}}]{review_cavity2022}%
  \BibitemOpen
  \bibfield  {author} {\bibinfo {author} {\bibfnamefont {F.}~\bibnamefont
  {Schlawin}}, \bibinfo {author} {\bibfnamefont {D.~M.}\ \bibnamefont
  {Kennes}},\ and\ \bibinfo {author} {\bibfnamefont {M.~A.}\ \bibnamefont
  {Sentef}},\ }\bibfield  {title} {\bibinfo {title} {{Cavity quantum
  materials}},\ }\href {https://doi.org/10.1063/5.0083825} {\bibfield
  {journal} {\bibinfo  {journal} {Applied Physics Reviews}\ }\textbf {\bibinfo
  {volume} {9}},\ \bibinfo {pages} {011312} (\bibinfo {year}
  {2022})}\BibitemShut {NoStop}%
\bibitem [{\citenamefont {Amelio}\ \emph {et~al.}(2021)\citenamefont {Amelio},
  \citenamefont {Korosec}, \citenamefont {Carusotto},\ and\ \citenamefont
  {Mazza}}]{amelio2021optical}%
  \BibitemOpen
  \bibfield  {author} {\bibinfo {author} {\bibfnamefont {I.}~\bibnamefont
  {Amelio}}, \bibinfo {author} {\bibfnamefont {L.}~\bibnamefont {Korosec}},
  \bibinfo {author} {\bibfnamefont {I.}~\bibnamefont {Carusotto}},\ and\
  \bibinfo {author} {\bibfnamefont {G.}~\bibnamefont {Mazza}},\ }\bibfield
  {title} {\bibinfo {title} {Optical dressing of the electronic response of
  two-dimensional semiconductors in quantum and classical descriptions of
  cavity electrodynamics},\ }\href@noop {} {\bibfield  {journal} {\bibinfo
  {journal} {Physical Review B}\ }\textbf {\bibinfo {volume} {104}},\ \bibinfo
  {pages} {235120} (\bibinfo {year} {2021})}\BibitemShut {NoStop}%
\bibitem [{\citenamefont {Kiffner}\ \emph {et~al.}(2019)\citenamefont
  {Kiffner}, \citenamefont {Coulthard}, \citenamefont {Schlawin}, \citenamefont
  {Ardavan},\ and\ \citenamefont {Jaksch}}]{kiffner2019manipulating}%
  \BibitemOpen
  \bibfield  {author} {\bibinfo {author} {\bibfnamefont {M.}~\bibnamefont
  {Kiffner}}, \bibinfo {author} {\bibfnamefont {J.~R.}\ \bibnamefont
  {Coulthard}}, \bibinfo {author} {\bibfnamefont {F.}~\bibnamefont {Schlawin}},
  \bibinfo {author} {\bibfnamefont {A.}~\bibnamefont {Ardavan}},\ and\ \bibinfo
  {author} {\bibfnamefont {D.}~\bibnamefont {Jaksch}},\ }\bibfield  {title}
  {\bibinfo {title} {Manipulating quantum materials with quantum light},\
  }\href@noop {} {\bibfield  {journal} {\bibinfo  {journal} {Physical Review
  B}\ }\textbf {\bibinfo {volume} {99}},\ \bibinfo {pages} {085116} (\bibinfo
  {year} {2019})}\BibitemShut {NoStop}%
\bibitem [{\citenamefont {Wang}\ \emph {et~al.}(2013)\citenamefont {Wang},
  \citenamefont {Steinberg}, \citenamefont {Jarillo-Herrero},\ and\
  \citenamefont {Gedik}}]{wang2013observation}%
  \BibitemOpen
  \bibfield  {author} {\bibinfo {author} {\bibfnamefont {Y.}~\bibnamefont
  {Wang}}, \bibinfo {author} {\bibfnamefont {H.}~\bibnamefont {Steinberg}},
  \bibinfo {author} {\bibfnamefont {P.}~\bibnamefont {Jarillo-Herrero}},\ and\
  \bibinfo {author} {\bibfnamefont {N.}~\bibnamefont {Gedik}},\ }\bibfield
  {title} {\bibinfo {title} {Observation of floquet-bloch states on the surface
  of a topological insulator},\ }\href@noop {} {\bibfield  {journal} {\bibinfo
  {journal} {Science}\ }\textbf {\bibinfo {volume} {342}},\ \bibinfo {pages}
  {453} (\bibinfo {year} {2013})}\BibitemShut {NoStop}%
\bibitem [{\citenamefont {Kibis}\ \emph {et~al.}(2011)\citenamefont {Kibis},
  \citenamefont {Kyriienko},\ and\ \citenamefont
  {Shelykh}}]{cavitygraphene2011}%
  \BibitemOpen
  \bibfield  {author} {\bibinfo {author} {\bibfnamefont {O.~V.}\ \bibnamefont
  {Kibis}}, \bibinfo {author} {\bibfnamefont {O.}~\bibnamefont {Kyriienko}},\
  and\ \bibinfo {author} {\bibfnamefont {I.~A.}\ \bibnamefont {Shelykh}},\
  }\bibfield  {title} {\bibinfo {title} {Band gap in graphene induced by vacuum
  fluctuations},\ }\href {https://doi.org/10.1103/PhysRevB.84.195413}
  {\bibfield  {journal} {\bibinfo  {journal} {Phys. Rev. B}\ }\textbf {\bibinfo
  {volume} {84}},\ \bibinfo {pages} {195413} (\bibinfo {year}
  {2011})}\BibitemShut {NoStop}%
\bibitem [{\citenamefont {Tokatly}\ \emph {et~al.}(2021)\citenamefont
  {Tokatly}, \citenamefont {Gulevich},\ and\ \citenamefont
  {Iorsh}}]{cavityHall2021}%
  \BibitemOpen
  \bibfield  {author} {\bibinfo {author} {\bibfnamefont {I.~V.}\ \bibnamefont
  {Tokatly}}, \bibinfo {author} {\bibfnamefont {D.~R.}\ \bibnamefont
  {Gulevich}},\ and\ \bibinfo {author} {\bibfnamefont {I.}~\bibnamefont
  {Iorsh}},\ }\bibfield  {title} {\bibinfo {title} {Vacuum anomalous hall
  effect in gyrotropic cavity},\ }\href
  {https://doi.org/10.1103/PhysRevB.104.L081408} {\bibfield  {journal}
  {\bibinfo  {journal} {Phys. Rev. B}\ }\textbf {\bibinfo {volume} {104}},\
  \bibinfo {pages} {L081408} (\bibinfo {year} {2021})}\BibitemShut {NoStop}%
\bibitem [{\citenamefont {Masuki}\ and\ \citenamefont
  {Ashida}(2023)}]{Ashida2023}%
  \BibitemOpen
  \bibfield  {author} {\bibinfo {author} {\bibfnamefont {K.}~\bibnamefont
  {Masuki}}\ and\ \bibinfo {author} {\bibfnamefont {Y.}~\bibnamefont
  {Ashida}},\ }\bibfield  {title} {\bibinfo {title} {Berry phase and topology
  in ultrastrongly coupled quantum light-matter systems},\ }\href
  {https://doi.org/10.1103/PhysRevB.107.195104} {\bibfield  {journal} {\bibinfo
   {journal} {Phys. Rev. B}\ }\textbf {\bibinfo {volume} {107}},\ \bibinfo
  {pages} {195104} (\bibinfo {year} {2023})}\BibitemShut {NoStop}%
\bibitem [{\citenamefont {Jiang}\ \emph {et~al.}(2023)\citenamefont {Jiang},
  \citenamefont {Baggioli},\ and\ \citenamefont
  {Jiang}}]{jiang2023engineering}%
  \BibitemOpen
  \bibfield  {author} {\bibinfo {author} {\bibfnamefont {C.}~\bibnamefont
  {Jiang}}, \bibinfo {author} {\bibfnamefont {M.}~\bibnamefont {Baggioli}},\
  and\ \bibinfo {author} {\bibfnamefont {Q.-D.}\ \bibnamefont {Jiang}},\
  }\href@noop {} {\bibinfo {title} {Engineering flat bands in twisted-bilayer
  graphene away from the magic angle with chiral optical cavities}} (\bibinfo
  {year} {2023}),\ \Eprint {https://arxiv.org/abs/2306.05149} {arXiv:2306.05149
  [cond-mat.mes-hall]} \BibitemShut {NoStop}%
\bibitem [{\citenamefont {Moddel}\ \emph {et~al.}(2021)\citenamefont {Moddel},
  \citenamefont {Weerakkody}, \citenamefont {Doroski},\ and\ \citenamefont
  {Bartusiak}}]{cavityConductance2021}%
  \BibitemOpen
  \bibfield  {author} {\bibinfo {author} {\bibfnamefont {G.}~\bibnamefont
  {Moddel}}, \bibinfo {author} {\bibfnamefont {A.}~\bibnamefont {Weerakkody}},
  \bibinfo {author} {\bibfnamefont {D.}~\bibnamefont {Doroski}},\ and\ \bibinfo
  {author} {\bibfnamefont {D.}~\bibnamefont {Bartusiak}},\ }\bibfield  {title}
  {\bibinfo {title} {Casimir-cavity-induced conductance changes},\ }\href
  {https://doi.org/10.1103/PhysRevResearch.3.L022007} {\bibfield  {journal}
  {\bibinfo  {journal} {Phys. Rev. Res.}\ }\textbf {\bibinfo {volume} {3}},\
  \bibinfo {pages} {L022007} (\bibinfo {year} {2021})}\BibitemShut {NoStop}%
\bibitem [{\citenamefont {Eckhardt}\ \emph {et~al.}(2022)\citenamefont
  {Eckhardt}, \citenamefont {Passetti}, \citenamefont {Othman}, \citenamefont
  {Karrasch}, \citenamefont {Cavaliere}, \citenamefont {Sentef},\ and\
  \citenamefont {Kennes}}]{eckhardt2022quantum}%
  \BibitemOpen
  \bibfield  {author} {\bibinfo {author} {\bibfnamefont {C.~J.}\ \bibnamefont
  {Eckhardt}}, \bibinfo {author} {\bibfnamefont {G.}~\bibnamefont {Passetti}},
  \bibinfo {author} {\bibfnamefont {M.}~\bibnamefont {Othman}}, \bibinfo
  {author} {\bibfnamefont {C.}~\bibnamefont {Karrasch}}, \bibinfo {author}
  {\bibfnamefont {F.}~\bibnamefont {Cavaliere}}, \bibinfo {author}
  {\bibfnamefont {M.~A.}\ \bibnamefont {Sentef}},\ and\ \bibinfo {author}
  {\bibfnamefont {D.~M.}\ \bibnamefont {Kennes}},\ }\bibfield  {title}
  {\bibinfo {title} {Quantum floquet engineering with an exactly solvable
  tight-binding chain in a cavity},\ }\href@noop {} {\bibfield  {journal}
  {\bibinfo  {journal} {Communications Physics}\ }\textbf {\bibinfo {volume}
  {5}},\ \bibinfo {pages} {122} (\bibinfo {year} {2022})}\BibitemShut {NoStop}%
\bibitem [{\citenamefont {Jarc}\ \emph
  {et~al.}(2023{\natexlab{b}})\citenamefont {Jarc}, \citenamefont
  {Mathengattil}, \citenamefont {Montanaro}, \citenamefont {Giusti},
  \citenamefont {Rigoni}, \citenamefont {Sergo}, \citenamefont {Fassioli},
  \citenamefont {Winnerl}, \citenamefont {Dal~Zilio}, \citenamefont
  {Mihailovic}, \citenamefont {Prelov{\v s}ek}, \citenamefont {Eckstein},\ and\
  \citenamefont {Fausti}}]{cavityTransition2023}%
  \BibitemOpen
  \bibfield  {author} {\bibinfo {author} {\bibfnamefont {G.}~\bibnamefont
  {Jarc}}, \bibinfo {author} {\bibfnamefont {S.~Y.}\ \bibnamefont
  {Mathengattil}}, \bibinfo {author} {\bibfnamefont {A.}~\bibnamefont
  {Montanaro}}, \bibinfo {author} {\bibfnamefont {F.}~\bibnamefont {Giusti}},
  \bibinfo {author} {\bibfnamefont {E.~M.}\ \bibnamefont {Rigoni}}, \bibinfo
  {author} {\bibfnamefont {R.}~\bibnamefont {Sergo}}, \bibinfo {author}
  {\bibfnamefont {F.}~\bibnamefont {Fassioli}}, \bibinfo {author}
  {\bibfnamefont {S.}~\bibnamefont {Winnerl}}, \bibinfo {author} {\bibfnamefont
  {S.}~\bibnamefont {Dal~Zilio}}, \bibinfo {author} {\bibfnamefont
  {D.}~\bibnamefont {Mihailovic}}, \bibinfo {author} {\bibfnamefont
  {P.}~\bibnamefont {Prelov{\v s}ek}}, \bibinfo {author} {\bibfnamefont
  {M.}~\bibnamefont {Eckstein}},\ and\ \bibinfo {author} {\bibfnamefont
  {D.}~\bibnamefont {Fausti}},\ }\bibfield  {title} {\bibinfo {title}
  {Cavity-mediated thermal control of metal-to-insulator transition in
  1t-tas2},\ }\href {https://doi.org/10.1038/s41586-023-06596-2} {\bibfield
  {journal} {\bibinfo  {journal} {Nature}\ }\textbf {\bibinfo {volume} {622}},\
  \bibinfo {pages} {487} (\bibinfo {year} {2023}{\natexlab{b}})}\BibitemShut
  {NoStop}%
\bibitem [{\citenamefont {Sentef}\ \emph {et~al.}(2018)\citenamefont {Sentef},
  \citenamefont {Ruggenthaler},\ and\ \citenamefont {Rubio}}]{cavitySC2018}%
  \BibitemOpen
  \bibfield  {author} {\bibinfo {author} {\bibfnamefont {M.~A.}\ \bibnamefont
  {Sentef}}, \bibinfo {author} {\bibfnamefont {M.}~\bibnamefont
  {Ruggenthaler}},\ and\ \bibinfo {author} {\bibfnamefont {A.}~\bibnamefont
  {Rubio}},\ }\bibfield  {title} {\bibinfo {title} {Cavity
  quantum-electrodynamical polaritonically enhanced electron-phonon coupling
  and its influence on superconductivity},\ }\href
  {https://doi.org/10.1126/sciadv.aau6969} {\bibfield  {journal} {\bibinfo
  {journal} {Science Advances}\ }\textbf {\bibinfo {volume} {4}},\ \bibinfo
  {pages} {eaau6969} (\bibinfo {year} {2018})}\BibitemShut {NoStop}%
\bibitem [{\citenamefont {Schlawin}\ \emph {et~al.}(2019)\citenamefont
  {Schlawin}, \citenamefont {Cavalleri},\ and\ \citenamefont
  {Jaksch}}]{cavityMediated2019}%
  \BibitemOpen
  \bibfield  {author} {\bibinfo {author} {\bibfnamefont {F.}~\bibnamefont
  {Schlawin}}, \bibinfo {author} {\bibfnamefont {A.}~\bibnamefont
  {Cavalleri}},\ and\ \bibinfo {author} {\bibfnamefont {D.}~\bibnamefont
  {Jaksch}},\ }\bibfield  {title} {\bibinfo {title} {Cavity-mediated
  electron-photon superconductivity},\ }\href
  {https://doi.org/10.1103/PhysRevLett.122.133602} {\bibfield  {journal}
  {\bibinfo  {journal} {Phys. Rev. Lett.}\ }\textbf {\bibinfo {volume} {122}},\
  \bibinfo {pages} {133602} (\bibinfo {year} {2019})}\BibitemShut {NoStop}%
\bibitem [{\citenamefont {Curtis}\ \emph {et~al.}(2019)\citenamefont {Curtis},
  \citenamefont {Raines}, \citenamefont {Allocca}, \citenamefont {Hafezi},\
  and\ \citenamefont {Galitski}}]{cavitySC2019}%
  \BibitemOpen
  \bibfield  {author} {\bibinfo {author} {\bibfnamefont {J.~B.}\ \bibnamefont
  {Curtis}}, \bibinfo {author} {\bibfnamefont {Z.~M.}\ \bibnamefont {Raines}},
  \bibinfo {author} {\bibfnamefont {A.~A.}\ \bibnamefont {Allocca}}, \bibinfo
  {author} {\bibfnamefont {M.}~\bibnamefont {Hafezi}},\ and\ \bibinfo {author}
  {\bibfnamefont {V.~M.}\ \bibnamefont {Galitski}},\ }\bibfield  {title}
  {\bibinfo {title} {Cavity quantum eliashberg enhancement of
  superconductivity},\ }\href {https://doi.org/10.1103/PhysRevLett.122.167002}
  {\bibfield  {journal} {\bibinfo  {journal} {Phys. Rev. Lett.}\ }\textbf
  {\bibinfo {volume} {122}},\ \bibinfo {pages} {167002} (\bibinfo {year}
  {2019})}\BibitemShut {NoStop}%
\bibitem [{\citenamefont {Li}\ and\ \citenamefont
  {Eckstein}(2020)}]{cavitySC2020}%
  \BibitemOpen
  \bibfield  {author} {\bibinfo {author} {\bibfnamefont {J.}~\bibnamefont
  {Li}}\ and\ \bibinfo {author} {\bibfnamefont {M.}~\bibnamefont {Eckstein}},\
  }\bibfield  {title} {\bibinfo {title} {Manipulating intertwined orders in
  solids with quantum light},\ }\href
  {https://doi.org/10.1103/PhysRevLett.125.217402} {\bibfield  {journal}
  {\bibinfo  {journal} {Phys. Rev. Lett.}\ }\textbf {\bibinfo {volume} {125}},\
  \bibinfo {pages} {217402} (\bibinfo {year} {2020})}\BibitemShut {NoStop}%
\bibitem [{\citenamefont {Curtis}\ \emph {et~al.}(2022)\citenamefont {Curtis},
  \citenamefont {Grankin}, \citenamefont {Poniatowski}, \citenamefont
  {Galitski}, \citenamefont {Narang},\ and\ \citenamefont
  {Demler}}]{cavitySC2022}%
  \BibitemOpen
  \bibfield  {author} {\bibinfo {author} {\bibfnamefont {J.~B.}\ \bibnamefont
  {Curtis}}, \bibinfo {author} {\bibfnamefont {A.}~\bibnamefont {Grankin}},
  \bibinfo {author} {\bibfnamefont {N.~R.}\ \bibnamefont {Poniatowski}},
  \bibinfo {author} {\bibfnamefont {V.~M.}\ \bibnamefont {Galitski}}, \bibinfo
  {author} {\bibfnamefont {P.}~\bibnamefont {Narang}},\ and\ \bibinfo {author}
  {\bibfnamefont {E.}~\bibnamefont {Demler}},\ }\bibfield  {title} {\bibinfo
  {title} {Cavity magnon-polaritons in cuprate parent compounds},\ }\href
  {https://doi.org/10.1103/PhysRevResearch.4.013101} {\bibfield  {journal}
  {\bibinfo  {journal} {Phys. Rev. Res.}\ }\textbf {\bibinfo {volume} {4}},\
  \bibinfo {pages} {013101} (\bibinfo {year} {2022})}\BibitemShut {NoStop}%
\bibitem [{\citenamefont {Dag}\ and\ \citenamefont
  {Rokaj}(2023)}]{Rokaj2023cavity}%
  \BibitemOpen
  \bibfield  {author} {\bibinfo {author} {\bibfnamefont {C.~B.}\ \bibnamefont
  {Dag}}\ and\ \bibinfo {author} {\bibfnamefont {V.}~\bibnamefont {Rokaj}},\
  }\href@noop {} {\bibinfo {title} {Cavity induced topology in graphene}}
  (\bibinfo {year} {2023}),\ \Eprint {https://arxiv.org/abs/2311.02806}
  {arXiv:2311.02806 [cond-mat.mes-hall]} \BibitemShut {NoStop}%
\bibitem [{\citenamefont {Plum}\ and\ \citenamefont
  {Zheludev}(2015)}]{chiralmirrors2015}%
  \BibitemOpen
  \bibfield  {author} {\bibinfo {author} {\bibfnamefont {E.}~\bibnamefont
  {Plum}}\ and\ \bibinfo {author} {\bibfnamefont {N.~I.}\ \bibnamefont
  {Zheludev}},\ }\bibfield  {title} {\bibinfo {title} {{Chiral mirrors}},\
  }\href {https://doi.org/10.1063/1.4921969} {\bibfield  {journal} {\bibinfo
  {journal} {Applied Physics Letters}\ }\textbf {\bibinfo {volume} {106}},\
  \bibinfo {pages} {221901} (\bibinfo {year} {2015})}\BibitemShut {NoStop}%
\bibitem [{\citenamefont {Baranov}\ \emph
  {et~al.}(2020{\natexlab{a}})\citenamefont {Baranov}, \citenamefont
  {Munkhbat}, \citenamefont {Länk}, \citenamefont {Verre}, \citenamefont
  {Käll},\ and\ \citenamefont {Shegai}}]{Baranov2020}%
  \BibitemOpen
  \bibfield  {author} {\bibinfo {author} {\bibfnamefont {D.~G.}\ \bibnamefont
  {Baranov}}, \bibinfo {author} {\bibfnamefont {B.}~\bibnamefont {Munkhbat}},
  \bibinfo {author} {\bibfnamefont {N.~O.}\ \bibnamefont {Länk}}, \bibinfo
  {author} {\bibfnamefont {R.}~\bibnamefont {Verre}}, \bibinfo {author}
  {\bibfnamefont {M.}~\bibnamefont {Käll}},\ and\ \bibinfo {author}
  {\bibfnamefont {T.}~\bibnamefont {Shegai}},\ }\bibfield  {title} {\bibinfo
  {title} {Circular dichroism mode splitting and bounds to its enhancement with
  cavity-plasmon-polaritons},\ }\href
  {https://doi.org/doi:10.1515/nanoph-2019-0372} {\bibfield  {journal}
  {\bibinfo  {journal} {Nanophotonics}\ }\textbf {\bibinfo {volume} {9}},\
  \bibinfo {pages} {283} (\bibinfo {year} {2020}{\natexlab{a}})}\BibitemShut
  {NoStop}%
\bibitem [{\citenamefont {Baranov}\ \emph
  {et~al.}(2020{\natexlab{b}})\citenamefont {Baranov}, \citenamefont
  {Munkhbat}, \citenamefont {Zhukova}, \citenamefont {Bisht}, \citenamefont
  {Canales}, \citenamefont {Rousseaux}, \citenamefont {Johansson},
  \citenamefont {Antosiewicz},\ and\ \citenamefont
  {Shegai}}]{Baranov2020nature}%
  \BibitemOpen
  \bibfield  {author} {\bibinfo {author} {\bibfnamefont {D.~G.}\ \bibnamefont
  {Baranov}}, \bibinfo {author} {\bibfnamefont {B.}~\bibnamefont {Munkhbat}},
  \bibinfo {author} {\bibfnamefont {E.}~\bibnamefont {Zhukova}}, \bibinfo
  {author} {\bibfnamefont {A.}~\bibnamefont {Bisht}}, \bibinfo {author}
  {\bibfnamefont {A.}~\bibnamefont {Canales}}, \bibinfo {author} {\bibfnamefont
  {B.}~\bibnamefont {Rousseaux}}, \bibinfo {author} {\bibfnamefont
  {G.}~\bibnamefont {Johansson}}, \bibinfo {author} {\bibfnamefont {T.~J.}\
  \bibnamefont {Antosiewicz}},\ and\ \bibinfo {author} {\bibfnamefont
  {T.}~\bibnamefont {Shegai}},\ }\bibfield  {title} {\bibinfo {title}
  {Ultrastrong coupling between nanoparticle plasmons and cavity photons at
  ambient conditions},\ }\href {https://doi.org/10.1038/s41467-020-16524-x}
  {\bibfield  {journal} {\bibinfo  {journal} {Nature Communications}\ }\textbf
  {\bibinfo {volume} {11}},\ \bibinfo {pages} {2715} (\bibinfo {year}
  {2020}{\natexlab{b}})}\BibitemShut {NoStop}%
\bibitem [{\citenamefont {Owens}\ \emph {et~al.}(2022)\citenamefont {Owens},
  \citenamefont {Panetta}, \citenamefont {Saxberg}, \citenamefont {Roberts},
  \citenamefont {Chakram}, \citenamefont {Ma}, \citenamefont {Vrajitoarea},
  \citenamefont {Simon},\ and\ \citenamefont {Schuster}}]{chiralQED2022}%
  \BibitemOpen
  \bibfield  {author} {\bibinfo {author} {\bibfnamefont {J.~C.}\ \bibnamefont
  {Owens}}, \bibinfo {author} {\bibfnamefont {M.~G.}\ \bibnamefont {Panetta}},
  \bibinfo {author} {\bibfnamefont {B.}~\bibnamefont {Saxberg}}, \bibinfo
  {author} {\bibfnamefont {G.}~\bibnamefont {Roberts}}, \bibinfo {author}
  {\bibfnamefont {S.}~\bibnamefont {Chakram}}, \bibinfo {author} {\bibfnamefont
  {R.}~\bibnamefont {Ma}}, \bibinfo {author} {\bibfnamefont {A.}~\bibnamefont
  {Vrajitoarea}}, \bibinfo {author} {\bibfnamefont {J.}~\bibnamefont {Simon}},\
  and\ \bibinfo {author} {\bibfnamefont {D.~I.}\ \bibnamefont {Schuster}},\
  }\bibfield  {title} {\bibinfo {title} {Chiral cavity quantum
  electrodynamics},\ }\href {https://doi.org/10.1038/s41567-022-01671-3}
  {\bibfield  {journal} {\bibinfo  {journal} {Nature Physics}\ }\textbf
  {\bibinfo {volume} {18}},\ \bibinfo {pages} {1048} (\bibinfo {year}
  {2022})}\BibitemShut {NoStop}%
\bibitem [{\citenamefont {Jiang}\ and\ \citenamefont
  {Wilczek}(2019{\natexlab{a}})}]{jiang2019atmospherics}%
  \BibitemOpen
  \bibfield  {author} {\bibinfo {author} {\bibfnamefont {Q.-D.}\ \bibnamefont
  {Jiang}}\ and\ \bibinfo {author} {\bibfnamefont {F.}~\bibnamefont
  {Wilczek}},\ }\bibfield  {title} {\bibinfo {title} {Quantum atmospherics for
  materials diagnosis},\ }\href {https://doi.org/10.1103/PhysRevB.99.201104}
  {\bibfield  {journal} {\bibinfo  {journal} {Phys. Rev. B}\ }\textbf {\bibinfo
  {volume} {99}},\ \bibinfo {pages} {201104} (\bibinfo {year}
  {2019}{\natexlab{a}})}\BibitemShut {NoStop}%
\bibitem [{\citenamefont {Butcher}\ \emph {et~al.}(2012)\citenamefont
  {Butcher}, \citenamefont {Buhmann},\ and\ \citenamefont
  {Scheel}}]{butcher2012casimir}%
  \BibitemOpen
  \bibfield  {author} {\bibinfo {author} {\bibfnamefont {D.~T.}\ \bibnamefont
  {Butcher}}, \bibinfo {author} {\bibfnamefont {S.~Y.}\ \bibnamefont
  {Buhmann}},\ and\ \bibinfo {author} {\bibfnamefont {S.}~\bibnamefont
  {Scheel}},\ }\bibfield  {title} {\bibinfo {title} {Casimir--polder forces
  between chiral objects},\ }\href@noop {} {\bibfield  {journal} {\bibinfo
  {journal} {New Journal of Physics}\ }\textbf {\bibinfo {volume} {14}},\
  \bibinfo {pages} {113013} (\bibinfo {year} {2012})}\BibitemShut {NoStop}%
\bibitem [{\citenamefont {Jiang}\ and\ \citenamefont
  {Wilczek}(2019{\natexlab{b}})}]{jiang2019axial}%
  \BibitemOpen
  \bibfield  {author} {\bibinfo {author} {\bibfnamefont {Q.-D.}\ \bibnamefont
  {Jiang}}\ and\ \bibinfo {author} {\bibfnamefont {F.}~\bibnamefont
  {Wilczek}},\ }\bibfield  {title} {\bibinfo {title} {Axial casimir force},\
  }\href@noop {} {\bibfield  {journal} {\bibinfo  {journal} {Physical Review
  B}\ }\textbf {\bibinfo {volume} {99}},\ \bibinfo {pages} {165402} (\bibinfo
  {year} {2019}{\natexlab{b}})}\BibitemShut {NoStop}%
\bibitem [{\citenamefont {Ke}\ \emph {et~al.}(2023)\citenamefont {Ke},
  \citenamefont {Song},\ and\ \citenamefont {Jiang}}]{ke2023vacuum}%
  \BibitemOpen
  \bibfield  {author} {\bibinfo {author} {\bibfnamefont {Y.}~\bibnamefont
  {Ke}}, \bibinfo {author} {\bibfnamefont {Z.}~\bibnamefont {Song}},\ and\
  \bibinfo {author} {\bibfnamefont {Q.-D.}\ \bibnamefont {Jiang}},\ }\bibfield
  {title} {\bibinfo {title} {Vacuum-induced symmetry breaking of chiral
  enantiomer formation in chemical reactions},\ }\href@noop {} {\bibfield
  {journal} {\bibinfo  {journal} {Physical Review Letters}\ }\textbf {\bibinfo
  {volume} {131}},\ \bibinfo {pages} {223601} (\bibinfo {year}
  {2023})}\BibitemShut {NoStop}%
\bibitem [{\citenamefont {Espinosa-Ortega}\ \emph {et~al.}(2014)\citenamefont
  {Espinosa-Ortega}, \citenamefont {Kyriienko}, \citenamefont {Kibis},\ and\
  \citenamefont {Shelykh}}]{espinosa2014semiconductor}%
  \BibitemOpen
  \bibfield  {author} {\bibinfo {author} {\bibfnamefont {T.}~\bibnamefont
  {Espinosa-Ortega}}, \bibinfo {author} {\bibfnamefont {O.}~\bibnamefont
  {Kyriienko}}, \bibinfo {author} {\bibfnamefont {O.}~\bibnamefont {Kibis}},\
  and\ \bibinfo {author} {\bibfnamefont {I.}~\bibnamefont {Shelykh}},\
  }\bibfield  {title} {\bibinfo {title} {Semiconductor cavity qed: Band gap
  induced by vacuum fluctuations},\ }\href@noop {} {\bibfield  {journal}
  {\bibinfo  {journal} {Physical Review A}\ }\textbf {\bibinfo {volume} {89}},\
  \bibinfo {pages} {062115} (\bibinfo {year} {2014})}\BibitemShut {NoStop}%
\bibitem [{\citenamefont {Wang}\ \emph {et~al.}(2019)\citenamefont {Wang},
  \citenamefont {Ronca},\ and\ \citenamefont {Sentef}}]{CavityChern2019}%
  \BibitemOpen
  \bibfield  {author} {\bibinfo {author} {\bibfnamefont {X.}~\bibnamefont
  {Wang}}, \bibinfo {author} {\bibfnamefont {E.}~\bibnamefont {Ronca}},\ and\
  \bibinfo {author} {\bibfnamefont {M.~A.}\ \bibnamefont {Sentef}},\ }\bibfield
   {title} {\bibinfo {title} {Cavity quantum electrodynamical chern insulator:
  Towards light-induced quantized anomalous hall effect in graphene},\ }\href
  {https://doi.org/10.1103/PhysRevB.99.235156} {\bibfield  {journal} {\bibinfo
  {journal} {Phys. Rev. B}\ }\textbf {\bibinfo {volume} {99}},\ \bibinfo
  {pages} {235156} (\bibinfo {year} {2019})}\BibitemShut {NoStop}%
\bibitem [{\citenamefont {Ashida}\ \emph {et~al.}(2021)\citenamefont {Ashida},
  \citenamefont {\ifmmode \dot{I}\else \.{I}\fi{}mamo\ifmmode~\breve{g}\else
  \u{g}\fi{}lu},\ and\ \citenamefont {Demler}}]{CavityQED2021}%
  \BibitemOpen
  \bibfield  {author} {\bibinfo {author} {\bibfnamefont {Y.}~\bibnamefont
  {Ashida}}, \bibinfo {author} {\bibfnamefont {A.~m.~c.}\ \bibnamefont
  {\ifmmode \dot{I}\else \.{I}\fi{}mamo\ifmmode~\breve{g}\else \u{g}\fi{}lu}},\
  and\ \bibinfo {author} {\bibfnamefont {E.}~\bibnamefont {Demler}},\
  }\bibfield  {title} {\bibinfo {title} {Cavity quantum electrodynamics at
  arbitrary light-matter coupling strengths},\ }\href
  {https://doi.org/10.1103/PhysRevLett.126.153603} {\bibfield  {journal}
  {\bibinfo  {journal} {Phys. Rev. Lett.}\ }\textbf {\bibinfo {volume} {126}},\
  \bibinfo {pages} {153603} (\bibinfo {year} {2021})}\BibitemShut {NoStop}%
\bibitem [{\citenamefont {Anappara}\ \emph {et~al.}(2009)\citenamefont
  {Anappara}, \citenamefont {De~Liberato}, \citenamefont {Tredicucci},
  \citenamefont {Ciuti}, \citenamefont {Biasiol}, \citenamefont {Sorba},\ and\
  \citenamefont {Beltram}}]{ultrastrong2009}%
  \BibitemOpen
  \bibfield  {author} {\bibinfo {author} {\bibfnamefont {A.~A.}\ \bibnamefont
  {Anappara}}, \bibinfo {author} {\bibfnamefont {S.}~\bibnamefont
  {De~Liberato}}, \bibinfo {author} {\bibfnamefont {A.}~\bibnamefont
  {Tredicucci}}, \bibinfo {author} {\bibfnamefont {C.}~\bibnamefont {Ciuti}},
  \bibinfo {author} {\bibfnamefont {G.}~\bibnamefont {Biasiol}}, \bibinfo
  {author} {\bibfnamefont {L.}~\bibnamefont {Sorba}},\ and\ \bibinfo {author}
  {\bibfnamefont {F.}~\bibnamefont {Beltram}},\ }\bibfield  {title} {\bibinfo
  {title} {Signatures of the ultrastrong light-matter coupling regime},\ }\href
  {https://doi.org/10.1103/PhysRevB.79.201303} {\bibfield  {journal} {\bibinfo
  {journal} {Phys. Rev. B}\ }\textbf {\bibinfo {volume} {79}},\ \bibinfo
  {pages} {201303} (\bibinfo {year} {2009})}\BibitemShut {NoStop}%
\bibitem [{\citenamefont {Scalari}\ \emph {et~al.}(2012)\citenamefont
  {Scalari}, \citenamefont {Maissen}, \citenamefont {Turčinková},
  \citenamefont {Hagenmüller}, \citenamefont {Liberato}, \citenamefont
  {Ciuti}, \citenamefont {Reichl}, \citenamefont {Schuh}, \citenamefont
  {Wegscheider}, \citenamefont {Beck},\ and\ \citenamefont
  {Faist}}]{ultrastrong2012}%
  \BibitemOpen
  \bibfield  {author} {\bibinfo {author} {\bibfnamefont {G.}~\bibnamefont
  {Scalari}}, \bibinfo {author} {\bibfnamefont {C.}~\bibnamefont {Maissen}},
  \bibinfo {author} {\bibfnamefont {D.}~\bibnamefont {Turčinková}}, \bibinfo
  {author} {\bibfnamefont {D.}~\bibnamefont {Hagenmüller}}, \bibinfo {author}
  {\bibfnamefont {S.~D.}\ \bibnamefont {Liberato}}, \bibinfo {author}
  {\bibfnamefont {C.}~\bibnamefont {Ciuti}}, \bibinfo {author} {\bibfnamefont
  {C.}~\bibnamefont {Reichl}}, \bibinfo {author} {\bibfnamefont
  {D.}~\bibnamefont {Schuh}}, \bibinfo {author} {\bibfnamefont
  {W.}~\bibnamefont {Wegscheider}}, \bibinfo {author} {\bibfnamefont
  {M.}~\bibnamefont {Beck}},\ and\ \bibinfo {author} {\bibfnamefont
  {J.}~\bibnamefont {Faist}},\ }\bibfield  {title} {\bibinfo {title}
  {Ultrastrong coupling of the cyclotron transition of a 2d electron gas to a
  thz metamaterial},\ }\href {https://doi.org/10.1126/science.1216022}
  {\bibfield  {journal} {\bibinfo  {journal} {Science}\ }\textbf {\bibinfo
  {volume} {335}},\ \bibinfo {pages} {1323} (\bibinfo {year}
  {2012})}\BibitemShut {NoStop}%
\bibitem [{\citenamefont {Chikkaraddy}\ \emph {et~al.}(2016)\citenamefont
  {Chikkaraddy}, \citenamefont {de~Nijs}, \citenamefont {Benz}, \citenamefont
  {Barrow}, \citenamefont {Scherman}, \citenamefont {Rosta}, \citenamefont
  {Demetriadou}, \citenamefont {Fox}, \citenamefont {Hess},\ and\ \citenamefont
  {Baumberg}}]{ultrastrong2016}%
  \BibitemOpen
  \bibfield  {author} {\bibinfo {author} {\bibfnamefont {R.}~\bibnamefont
  {Chikkaraddy}}, \bibinfo {author} {\bibfnamefont {B.}~\bibnamefont
  {de~Nijs}}, \bibinfo {author} {\bibfnamefont {F.}~\bibnamefont {Benz}},
  \bibinfo {author} {\bibfnamefont {S.~J.}\ \bibnamefont {Barrow}}, \bibinfo
  {author} {\bibfnamefont {O.~A.}\ \bibnamefont {Scherman}}, \bibinfo {author}
  {\bibfnamefont {E.}~\bibnamefont {Rosta}}, \bibinfo {author} {\bibfnamefont
  {A.}~\bibnamefont {Demetriadou}}, \bibinfo {author} {\bibfnamefont
  {P.}~\bibnamefont {Fox}}, \bibinfo {author} {\bibfnamefont {O.}~\bibnamefont
  {Hess}},\ and\ \bibinfo {author} {\bibfnamefont {J.~J.}\ \bibnamefont
  {Baumberg}},\ }\bibfield  {title} {\bibinfo {title} {Single-molecule strong
  coupling at room temperature in plasmonic nanocavities},\ }\href
  {https://doi.org/10.1038/nature17974} {\bibfield  {journal} {\bibinfo
  {journal} {Nature}\ }\textbf {\bibinfo {volume} {535}},\ \bibinfo {pages}
  {127} (\bibinfo {year} {2016})}\BibitemShut {NoStop}%
\bibitem [{\citenamefont {Yoshihara}\ \emph {et~al.}(2017)\citenamefont
  {Yoshihara}, \citenamefont {Fuse}, \citenamefont {Ashhab}, \citenamefont
  {Kakuyanagi}, \citenamefont {Saito},\ and\ \citenamefont
  {Semba}}]{ultrastrongNP2017}%
  \BibitemOpen
  \bibfield  {author} {\bibinfo {author} {\bibfnamefont {F.}~\bibnamefont
  {Yoshihara}}, \bibinfo {author} {\bibfnamefont {T.}~\bibnamefont {Fuse}},
  \bibinfo {author} {\bibfnamefont {S.}~\bibnamefont {Ashhab}}, \bibinfo
  {author} {\bibfnamefont {K.}~\bibnamefont {Kakuyanagi}}, \bibinfo {author}
  {\bibfnamefont {S.}~\bibnamefont {Saito}},\ and\ \bibinfo {author}
  {\bibfnamefont {K.}~\bibnamefont {Semba}},\ }\bibfield  {title} {\bibinfo
  {title} {Superconducting qubit--oscillator circuit beyond the
  ultrastrong-coupling regime},\ }\href {https://doi.org/10.1038/nphys3906}
  {\bibfield  {journal} {\bibinfo  {journal} {Nature Physics}\ }\textbf
  {\bibinfo {volume} {13}},\ \bibinfo {pages} {44} (\bibinfo {year}
  {2017})}\BibitemShut {NoStop}%
\bibitem [{\citenamefont {Flick}\ \emph {et~al.}(2017)\citenamefont {Flick},
  \citenamefont {Ruggenthaler}, \citenamefont {Appel},\ and\ \citenamefont
  {Rubio}}]{ultrastrong2017}%
  \BibitemOpen
  \bibfield  {author} {\bibinfo {author} {\bibfnamefont {J.}~\bibnamefont
  {Flick}}, \bibinfo {author} {\bibfnamefont {M.}~\bibnamefont {Ruggenthaler}},
  \bibinfo {author} {\bibfnamefont {H.}~\bibnamefont {Appel}},\ and\ \bibinfo
  {author} {\bibfnamefont {A.}~\bibnamefont {Rubio}},\ }\bibfield  {title}
  {\bibinfo {title} {Atoms and molecules in cavities, from weak to strong
  coupling in quantum-electrodynamics (qed) chemistry},\ }\href
  {https://doi.org/10.1073/pnas.1615509114} {\bibfield  {journal} {\bibinfo
  {journal} {Proceedings of the National Academy of Sciences}\ }\textbf
  {\bibinfo {volume} {114}},\ \bibinfo {pages} {3026} (\bibinfo {year}
  {2017})}\BibitemShut {NoStop}%
\bibitem [{\citenamefont {Frisk~Kockum}\ \emph {et~al.}(2019)\citenamefont
  {Frisk~Kockum}, \citenamefont {Miranowicz}, \citenamefont {De~Liberato},
  \citenamefont {Savasta},\ and\ \citenamefont {Nori}}]{ultrastrong2019}%
  \BibitemOpen
  \bibfield  {author} {\bibinfo {author} {\bibfnamefont {A.}~\bibnamefont
  {Frisk~Kockum}}, \bibinfo {author} {\bibfnamefont {A.}~\bibnamefont
  {Miranowicz}}, \bibinfo {author} {\bibfnamefont {S.}~\bibnamefont
  {De~Liberato}}, \bibinfo {author} {\bibfnamefont {S.}~\bibnamefont
  {Savasta}},\ and\ \bibinfo {author} {\bibfnamefont {F.}~\bibnamefont
  {Nori}},\ }\bibfield  {title} {\bibinfo {title} {Ultrastrong coupling between
  light and matter},\ }\href {https://doi.org/10.1038/s42254-018-0006-2}
  {\bibfield  {journal} {\bibinfo  {journal} {Nature Reviews Physics}\ }\textbf
  {\bibinfo {volume} {1}},\ \bibinfo {pages} {19} (\bibinfo {year}
  {2019})}\BibitemShut {NoStop}%
\bibitem [{\citenamefont {Forn-D\'{\i}az}\ \emph {et~al.}(2019)\citenamefont
  {Forn-D\'{\i}az}, \citenamefont {Lamata}, \citenamefont {Rico}, \citenamefont
  {Kono},\ and\ \citenamefont {Solano}}]{ultrastrongRMP2019}%
  \BibitemOpen
  \bibfield  {author} {\bibinfo {author} {\bibfnamefont {P.}~\bibnamefont
  {Forn-D\'{\i}az}}, \bibinfo {author} {\bibfnamefont {L.}~\bibnamefont
  {Lamata}}, \bibinfo {author} {\bibfnamefont {E.}~\bibnamefont {Rico}},
  \bibinfo {author} {\bibfnamefont {J.}~\bibnamefont {Kono}},\ and\ \bibinfo
  {author} {\bibfnamefont {E.}~\bibnamefont {Solano}},\ }\bibfield  {title}
  {\bibinfo {title} {Ultrastrong coupling regimes of light-matter
  interaction},\ }\href {https://doi.org/10.1103/RevModPhys.91.025005}
  {\bibfield  {journal} {\bibinfo  {journal} {Rev. Mod. Phys.}\ }\textbf
  {\bibinfo {volume} {91}},\ \bibinfo {pages} {025005} (\bibinfo {year}
  {2019})}\BibitemShut {NoStop}%
\bibitem [{\citenamefont {Halbhuber}\ \emph {et~al.}(2020)\citenamefont
  {Halbhuber}, \citenamefont {Mornhinweg}, \citenamefont {Zeller},
  \citenamefont {Ciuti}, \citenamefont {Bougeard}, \citenamefont {Huber},\ and\
  \citenamefont {Lange}}]{ultrastrong2020}%
  \BibitemOpen
  \bibfield  {author} {\bibinfo {author} {\bibfnamefont {M.}~\bibnamefont
  {Halbhuber}}, \bibinfo {author} {\bibfnamefont {J.}~\bibnamefont
  {Mornhinweg}}, \bibinfo {author} {\bibfnamefont {V.}~\bibnamefont {Zeller}},
  \bibinfo {author} {\bibfnamefont {C.}~\bibnamefont {Ciuti}}, \bibinfo
  {author} {\bibfnamefont {D.}~\bibnamefont {Bougeard}}, \bibinfo {author}
  {\bibfnamefont {R.}~\bibnamefont {Huber}},\ and\ \bibinfo {author}
  {\bibfnamefont {C.}~\bibnamefont {Lange}},\ }\bibfield  {title} {\bibinfo
  {title} {Non-adiabatic stripping of a cavity field from electrons in the
  deep-strong coupling regime},\ }\href
  {https://doi.org/10.1038/s41566-020-0673-2} {\bibfield  {journal} {\bibinfo
  {journal} {Nature Photonics}\ }\textbf {\bibinfo {volume} {14}},\ \bibinfo
  {pages} {675} (\bibinfo {year} {2020})}\BibitemShut {NoStop}%
\bibitem [{\citenamefont {Lee}\ \emph {et~al.}(1953)\citenamefont {Lee},
  \citenamefont {Low},\ and\ \citenamefont {Pines}}]{LeeTrans1953}%
  \BibitemOpen
  \bibfield  {author} {\bibinfo {author} {\bibfnamefont {T.~D.}\ \bibnamefont
  {Lee}}, \bibinfo {author} {\bibfnamefont {F.~E.}\ \bibnamefont {Low}},\ and\
  \bibinfo {author} {\bibfnamefont {D.}~\bibnamefont {Pines}},\ }\bibfield
  {title} {\bibinfo {title} {The motion of slow electrons in a polar crystal},\
  }\href {https://doi.org/10.1103/PhysRev.90.297} {\bibfield  {journal}
  {\bibinfo  {journal} {Phys. Rev.}\ }\textbf {\bibinfo {volume} {90}},\
  \bibinfo {pages} {297} (\bibinfo {year} {1953})}\BibitemShut {NoStop}%
\bibitem [{\citenamefont {Reich}\ \emph {et~al.}(2002)\citenamefont {Reich},
  \citenamefont {Maultzsch}, \citenamefont {Thomsen},\ and\ \citenamefont
  {Ordej\'on}}]{TBgraphene2002}%
  \BibitemOpen
  \bibfield  {author} {\bibinfo {author} {\bibfnamefont {S.}~\bibnamefont
  {Reich}}, \bibinfo {author} {\bibfnamefont {J.}~\bibnamefont {Maultzsch}},
  \bibinfo {author} {\bibfnamefont {C.}~\bibnamefont {Thomsen}},\ and\ \bibinfo
  {author} {\bibfnamefont {P.}~\bibnamefont {Ordej\'on}},\ }\bibfield  {title}
  {\bibinfo {title} {Tight-binding description of graphene},\ }\href
  {https://doi.org/10.1103/PhysRevB.66.035412} {\bibfield  {journal} {\bibinfo
  {journal} {Phys. Rev. B}\ }\textbf {\bibinfo {volume} {66}},\ \bibinfo
  {pages} {035412} (\bibinfo {year} {2002})}\BibitemShut {NoStop}%
\bibitem [{SM(2022)}]{SM}%
  \BibitemOpen
  \href@noop {} {\bibinfo {title} {See supplemental material at [url will be
  inserted by publisher] for details of the calculations}} (\bibinfo {year}
  {2022})\BibitemShut {NoStop}%
\bibitem [{\citenamefont {Lanneb\`ere}\ and\ \citenamefont
  {Silveirinha}(2018)}]{HaldaneModel2018}%
  \BibitemOpen
  \bibfield  {author} {\bibinfo {author} {\bibfnamefont {S.}~\bibnamefont
  {Lanneb\`ere}}\ and\ \bibinfo {author} {\bibfnamefont {M.~G.}\ \bibnamefont
  {Silveirinha}},\ }\bibfield  {title} {\bibinfo {title} {Link between the
  photonic and electronic topological phases in artificial graphene},\ }\href
  {https://doi.org/10.1103/PhysRevB.97.165128} {\bibfield  {journal} {\bibinfo
  {journal} {Phys. Rev. B}\ }\textbf {\bibinfo {volume} {97}},\ \bibinfo
  {pages} {165128} (\bibinfo {year} {2018})}\BibitemShut {NoStop}%
\bibitem [{\citenamefont {Peierls}(1933)}]{Peierlsub}%
  \BibitemOpen
  \bibfield  {author} {\bibinfo {author} {\bibfnamefont {R.}~\bibnamefont
  {Peierls}},\ }\bibfield  {title} {\bibinfo {title} {Zur theorie des
  diamagnetismus von leitungselektronen},\ }\href
  {https://doi.org/10.1007/BF01342591} {\bibfield  {journal} {\bibinfo
  {journal} {Zeitschrift f{\"u}r Physik}\ }\textbf {\bibinfo {volume} {80}},\
  \bibinfo {pages} {763} (\bibinfo {year} {1933})}\BibitemShut {NoStop}%
\bibitem [{\citenamefont {Resta}(1994)}]{RMP_Resta1994}%
  \BibitemOpen
  \bibfield  {author} {\bibinfo {author} {\bibfnamefont {R.}~\bibnamefont
  {Resta}},\ }\bibfield  {title} {\bibinfo {title} {Macroscopic polarization in
  crystalline dielectrics: the geometric phase approach},\ }\href
  {https://doi.org/10.1103/RevModPhys.66.899} {\bibfield  {journal} {\bibinfo
  {journal} {Rev. Mod. Phys.}\ }\textbf {\bibinfo {volume} {66}},\ \bibinfo
  {pages} {899} (\bibinfo {year} {1994})}\BibitemShut {NoStop}%
\bibitem [{\citenamefont {Resta}(2006)}]{Resta2006}%
  \BibitemOpen
  \bibfield  {author} {\bibinfo {author} {\bibfnamefont {R.}~\bibnamefont
  {Resta}},\ }\bibfield  {title} {\bibinfo {title} {{Kohn’s theory of the
  insulating state: A quantum-chemistry viewpoint}},\ }\href
  {https://doi.org/10.1063/1.2176604} {\bibfield  {journal} {\bibinfo
  {journal} {The Journal of Chemical Physics}\ }\textbf {\bibinfo {volume}
  {124}},\ \bibinfo {pages} {104104} (\bibinfo {year} {2006})}\BibitemShut
  {NoStop}%
\bibitem [{\citenamefont {Resta}(2011)}]{Resta2011}%
  \BibitemOpen
  \bibfield  {author} {\bibinfo {author} {\bibfnamefont {R.}~\bibnamefont
  {Resta}},\ }\bibfield  {title} {\bibinfo {title} {The insulating state of
  matter: a geometrical theory},\ }\href
  {https://doi.org/10.1140/epjb/e2010-10874-4} {\bibfield  {journal} {\bibinfo
  {journal} {The European Physical Journal B}\ }\textbf {\bibinfo {volume}
  {79}},\ \bibinfo {pages} {121} (\bibinfo {year} {2011})}\BibitemShut
  {NoStop}%
\bibitem [{\citenamefont {Roy}(2014)}]{bandgeometry2014}%
  \BibitemOpen
  \bibfield  {author} {\bibinfo {author} {\bibfnamefont {R.}~\bibnamefont
  {Roy}},\ }\bibfield  {title} {\bibinfo {title} {Band geometry of fractional
  topological insulators},\ }\href {https://doi.org/10.1103/PhysRevB.90.165139}
  {\bibfield  {journal} {\bibinfo  {journal} {Phys. Rev. B}\ }\textbf {\bibinfo
  {volume} {90}},\ \bibinfo {pages} {165139} (\bibinfo {year}
  {2014})}\BibitemShut {NoStop}%
\bibitem [{\citenamefont {Liang}\ \emph {et~al.}(2017)\citenamefont {Liang},
  \citenamefont {Vanhala}, \citenamefont {Peotta}, \citenamefont {Siro},
  \citenamefont {Harju},\ and\ \citenamefont {T\"orm\"a}}]{Bandgeometry2017}%
  \BibitemOpen
  \bibfield  {author} {\bibinfo {author} {\bibfnamefont {L.}~\bibnamefont
  {Liang}}, \bibinfo {author} {\bibfnamefont {T.~I.}\ \bibnamefont {Vanhala}},
  \bibinfo {author} {\bibfnamefont {S.}~\bibnamefont {Peotta}}, \bibinfo
  {author} {\bibfnamefont {T.}~\bibnamefont {Siro}}, \bibinfo {author}
  {\bibfnamefont {A.}~\bibnamefont {Harju}},\ and\ \bibinfo {author}
  {\bibfnamefont {P.}~\bibnamefont {T\"orm\"a}},\ }\bibfield  {title} {\bibinfo
  {title} {Band geometry, berry curvature, and superfluid weight},\ }\href
  {https://doi.org/10.1103/PhysRevB.95.024515} {\bibfield  {journal} {\bibinfo
  {journal} {Phys. Rev. B}\ }\textbf {\bibinfo {volume} {95}},\ \bibinfo
  {pages} {024515} (\bibinfo {year} {2017})}\BibitemShut {NoStop}%
\bibitem [{\citenamefont {Xie}\ \emph {et~al.}(2020)\citenamefont {Xie},
  \citenamefont {Song}, \citenamefont {Lian},\ and\ \citenamefont
  {Bernevig}}]{topobound2020}%
  \BibitemOpen
  \bibfield  {author} {\bibinfo {author} {\bibfnamefont {F.}~\bibnamefont
  {Xie}}, \bibinfo {author} {\bibfnamefont {Z.}~\bibnamefont {Song}}, \bibinfo
  {author} {\bibfnamefont {B.}~\bibnamefont {Lian}},\ and\ \bibinfo {author}
  {\bibfnamefont {B.~A.}\ \bibnamefont {Bernevig}},\ }\bibfield  {title}
  {\bibinfo {title} {Topology-bounded superfluid weight in twisted bilayer
  graphene},\ }\href {https://doi.org/10.1103/PhysRevLett.124.167002}
  {\bibfield  {journal} {\bibinfo  {journal} {Phys. Rev. Lett.}\ }\textbf
  {\bibinfo {volume} {124}},\ \bibinfo {pages} {167002} (\bibinfo {year}
  {2020})}\BibitemShut {NoStop}%
\bibitem [{\citenamefont {Ozawa}\ and\ \citenamefont
  {Mera}(2021)}]{relation2021}%
  \BibitemOpen
  \bibfield  {author} {\bibinfo {author} {\bibfnamefont {T.}~\bibnamefont
  {Ozawa}}\ and\ \bibinfo {author} {\bibfnamefont {B.}~\bibnamefont {Mera}},\
  }\bibfield  {title} {\bibinfo {title} {Relations between topology and the
  quantum metric for chern insulators},\ }\href
  {https://doi.org/10.1103/PhysRevB.104.045103} {\bibfield  {journal} {\bibinfo
   {journal} {Phys. Rev. B}\ }\textbf {\bibinfo {volume} {104}},\ \bibinfo
  {pages} {045103} (\bibinfo {year} {2021})}\BibitemShut {NoStop}%
\bibitem [{\citenamefont {Hu}\ \emph {et~al.}(2023)\citenamefont {Hu},
  \citenamefont {Chen},\ and\ \citenamefont {Law}}]{hu2023anomalous}%
  \BibitemOpen
  \bibfield  {author} {\bibinfo {author} {\bibfnamefont {J.-X.}\ \bibnamefont
  {Hu}}, \bibinfo {author} {\bibfnamefont {S.~A.}\ \bibnamefont {Chen}},\ and\
  \bibinfo {author} {\bibfnamefont {K.~T.}\ \bibnamefont {Law}},\ }\href@noop
  {} {\bibinfo {title} {Anomalous coherence length in superconductors with
  quantum metric}} (\bibinfo {year} {2023}),\ \Eprint
  {https://arxiv.org/abs/2308.05686} {arXiv:2308.05686 [cond-mat.supr-con]}
  \BibitemShut {NoStop}%
\bibitem [{\citenamefont {Ledwith}\ \emph {et~al.}(2020)\citenamefont
  {Ledwith}, \citenamefont {Tarnopolsky}, \citenamefont {Khalaf},\ and\
  \citenamefont {Vishwanath}}]{fracTBG2020}%
  \BibitemOpen
  \bibfield  {author} {\bibinfo {author} {\bibfnamefont {P.~J.}\ \bibnamefont
  {Ledwith}}, \bibinfo {author} {\bibfnamefont {G.}~\bibnamefont
  {Tarnopolsky}}, \bibinfo {author} {\bibfnamefont {E.}~\bibnamefont
  {Khalaf}},\ and\ \bibinfo {author} {\bibfnamefont {A.}~\bibnamefont
  {Vishwanath}},\ }\bibfield  {title} {\bibinfo {title} {Fractional chern
  insulator states in twisted bilayer graphene: An analytical approach},\
  }\href {https://doi.org/10.1103/PhysRevResearch.2.023237} {\bibfield
  {journal} {\bibinfo  {journal} {Phys. Rev. Res.}\ }\textbf {\bibinfo {volume}
  {2}},\ \bibinfo {pages} {023237} (\bibinfo {year} {2020})}\BibitemShut
  {NoStop}%
\bibitem [{\citenamefont {Yao}\ \emph {et~al.}(2008)\citenamefont {Yao},
  \citenamefont {Xiao},\ and\ \citenamefont {Niu}}]{valley_symmetry2008}%
  \BibitemOpen
  \bibfield  {author} {\bibinfo {author} {\bibfnamefont {W.}~\bibnamefont
  {Yao}}, \bibinfo {author} {\bibfnamefont {D.}~\bibnamefont {Xiao}},\ and\
  \bibinfo {author} {\bibfnamefont {Q.}~\bibnamefont {Niu}},\ }\bibfield
  {title} {\bibinfo {title} {Valley-dependent optoelectronics from inversion
  symmetry breaking},\ }\href {https://doi.org/10.1103/PhysRevB.77.235406}
  {\bibfield  {journal} {\bibinfo  {journal} {Phys. Rev. B}\ }\textbf {\bibinfo
  {volume} {77}},\ \bibinfo {pages} {235406} (\bibinfo {year}
  {2008})}\BibitemShut {NoStop}%
\bibitem [{\citenamefont {Xiao}\ \emph {et~al.}(2012)\citenamefont {Xiao},
  \citenamefont {Liu}, \citenamefont {Feng}, \citenamefont {Xu},\ and\
  \citenamefont {Yao}}]{Valley_selection2012}%
  \BibitemOpen
  \bibfield  {author} {\bibinfo {author} {\bibfnamefont {D.}~\bibnamefont
  {Xiao}}, \bibinfo {author} {\bibfnamefont {G.-B.}\ \bibnamefont {Liu}},
  \bibinfo {author} {\bibfnamefont {W.}~\bibnamefont {Feng}}, \bibinfo {author}
  {\bibfnamefont {X.}~\bibnamefont {Xu}},\ and\ \bibinfo {author}
  {\bibfnamefont {W.}~\bibnamefont {Yao}},\ }\bibfield  {title} {\bibinfo
  {title} {Coupled spin and valley physics in monolayers of
  ${\mathrm{mos}}_{2}$ and other group-vi dichalcogenides},\ }\href
  {https://doi.org/10.1103/PhysRevLett.108.196802} {\bibfield  {journal}
  {\bibinfo  {journal} {Phys. Rev. Lett.}\ }\textbf {\bibinfo {volume} {108}},\
  \bibinfo {pages} {196802} (\bibinfo {year} {2012})}\BibitemShut {NoStop}%
\bibitem [{\citenamefont {Zeng}\ \emph {et~al.}(2012)\citenamefont {Zeng},
  \citenamefont {Dai}, \citenamefont {Yao}, \citenamefont {Xiao},\ and\
  \citenamefont {Cui}}]{Valleyoptical2012}%
  \BibitemOpen
  \bibfield  {author} {\bibinfo {author} {\bibfnamefont {H.}~\bibnamefont
  {Zeng}}, \bibinfo {author} {\bibfnamefont {J.}~\bibnamefont {Dai}}, \bibinfo
  {author} {\bibfnamefont {W.}~\bibnamefont {Yao}}, \bibinfo {author}
  {\bibfnamefont {D.}~\bibnamefont {Xiao}},\ and\ \bibinfo {author}
  {\bibfnamefont {X.}~\bibnamefont {Cui}},\ }\bibfield  {title} {\bibinfo
  {title} {Valley polarization in mos2 monolayers by optical pumping},\ }\href
  {https://doi.org/10.1038/nnano.2012.95} {\bibfield  {journal} {\bibinfo
  {journal} {Nature Nanotechnology}\ }\textbf {\bibinfo {volume} {7}},\
  \bibinfo {pages} {490} (\bibinfo {year} {2012})}\BibitemShut {NoStop}%
\bibitem [{\citenamefont {Mak}\ \emph {et~al.}(2012)\citenamefont {Mak},
  \citenamefont {He}, \citenamefont {Shan},\ and\ \citenamefont
  {Heinz}}]{Valley_helicity2012}%
  \BibitemOpen
  \bibfield  {author} {\bibinfo {author} {\bibfnamefont {K.~F.}\ \bibnamefont
  {Mak}}, \bibinfo {author} {\bibfnamefont {K.}~\bibnamefont {He}}, \bibinfo
  {author} {\bibfnamefont {J.}~\bibnamefont {Shan}},\ and\ \bibinfo {author}
  {\bibfnamefont {T.~F.}\ \bibnamefont {Heinz}},\ }\bibfield  {title} {\bibinfo
  {title} {Control of valley polarization in monolayer mos2 by optical
  helicity},\ }\href {https://doi.org/10.1038/nnano.2012.96} {\bibfield
  {journal} {\bibinfo  {journal} {Nature Nanotechnology}\ }\textbf {\bibinfo
  {volume} {7}},\ \bibinfo {pages} {494} (\bibinfo {year} {2012})}\BibitemShut
  {NoStop}%
\bibitem [{\citenamefont {Schaibley}\ \emph {et~al.}(2016)\citenamefont
  {Schaibley}, \citenamefont {Yu}, \citenamefont {Clark}, \citenamefont
  {Rivera}, \citenamefont {Ross}, \citenamefont {Seyler}, \citenamefont {Yao},\
  and\ \citenamefont {Xu}}]{ValleytronicsView2016}%
  \BibitemOpen
  \bibfield  {author} {\bibinfo {author} {\bibfnamefont {J.~R.}\ \bibnamefont
  {Schaibley}}, \bibinfo {author} {\bibfnamefont {H.}~\bibnamefont {Yu}},
  \bibinfo {author} {\bibfnamefont {G.}~\bibnamefont {Clark}}, \bibinfo
  {author} {\bibfnamefont {P.}~\bibnamefont {Rivera}}, \bibinfo {author}
  {\bibfnamefont {J.~S.}\ \bibnamefont {Ross}}, \bibinfo {author}
  {\bibfnamefont {K.~L.}\ \bibnamefont {Seyler}}, \bibinfo {author}
  {\bibfnamefont {W.}~\bibnamefont {Yao}},\ and\ \bibinfo {author}
  {\bibfnamefont {X.}~\bibnamefont {Xu}},\ }\bibfield  {title} {\bibinfo
  {title} {Valleytronics in 2d materials},\ }\href
  {https://doi.org/10.1038/natrevmats.2016.55} {\bibfield  {journal} {\bibinfo
  {journal} {Nature Reviews Materials}\ }\textbf {\bibinfo {volume} {1}},\
  \bibinfo {pages} {16055} (\bibinfo {year} {2016})}\BibitemShut {NoStop}%
\bibitem [{\citenamefont {Karch}\ \emph {et~al.}(2010)\citenamefont {Karch},
  \citenamefont {Olbrich}, \citenamefont {Schmalzbauer}, \citenamefont {Zoth},
  \citenamefont {Brinsteiner}, \citenamefont {Fehrenbacher}, \citenamefont
  {Wurstbauer}, \citenamefont {Glazov}, \citenamefont {Tarasenko},
  \citenamefont {Ivchenko}, \citenamefont {Weiss}, \citenamefont {Eroms},
  \citenamefont {Yakimova}, \citenamefont {Lara-Avila}, \citenamefont
  {Kubatkin},\ and\ \citenamefont {Ganichev}}]{lightHall2010}%
  \BibitemOpen
  \bibfield  {author} {\bibinfo {author} {\bibfnamefont {J.}~\bibnamefont
  {Karch}}, \bibinfo {author} {\bibfnamefont {P.}~\bibnamefont {Olbrich}},
  \bibinfo {author} {\bibfnamefont {M.}~\bibnamefont {Schmalzbauer}}, \bibinfo
  {author} {\bibfnamefont {C.}~\bibnamefont {Zoth}}, \bibinfo {author}
  {\bibfnamefont {C.}~\bibnamefont {Brinsteiner}}, \bibinfo {author}
  {\bibfnamefont {M.}~\bibnamefont {Fehrenbacher}}, \bibinfo {author}
  {\bibfnamefont {U.}~\bibnamefont {Wurstbauer}}, \bibinfo {author}
  {\bibfnamefont {M.~M.}\ \bibnamefont {Glazov}}, \bibinfo {author}
  {\bibfnamefont {S.~A.}\ \bibnamefont {Tarasenko}}, \bibinfo {author}
  {\bibfnamefont {E.~L.}\ \bibnamefont {Ivchenko}}, \bibinfo {author}
  {\bibfnamefont {D.}~\bibnamefont {Weiss}}, \bibinfo {author} {\bibfnamefont
  {J.}~\bibnamefont {Eroms}}, \bibinfo {author} {\bibfnamefont
  {R.}~\bibnamefont {Yakimova}}, \bibinfo {author} {\bibfnamefont
  {S.}~\bibnamefont {Lara-Avila}}, \bibinfo {author} {\bibfnamefont
  {S.}~\bibnamefont {Kubatkin}},\ and\ \bibinfo {author} {\bibfnamefont
  {S.~D.}\ \bibnamefont {Ganichev}},\ }\bibfield  {title} {\bibinfo {title}
  {Dynamic hall effect driven by circularly polarized light in a graphene
  layer},\ }\href {https://doi.org/10.1103/PhysRevLett.105.227402} {\bibfield
  {journal} {\bibinfo  {journal} {Phys. Rev. Lett.}\ }\textbf {\bibinfo
  {volume} {105}},\ \bibinfo {pages} {227402} (\bibinfo {year}
  {2010})}\BibitemShut {NoStop}%
\bibitem [{\citenamefont {Sato}\ \emph {et~al.}(2019)\citenamefont {Sato},
  \citenamefont {McIver}, \citenamefont {Nuske}, \citenamefont {Tang},
  \citenamefont {Jotzu}, \citenamefont {Schulte}, \citenamefont {H\"ubener},
  \citenamefont {De~Giovannini}, \citenamefont {Mathey}, \citenamefont
  {Sentef}, \citenamefont {Cavalleri},\ and\ \citenamefont
  {Rubio}}]{lightHall2019}%
  \BibitemOpen
  \bibfield  {author} {\bibinfo {author} {\bibfnamefont {S.~A.}\ \bibnamefont
  {Sato}}, \bibinfo {author} {\bibfnamefont {J.~W.}\ \bibnamefont {McIver}},
  \bibinfo {author} {\bibfnamefont {M.}~\bibnamefont {Nuske}}, \bibinfo
  {author} {\bibfnamefont {P.}~\bibnamefont {Tang}}, \bibinfo {author}
  {\bibfnamefont {G.}~\bibnamefont {Jotzu}}, \bibinfo {author} {\bibfnamefont
  {B.}~\bibnamefont {Schulte}}, \bibinfo {author} {\bibfnamefont
  {H.}~\bibnamefont {H\"ubener}}, \bibinfo {author} {\bibfnamefont
  {U.}~\bibnamefont {De~Giovannini}}, \bibinfo {author} {\bibfnamefont
  {L.}~\bibnamefont {Mathey}}, \bibinfo {author} {\bibfnamefont {M.~A.}\
  \bibnamefont {Sentef}}, \bibinfo {author} {\bibfnamefont {A.}~\bibnamefont
  {Cavalleri}},\ and\ \bibinfo {author} {\bibfnamefont {A.}~\bibnamefont
  {Rubio}},\ }\bibfield  {title} {\bibinfo {title} {Microscopic theory for the
  light-induced anomalous hall effect in graphene},\ }\href
  {https://doi.org/10.1103/PhysRevB.99.214302} {\bibfield  {journal} {\bibinfo
  {journal} {Phys. Rev. B}\ }\textbf {\bibinfo {volume} {99}},\ \bibinfo
  {pages} {214302} (\bibinfo {year} {2019})}\BibitemShut {NoStop}%
\bibitem [{\citenamefont {McIver}\ \emph
  {et~al.}(2020{\natexlab{b}})\citenamefont {McIver}, \citenamefont {Schulte},
  \citenamefont {Stein}, \citenamefont {Matsuyama}, \citenamefont {Jotzu},
  \citenamefont {Meier},\ and\ \citenamefont {Cavalleri}}]{lightHall2020}%
  \BibitemOpen
  \bibfield  {author} {\bibinfo {author} {\bibfnamefont {J.~W.}\ \bibnamefont
  {McIver}}, \bibinfo {author} {\bibfnamefont {B.}~\bibnamefont {Schulte}},
  \bibinfo {author} {\bibfnamefont {F.~U.}\ \bibnamefont {Stein}}, \bibinfo
  {author} {\bibfnamefont {T.}~\bibnamefont {Matsuyama}}, \bibinfo {author}
  {\bibfnamefont {G.}~\bibnamefont {Jotzu}}, \bibinfo {author} {\bibfnamefont
  {G.}~\bibnamefont {Meier}},\ and\ \bibinfo {author} {\bibfnamefont
  {A.}~\bibnamefont {Cavalleri}},\ }\bibfield  {title} {\bibinfo {title}
  {Light-induced anomalous hall effect in graphene},\ }\href
  {https://doi.org/10.1038/s41567-019-0698-y} {\bibfield  {journal} {\bibinfo
  {journal} {Nature Physics}\ }\textbf {\bibinfo {volume} {16}},\ \bibinfo
  {pages} {38} (\bibinfo {year} {2020}{\natexlab{b}})}\BibitemShut {NoStop}%
\bibitem [{\citenamefont {Koksma}\ \emph {et~al.}(2011)\citenamefont {Koksma},
  \citenamefont {Prokopec},\ and\ \citenamefont {Schmidt}}]{decoherence2011}%
  \BibitemOpen
  \bibfield  {author} {\bibinfo {author} {\bibfnamefont {J.~F.}\ \bibnamefont
  {Koksma}}, \bibinfo {author} {\bibfnamefont {T.}~\bibnamefont {Prokopec}},\
  and\ \bibinfo {author} {\bibfnamefont {M.~G.}\ \bibnamefont {Schmidt}},\
  }\bibfield  {title} {\bibinfo {title} {Decoherence in an interacting quantum
  field theory: Thermal case},\ }\href
  {https://doi.org/10.1103/PhysRevD.83.085011} {\bibfield  {journal} {\bibinfo
  {journal} {Phys. Rev. D}\ }\textbf {\bibinfo {volume} {83}},\ \bibinfo
  {pages} {085011} (\bibinfo {year} {2011})}\BibitemShut {NoStop}%
\bibitem [{\citenamefont {Arg\"uello-Luengo}\ \emph {et~al.}(2015)\citenamefont
  {Arg\"uello-Luengo}, \citenamefont {S\'anchez},\ and\ \citenamefont
  {L\'opez}}]{decoherence2015}%
  \BibitemOpen
  \bibfield  {author} {\bibinfo {author} {\bibfnamefont {J.}~\bibnamefont
  {Arg\"uello-Luengo}}, \bibinfo {author} {\bibfnamefont {D.}~\bibnamefont
  {S\'anchez}},\ and\ \bibinfo {author} {\bibfnamefont {R.}~\bibnamefont
  {L\'opez}},\ }\bibfield  {title} {\bibinfo {title} {Heat asymmetries in
  nanoscale conductors: The role of decoherence and inelasticity},\ }\href
  {https://doi.org/10.1103/PhysRevB.91.165431} {\bibfield  {journal} {\bibinfo
  {journal} {Phys. Rev. B}\ }\textbf {\bibinfo {volume} {91}},\ \bibinfo
  {pages} {165431} (\bibinfo {year} {2015})}\BibitemShut {NoStop}%
\bibitem [{\citenamefont {Matsumoto}\ \emph {et~al.}(2016)\citenamefont
  {Matsumoto}, \citenamefont {Komori}, \citenamefont {Ito}, \citenamefont
  {Michimura},\ and\ \citenamefont {Aso}}]{decoherence2016}%
  \BibitemOpen
  \bibfield  {author} {\bibinfo {author} {\bibfnamefont {N.}~\bibnamefont
  {Matsumoto}}, \bibinfo {author} {\bibfnamefont {K.}~\bibnamefont {Komori}},
  \bibinfo {author} {\bibfnamefont {S.}~\bibnamefont {Ito}}, \bibinfo {author}
  {\bibfnamefont {Y.}~\bibnamefont {Michimura}},\ and\ \bibinfo {author}
  {\bibfnamefont {Y.}~\bibnamefont {Aso}},\ }\bibfield  {title} {\bibinfo
  {title} {Direct measurement of optical-trap-induced decoherence},\ }\href
  {https://doi.org/10.1103/PhysRevA.94.033822} {\bibfield  {journal} {\bibinfo
  {journal} {Phys. Rev. A}\ }\textbf {\bibinfo {volume} {94}},\ \bibinfo
  {pages} {033822} (\bibinfo {year} {2016})}\BibitemShut {NoStop}%
\bibitem [{\citenamefont {Klembt}\ \emph {et~al.}(2018)\citenamefont {Klembt},
  \citenamefont {Stepanov}, \citenamefont {Klein}, \citenamefont {Minguzzi},\
  and\ \citenamefont {Richard}}]{decoherence2018}%
  \BibitemOpen
  \bibfield  {author} {\bibinfo {author} {\bibfnamefont {S.}~\bibnamefont
  {Klembt}}, \bibinfo {author} {\bibfnamefont {P.}~\bibnamefont {Stepanov}},
  \bibinfo {author} {\bibfnamefont {T.}~\bibnamefont {Klein}}, \bibinfo
  {author} {\bibfnamefont {A.}~\bibnamefont {Minguzzi}},\ and\ \bibinfo
  {author} {\bibfnamefont {M.}~\bibnamefont {Richard}},\ }\bibfield  {title}
  {\bibinfo {title} {Thermal decoherence of a nonequilibrium polariton fluid},\
  }\href {https://doi.org/10.1103/PhysRevLett.120.035301} {\bibfield  {journal}
  {\bibinfo  {journal} {Phys. Rev. Lett.}\ }\textbf {\bibinfo {volume} {120}},\
  \bibinfo {pages} {035301} (\bibinfo {year} {2018})}\BibitemShut {NoStop}%
\bibitem [{\citenamefont {Popovic}\ \emph {et~al.}(2023)\citenamefont
  {Popovic}, \citenamefont {Mitchison},\ and\ \citenamefont
  {Goold}}]{decoherence2023}%
  \BibitemOpen
  \bibfield  {author} {\bibinfo {author} {\bibfnamefont {M.}~\bibnamefont
  {Popovic}}, \bibinfo {author} {\bibfnamefont {M.~T.}\ \bibnamefont
  {Mitchison}},\ and\ \bibinfo {author} {\bibfnamefont {J.}~\bibnamefont
  {Goold}},\ }\bibfield  {title} {\bibinfo {title} {Thermodynamics of
  decoherence},\ }\href {https://doi.org/10.1098/rspa.2023.0040} {\bibfield
  {journal} {\bibinfo  {journal} {Proceedings of the Royal Society A:
  Mathematical, Physical and Engineering Sciences}\ }\textbf {\bibinfo {volume}
  {479}},\ \bibinfo {pages} {20230040} (\bibinfo {year} {2023})}\BibitemShut
  {NoStop}%
\bibitem [{\citenamefont {Nguyen}\ \emph {et~al.}(2023)\citenamefont {Nguyen},
  \citenamefont {Arwas}, \citenamefont {Lin}, \citenamefont {Yao},\ and\
  \citenamefont {Ciuti}}]{yaowang2023}%
  \BibitemOpen
  \bibfield  {author} {\bibinfo {author} {\bibfnamefont {D.-P.}\ \bibnamefont
  {Nguyen}}, \bibinfo {author} {\bibfnamefont {G.}~\bibnamefont {Arwas}},
  \bibinfo {author} {\bibfnamefont {Z.}~\bibnamefont {Lin}}, \bibinfo {author}
  {\bibfnamefont {W.}~\bibnamefont {Yao}},\ and\ \bibinfo {author}
  {\bibfnamefont {C.}~\bibnamefont {Ciuti}},\ }\bibfield  {title} {\bibinfo
  {title} {Electron-photon chern number in cavity-embedded 2d moir\'e
  materials},\ }\href {https://doi.org/10.1103/PhysRevLett.131.176602}
  {\bibfield  {journal} {\bibinfo  {journal} {Phys. Rev. Lett.}\ }\textbf
  {\bibinfo {volume} {131}},\ \bibinfo {pages} {176602} (\bibinfo {year}
  {2023})}\BibitemShut {NoStop}%
\end{thebibliography}%


\begin{thebibliography}{16}%
\makeatletter
\providecommand \@ifxundefined [1]{%
 \@ifx{#1\undefined}
}%
\providecommand \@ifnum [1]{%
 \ifnum #1\expandafter \@firstoftwo
 \else \expandafter \@secondoftwo
 \fi
}%
\providecommand \@ifx [1]{%
 \ifx #1\expandafter \@firstoftwo
 \else \expandafter \@secondoftwo
 \fi
}%
\providecommand \natexlab [1]{#1}%
\providecommand \enquote  [1]{``#1''}%
\providecommand \bibnamefont  [1]{#1}%
\providecommand \bibfnamefont [1]{#1}%
\providecommand \citenamefont [1]{#1}%
\providecommand \href@noop [0]{\@secondoftwo}%
\providecommand \href [0]{\begingroup \@sanitize@url \@href}%
\providecommand \@href[1]{\@@startlink{#1}\@@href}%
\providecommand \@@href[1]{\endgroup#1\@@endlink}%
\providecommand \@sanitize@url [0]{\catcode `\\12\catcode `\$12\catcode
  `\&12\catcode `\#12\catcode `\^12\catcode `\_12\catcode `\%12\relax}%
\providecommand \@@startlink[1]{}%
\providecommand \@@endlink[0]{}%
\providecommand \url  [0]{\begingroup\@sanitize@url \@url }%
\providecommand \@url [1]{\endgroup\@href {#1}{\urlprefix }}%
\providecommand \urlprefix  [0]{URL }%
\providecommand \Eprint [0]{\href }%
\providecommand \doibase [0]{https://doi.org/}%
\providecommand \selectlanguage [0]{\@gobble}%
\providecommand \bibinfo  [0]{\@secondoftwo}%
\providecommand \bibfield  [0]{\@secondoftwo}%
\providecommand \translation [1]{[#1]}%
\providecommand \BibitemOpen [0]{}%
\providecommand \bibitemStop [0]{}%
\providecommand \bibitemNoStop [0]{.\EOS\space}%
\providecommand \EOS [0]{\spacefactor3000\relax}%
\providecommand \BibitemShut  [1]{\csname bibitem#1\endcsname}%
\let\auto@bib@innerbib\@empty
%</preamble>
\bibitem [{\citenamefont {Ashida}\ \emph {et~al.}(2021)\citenamefont {Ashida},
  \citenamefont {\ifmmode \dot{I}\else \.{I}\fi{}mamo\ifmmode~\breve{g}\else
  \u{g}\fi{}lu},\ and\ \citenamefont {Demler}}]{CavityQED2021}%
  \BibitemOpen
  \bibfield  {author} {\bibinfo {author} {\bibfnamefont {Y.}~\bibnamefont
  {Ashida}}, \bibinfo {author} {\bibfnamefont {A.~m.~c.}\ \bibnamefont
  {\ifmmode \dot{I}\else \.{I}\fi{}mamo\ifmmode~\breve{g}\else \u{g}\fi{}lu}},\
  and\ \bibinfo {author} {\bibfnamefont {E.}~\bibnamefont {Demler}},\
  }\bibfield  {title} {\bibinfo {title} {Cavity quantum electrodynamics at
  arbitrary light-matter coupling strengths},\ }\href
  {https://doi.org/10.1103/PhysRevLett.126.153603} {\bibfield  {journal}
  {\bibinfo  {journal} {Phys. Rev. Lett.}\ }\textbf {\bibinfo {volume} {126}},\
  \bibinfo {pages} {153603} (\bibinfo {year} {2021})}\BibitemShut {NoStop}%
\bibitem [{\citenamefont {Masuki}\ and\ \citenamefont
  {Ashida}(2023)}]{Ashida2023}%
  \BibitemOpen
  \bibfield  {author} {\bibinfo {author} {\bibfnamefont {K.}~\bibnamefont
  {Masuki}}\ and\ \bibinfo {author} {\bibfnamefont {Y.}~\bibnamefont
  {Ashida}},\ }\bibfield  {title} {\bibinfo {title} {Berry phase and topology
  in ultrastrongly coupled quantum light-matter systems},\ }\href
  {https://doi.org/10.1103/PhysRevB.107.195104} {\bibfield  {journal} {\bibinfo
   {journal} {Phys. Rev. B}\ }\textbf {\bibinfo {volume} {107}},\ \bibinfo
  {pages} {195104} (\bibinfo {year} {2023})}\BibitemShut {NoStop}%
\bibitem [{\citenamefont {Resta}(1994)}]{RMP_Resta1994}%
  \BibitemOpen
  \bibfield  {author} {\bibinfo {author} {\bibfnamefont {R.}~\bibnamefont
  {Resta}},\ }\bibfield  {title} {\bibinfo {title} {Macroscopic polarization in
  crystalline dielectrics: the geometric phase approach},\ }\href
  {https://doi.org/10.1103/RevModPhys.66.899} {\bibfield  {journal} {\bibinfo
  {journal} {Rev. Mod. Phys.}\ }\textbf {\bibinfo {volume} {66}},\ \bibinfo
  {pages} {899} (\bibinfo {year} {1994})}\BibitemShut {NoStop}%
\bibitem [{\citenamefont {Marzari}\ and\ \citenamefont
  {Vanderbilt}(1997)}]{Vanderbilt1997}%
  \BibitemOpen
  \bibfield  {author} {\bibinfo {author} {\bibfnamefont {N.}~\bibnamefont
  {Marzari}}\ and\ \bibinfo {author} {\bibfnamefont {D.}~\bibnamefont
  {Vanderbilt}},\ }\bibfield  {title} {\bibinfo {title} {Maximally localized
  generalized wannier functions for composite energy bands},\ }\href
  {https://doi.org/10.1103/PhysRevB.56.12847} {\bibfield  {journal} {\bibinfo
  {journal} {Phys. Rev. B}\ }\textbf {\bibinfo {volume} {56}},\ \bibinfo
  {pages} {12847} (\bibinfo {year} {1997})}\BibitemShut {NoStop}%
\bibitem [{\citenamefont {Resta}(2006)}]{Resta2006}%
  \BibitemOpen
  \bibfield  {author} {\bibinfo {author} {\bibfnamefont {R.}~\bibnamefont
  {Resta}},\ }\bibfield  {title} {\bibinfo {title} {{Kohn’s theory of the
  insulating state: A quantum-chemistry viewpoint}},\ }\href
  {https://doi.org/10.1063/1.2176604} {\bibfield  {journal} {\bibinfo
  {journal} {The Journal of Chemical Physics}\ }\textbf {\bibinfo {volume}
  {124}},\ \bibinfo {pages} {104104} (\bibinfo {year} {2006})}\BibitemShut
  {NoStop}%
\bibitem [{\citenamefont {Resta}(2011)}]{Resta2011}%
  \BibitemOpen
  \bibfield  {author} {\bibinfo {author} {\bibfnamefont {R.}~\bibnamefont
  {Resta}},\ }\bibfield  {title} {\bibinfo {title} {The insulating state of
  matter: a geometrical theory},\ }\href
  {https://doi.org/10.1140/epjb/e2010-10874-4} {\bibfield  {journal} {\bibinfo
  {journal} {The European Physical Journal B}\ }\textbf {\bibinfo {volume}
  {79}},\ \bibinfo {pages} {121} (\bibinfo {year} {2011})}\BibitemShut
  {NoStop}%
\bibitem [{\citenamefont {Ozawa}\ and\ \citenamefont
  {Mera}(2021)}]{relation2021}%
  \BibitemOpen
  \bibfield  {author} {\bibinfo {author} {\bibfnamefont {T.}~\bibnamefont
  {Ozawa}}\ and\ \bibinfo {author} {\bibfnamefont {B.}~\bibnamefont {Mera}},\
  }\bibfield  {title} {\bibinfo {title} {Relations between topology and the
  quantum metric for chern insulators},\ }\href
  {https://doi.org/10.1103/PhysRevB.104.045103} {\bibfield  {journal} {\bibinfo
   {journal} {Phys. Rev. B}\ }\textbf {\bibinfo {volume} {104}},\ \bibinfo
  {pages} {045103} (\bibinfo {year} {2021})}\BibitemShut {NoStop}%
\bibitem [{\citenamefont {Xie}\ \emph {et~al.}(2020)\citenamefont {Xie},
  \citenamefont {Song}, \citenamefont {Lian},\ and\ \citenamefont
  {Bernevig}}]{topobound2020}%
  \BibitemOpen
  \bibfield  {author} {\bibinfo {author} {\bibfnamefont {F.}~\bibnamefont
  {Xie}}, \bibinfo {author} {\bibfnamefont {Z.}~\bibnamefont {Song}}, \bibinfo
  {author} {\bibfnamefont {B.}~\bibnamefont {Lian}},\ and\ \bibinfo {author}
  {\bibfnamefont {B.~A.}\ \bibnamefont {Bernevig}},\ }\bibfield  {title}
  {\bibinfo {title} {Topology-bounded superfluid weight in twisted bilayer
  graphene},\ }\href {https://doi.org/10.1103/PhysRevLett.124.167002}
  {\bibfield  {journal} {\bibinfo  {journal} {Phys. Rev. Lett.}\ }\textbf
  {\bibinfo {volume} {124}},\ \bibinfo {pages} {167002} (\bibinfo {year}
  {2020})}\BibitemShut {NoStop}%
\bibitem [{\citenamefont {Qi}\ \emph {et~al.}(2006)\citenamefont {Qi},
  \citenamefont {Wu},\ and\ \citenamefont {Zhang}}]{QWZmodel2006}%
  \BibitemOpen
  \bibfield  {author} {\bibinfo {author} {\bibfnamefont {X.-L.}\ \bibnamefont
  {Qi}}, \bibinfo {author} {\bibfnamefont {Y.-S.}\ \bibnamefont {Wu}},\ and\
  \bibinfo {author} {\bibfnamefont {S.-C.}\ \bibnamefont {Zhang}},\ }\bibfield
  {title} {\bibinfo {title} {Topological quantization of the spin hall effect
  in two-dimensional paramagnetic semiconductors},\ }\href
  {https://doi.org/10.1103/PhysRevB.74.085308} {\bibfield  {journal} {\bibinfo
  {journal} {Phys. Rev. B}\ }\textbf {\bibinfo {volume} {74}},\ \bibinfo
  {pages} {085308} (\bibinfo {year} {2006})}\BibitemShut {NoStop}%
\bibitem [{\citenamefont {Thouless}\ \emph {et~al.}(1982)\citenamefont
  {Thouless}, \citenamefont {Kohmoto}, \citenamefont {Nightingale},\ and\
  \citenamefont {den Nijs}}]{TKNN}%
  \BibitemOpen
  \bibfield  {author} {\bibinfo {author} {\bibfnamefont {D.~J.}\ \bibnamefont
  {Thouless}}, \bibinfo {author} {\bibfnamefont {M.}~\bibnamefont {Kohmoto}},
  \bibinfo {author} {\bibfnamefont {M.~P.}\ \bibnamefont {Nightingale}},\ and\
  \bibinfo {author} {\bibfnamefont {M.}~\bibnamefont {den Nijs}},\ }\bibfield
  {title} {\bibinfo {title} {Quantized hall conductance in a two-dimensional
  periodic potential},\ }\href@noop {} {\bibfield  {journal} {\bibinfo
  {journal} {Phys. Rev. Lett.}\ }\textbf {\bibinfo {volume} {49}},\ \bibinfo
  {pages} {405} (\bibinfo {year} {1982})}\BibitemShut {NoStop}%
\bibitem [{\citenamefont {Ahn}\ \emph {et~al.}(2020)\citenamefont {Ahn},
  \citenamefont {Guo},\ and\ \citenamefont {Nagaosa}}]{guangyu2020}%
  \BibitemOpen
  \bibfield  {author} {\bibinfo {author} {\bibfnamefont {J.}~\bibnamefont
  {Ahn}}, \bibinfo {author} {\bibfnamefont {G.-Y.}\ \bibnamefont {Guo}},\ and\
  \bibinfo {author} {\bibfnamefont {N.}~\bibnamefont {Nagaosa}},\ }\bibfield
  {title} {\bibinfo {title} {Low-frequency divergence and quantum geometry of
  the bulk photovoltaic effect in topological semimetals},\ }\href
  {https://doi.org/10.1103/PhysRevX.10.041041} {\bibfield  {journal} {\bibinfo
  {journal} {Phys. Rev. X}\ }\textbf {\bibinfo {volume} {10}},\ \bibinfo
  {pages} {041041} (\bibinfo {year} {2020})}\BibitemShut {NoStop}%
\bibitem [{\citenamefont {Roy}(2014)}]{bandgeometry2014}%
  \BibitemOpen
  \bibfield  {author} {\bibinfo {author} {\bibfnamefont {R.}~\bibnamefont
  {Roy}},\ }\bibfield  {title} {\bibinfo {title} {Band geometry of fractional
  topological insulators},\ }\href {https://doi.org/10.1103/PhysRevB.90.165139}
  {\bibfield  {journal} {\bibinfo  {journal} {Phys. Rev. B}\ }\textbf {\bibinfo
  {volume} {90}},\ \bibinfo {pages} {165139} (\bibinfo {year}
  {2014})}\BibitemShut {NoStop}%
\bibitem [{\citenamefont {Ledwith}\ \emph {et~al.}(2020)\citenamefont
  {Ledwith}, \citenamefont {Tarnopolsky}, \citenamefont {Khalaf},\ and\
  \citenamefont {Vishwanath}}]{fracTBG2020}%
  \BibitemOpen
  \bibfield  {author} {\bibinfo {author} {\bibfnamefont {P.~J.}\ \bibnamefont
  {Ledwith}}, \bibinfo {author} {\bibfnamefont {G.}~\bibnamefont
  {Tarnopolsky}}, \bibinfo {author} {\bibfnamefont {E.}~\bibnamefont
  {Khalaf}},\ and\ \bibinfo {author} {\bibfnamefont {A.}~\bibnamefont
  {Vishwanath}},\ }\bibfield  {title} {\bibinfo {title} {Fractional chern
  insulator states in twisted bilayer graphene: An analytical approach},\
  }\href {https://doi.org/10.1103/PhysRevResearch.2.023237} {\bibfield
  {journal} {\bibinfo  {journal} {Phys. Rev. Res.}\ }\textbf {\bibinfo {volume}
  {2}},\ \bibinfo {pages} {023237} (\bibinfo {year} {2020})}\BibitemShut
  {NoStop}%
\bibitem [{\citenamefont {Haldane}(1988)}]{HaldaneModel}%
  \BibitemOpen
  \bibfield  {author} {\bibinfo {author} {\bibfnamefont {F.~D.~M.}\
  \bibnamefont {Haldane}},\ }\bibfield  {title} {\bibinfo {title} {Model for a
  quantum hall effect without landau levels: Condensed-matter realization of
  the "parity anomaly"},\ }\href {https://doi.org/10.1103/PhysRevLett.61.2015}
  {\bibfield  {journal} {\bibinfo  {journal} {Phys. Rev. Lett.}\ }\textbf
  {\bibinfo {volume} {61}},\ \bibinfo {pages} {2015} (\bibinfo {year}
  {1988})}\BibitemShut {NoStop}%
\bibitem [{\citenamefont {Kohn}(1959)}]{Kohn1959}%
  \BibitemOpen
  \bibfield  {author} {\bibinfo {author} {\bibfnamefont {W.}~\bibnamefont
  {Kohn}},\ }\bibfield  {title} {\bibinfo {title} {Theory of bloch electrons in
  a magnetic field: The effective hamiltonian},\ }\href
  {https://doi.org/10.1103/PhysRev.115.1460} {\bibfield  {journal} {\bibinfo
  {journal} {Phys. Rev.}\ }\textbf {\bibinfo {volume} {115}},\ \bibinfo {pages}
  {1460} (\bibinfo {year} {1959})}\BibitemShut {NoStop}%
\bibitem [{\citenamefont {Alexandradinata}\ \emph {et~al.}(2014)\citenamefont
  {Alexandradinata}, \citenamefont {Dai},\ and\ \citenamefont
  {Bernevig}}]{wilsonloop2014}%
  \BibitemOpen
  \bibfield  {author} {\bibinfo {author} {\bibfnamefont {A.}~\bibnamefont
  {Alexandradinata}}, \bibinfo {author} {\bibfnamefont {X.}~\bibnamefont
  {Dai}},\ and\ \bibinfo {author} {\bibfnamefont {B.~A.}\ \bibnamefont
  {Bernevig}},\ }\bibfield  {title} {\bibinfo {title} {Wilson-loop
  characterization of inversion-symmetric topological insulators},\ }\href
  {https://doi.org/10.1103/PhysRevB.89.155114} {\bibfield  {journal} {\bibinfo
  {journal} {Phys. Rev. B}\ }\textbf {\bibinfo {volume} {89}},\ \bibinfo
  {pages} {155114} (\bibinfo {year} {2014})}\BibitemShut {NoStop}%
\end{thebibliography}%

\end{document}

\title{Supplementary Material for ``Emergent Haldane model and photon number locking in chiral cavities"}
\author{Liu Yang} 
\affiliation{Tsung-Dao Lee Institute, Shanghai Jiao Tong University, Shanghai 200240, China}
\affiliation{School of Physics and Astronomy, Shanghai Jiao Tong University, Shanghai 200240, China}
%\email{}
\author{Qing-Dong Jiang}
\email{qingdong.jiang@sjtu.edu.cn}
\affiliation{Tsung-Dao Lee Institute, Shanghai Jiao Tong University, Shanghai 200240, China}
\affiliation{School of Physics and Astronomy, Shanghai Jiao Tong University, Shanghai 200240, China}
\affiliation{Shanghai Branch, Hefei National Laboratory, Shanghai 201315, China}
%\author{} 
%\email{}
%\affiliation{}
%\author{}
%\affiliation{}

\maketitle
\tableofcontents

\section{Effective Hamiltonian in the asymptotically decoupled frame }
The Hamiltonian for materials in a general single-mode cavity can be written as 
\begin{align} 
\hat{H}^C & =\frac{(\hat{\vec{p}}+e\hat{\mathbf{A}})^2}{2 m}+V(\vec{r})+\hbar \omega_c\left(\hat{a}^{\dagger} \hat{a}+\frac{1}{2}\right),\\
\hat{\mathbf{A}}&=A_0\left(\vec{f} \hat{a}+\vec{f}^* \hat{a}^{\dagger}\right).
\end{align}
Here, the vector potential $\hat{\mathbf{A}}$ is quantized in the Coulomb gauge. If we consider a generalized polarization as a superposition of right- and left-handed directions, $\vec{f}$ can be defined as 
\begin{align}
\vec{f}&=\vec{\epsilon}\cos{\theta}+\vec{\epsilon}^\ast\sin{\theta}\nonumber\\
   &=\vec{e}_x\cos(\frac{\pi}{4}-\theta)-i\vec{e}_y\sin(\frac{\pi}{4}-\theta),\\
   \vec{\epsilon}&=\frac{\vec{e}_x-\mathrm{i}\vec{e}_y}{\sqrt{2}}.\label{eq:Atheta}
\end{align}
When $\theta=0$ ($\pi/2$), $\vec{f}$ corresponds to the right-handed (left-handed) circular polarization. When $\theta=\pi/4$,  $\vec{f}=\vec{e}_x$.  

Then, we rewrite the gauge potential operator as
\begin{align}
    \hat{\mathbf{A}}&=A_0\left(c_x\vec{e}_x\hat{\phi}_x+c_y\vec{e}_y\hat{\phi}_y\right),\\
    \hat{\phi}_x&=\frac{a^\dagger+a}{\sqrt{2}},\quad\hat{\phi}_y=\mathrm{i}\frac{a^\dagger-a}{\sqrt{2}}\\
    c_x&=\sqrt{2}\cos{(\frac{\pi}{4}-\theta)},c_y=\sqrt{2}\sin{(\frac{\pi}{4}-\theta)}.
\end{align}
Here, $ \hat{\phi}_x$ and $ \hat{\phi}_y$ satisfy
\begin{align}
[ \hat{\phi}_x, \hat{\phi}_y]=\mathrm{i}.
\end{align}
The unitary transformation $\hat{U}$ that decouples the light-matter interaction asymptotically is
\begin{align}
    \hat{U}&=\exp\left[\mathrm{i}(\xi_x\hat{\phi}_y\hat{k}_x-\xi_y\hat{\phi}_x\hat{k}_y)\right],\label{eq:ADunitary}\\
    \xi_\mu&=\sqrt{\frac{\hbar}{m\omega_c}}\frac{c_\mu g/\omega_c}{1+c_\mu^2 g^2/\omega_c^2},\\
   \frac{g}{\omega_c}&=\frac{eA_0}{\sqrt{m\hbar\omega_c}}.
\end{align}
Here $\hat{k}_\mu=-\mathrm{i}\partial_{\mu}$. We then obtain the Hamiltonian in the asymptotically decoupled (AD) frame:
\begin{align}
     \hat{U}^\dagger\hat{H}^C\hat{U}&=\frac{\hat{p}_x^2}{2 m\gamma^2_x}+\frac{\hat{p}_y^2}{2 m\gamma^2_y}+V(\vec{r}+\hat{\vec{\tau}})+ \frac{\hbar\Omega}{2}\left(\hat{\Phi}_x^2+\hat{\Phi}_y^2\right),
   \end{align}
   where 
\begin{align}
    \Omega&= \omega_c\gamma_x\gamma_y,\\
    \gamma^2_\mu&=1+\frac{c_\mu^2g^2}{\omega_c^2},
\end{align}
and $\hat{\Phi}_x$ and $\hat{\Phi}_y$ are defined as
\begin{align}
    \hat{\Phi}_x&=\sqrt{\frac{\gamma_x}{\gamma_y}}\hat{\phi}_x, \hat{\Phi}_y=\sqrt{\frac{\gamma_y}{\gamma_x}}\hat{\phi}_y.
 \end{align}
 The commutation relation between $\hat{\Phi}_x$ and $\hat{\Phi}_y$ is normalized as 
\begin{align}
    [\hat{\Phi}_x,\hat{\Phi}_y]=\mathrm{i}.
\end{align}
We define a characteristic length
\begin{align}
    \xi&=\sqrt{\frac{\hbar}{m\omega_c}}\frac{g/\omega_c}{\gamma_x\gamma_y}.
\end{align}
Using $\xi$, we can express the operator $\hat{\vec{\tau}}$ as follow
\begin{align}
     \hat{\vec{\tau}}&=\frac{\xi^2}{2}\cos(2\theta)\hat{\vec{k}}\times\vec{e}_z+\xi\hat{\vec{\Pi}},
\end{align}
where the operator $\hat{\vec{\Pi}}$ is defined as 
\begin{align}
    \hat{\Pi}_x&=-c_x\sqrt{\frac{\gamma_y}{\gamma_x}}\hat{\Phi}_y,\\
    \hat{\Pi}_y&=c_y\sqrt{\frac{\gamma_x}{\gamma_y}}\hat{\Phi}_x.
\end{align}
Note that the commutation relation between the operators $\hat{\Pi}_x$ and $\hat{\Pi}_y$ is
\begin{align}
    [\hat{\Pi}_x,\hat{\Pi}_y]=\mathrm{i}\,\cos(2\theta).
\end{align}
By the above notations, the unitary transformation Eq.~(\ref{eq:ADunitary}) can be rewritten as 
\begin{align}
\hat{U}=\exp[-\mathrm{i}\xi\hat{\vec{\Pi}}\cdot\hat{\vec{k}}].
\end{align}
We now introduce the crystalline potential of the material by the Fourier series
\begin{align}
V(\vec{r})=\sum_{\vec{b}}V_{\vec{b}}e^{\mathrm{i}\vec{b}\cdot\vec{r}},
\end{align}
where $\vec{b}$ are the reciprocal lattice vectors. By the photon vacuum approximation in the AD frame, we derive the effective Hamiltonian as follows
\begin{align}
    \hat{H}^U&=\langle 0_\text{photon}|\hat{U}^\dagger\hat{H}^C\hat{U}|0_\text{photon}\rangle\nonumber\\
    &=\frac{\hat{p}_x^2}{2 m\gamma^2_x}+\frac{\hat{p}_y^2}{2 m\gamma^2_y}+\widetilde{V}\left(\vec{r}+\frac{\xi^2}{2}\cos(2\theta)\hat{\vec{k}}\times\vec{e}_z\right),\label{eq:Heff}
\end{align}
where the modified potential function is defined as 
\begin{align}
    \widetilde{V}(\vec{r})&=\sum_{\vec{b}}V_{\vec{b}}e^{\mathrm{i}\vec{b}\cdot\vec{r}}\langle 0_\text{photon}|e^{\mathrm{i}\xi\vec{b}\cdot\hat{\vec{\Pi}}}|0_\text{photon}\rangle+\frac{\hbar\Omega}{2}\nonumber\\
    &=\sum_{\vec{b}}\widetilde{V}_{\vec{b}}e^{\mathrm{i}\vec{b}\cdot\vec{r}}+\frac{\hbar\Omega}{2},\label{eq:tildeV}\\
    \widetilde{V}_{\vec{b}}&=V_{\vec{b}}e^{-\frac{\xi^2}{4}\left(b_x^2 c_x^2\frac{\gamma_y}{\gamma_x}+b_y^2c_y^2\frac{\gamma_x}{\gamma_y}\right)}.
\end{align}
For standard chiral cavities, $\theta=0$. The effective Hamiltonian (which we used in the main text) is
\begin{align}
    \hat{H}^U&=\langle 0_\text{photon}|\hat{U}^\dagger \hat{H}^C\hat{U}|0_\text{photon}\rangle\nonumber\\&=\frac{\hat{p}^2}{2m_{\text{eff}}}+\widetilde{V}(\boldsymbol{r}+\frac{\xi^2}{2\hbar}\hat{\boldsymbol{p}}\times \boldsymbol{e}_z),\label{eq:H_U}
    \end{align}
    where
    \begin{align}
        m_{\text{eff}}&=m(1+\frac{g^2}{\omega_c^2}),\\
        \xi&=\sqrt{\frac{\hbar}{m\omega_c}}\frac{g/\omega_c}{1+g^2/\omega_c^2}.
    \end{align}
As $\xi$ is usually small~\cite{CavityQED2021,Ashida2023}, we can expand this Hamiltonian to the first-order of $\xi$ and obtain the effective Hamiltonian as
\begin{align}
     \hat{H}^\text{eff}\approx\frac{[\hat{\boldsymbol{p}}+e\boldsymbol{A}(\boldsymbol{r})]^2}{2m_{\text{eff}}}+V_{\text{eff}}(\boldsymbol{r}),\label{eq:Heff_A}
\end{align}
where
\begin{align}
    \boldsymbol{A}(\boldsymbol{r})&=\frac{\beta}{2e\omega_c}\boldsymbol{e}_z\times\nabla \widetilde{V},\label{eq:Aeff}\\
    V_{\text{eff}}(\boldsymbol{r})&=\widetilde{V}(\boldsymbol{r})-\frac{\beta^3}{8m g^2}|\nabla\widetilde{V}(\boldsymbol{r})|^2.\label{eq:eff_V}
\end{align}

\section{The modified and effective potentials of graphene in chiral cavities}
\begin{figure}
    \centering
    \includegraphics[width=0.7\textwidth]{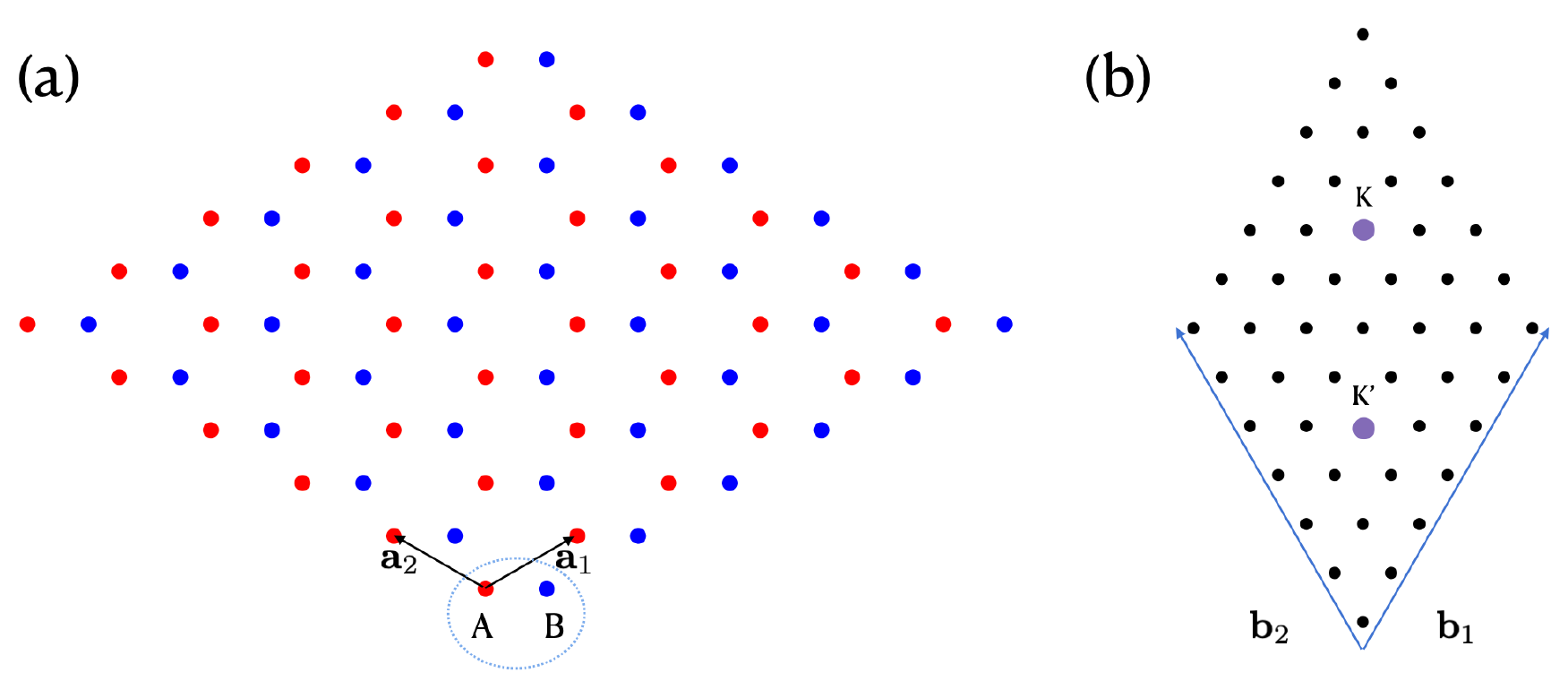}
    \caption{Lattices and reciprocal lattices of graphene.}\label{smfig:graphene}
\end{figure}

We show the graphene lattices in Fig.~\ref{smfig:graphene}, where the lattice and reciprocal lattice bases are defined as
\begin{align}
    \vec{a}_1&=\sqrt{3}a_0(\cos\frac{\pi}{6},\sin\frac{\pi}{6}),\nonumber\\
    \vec{a}_2&=\sqrt{3}a_0(\cos\frac{5\pi}{6},\sin\frac{5\pi}{6}),\nonumber\\
    \vec{b}_1&=\frac{4\pi}{3a_0}(\cos\frac{\pi}{3},\sin\frac{\pi}{3}),\nonumber\\
    \vec{b}_2&=\frac{4\pi}{3a_0}(\cos\frac{2\pi}{3},\sin\frac{2\pi}{3}),\label{eq:ab_bases}
\end{align}
where $a_0$ is the lattice constant, and we set it to be 1 for convenience. These vectors satisfy the orthonormalization relation
\begin{align}
\vec{a}_i\cdot\vec{b}_j=2\pi\delta_{ij}
\end{align} 
Any crystal momentum $\vec{k}$ in the Brillouin zone (BZ) can be expressed as
\begin{align}
    \vec{k}=\sum_{i=1,2}\frac{k_i}{2\pi}\vec{b}_i.\label{smeq:kv}
\end{align}
To cover the whole first BZ, the value of the coordinate $k_i$ should be taken from 0 to $2\pi$.

For the graphene continuum model we use in the main text, the potential function for the honeycomb lattices can be described by 
\begin{equation}
    V(\boldsymbol{r})=\sum_{i=1}^6\left(V_0+\frac{\Delta}{9}e^{-\mathrm{i}\boldsymbol{b}_i\cdot\boldsymbol{\delta}}\right)\exp(\mathrm{i}\boldsymbol{b}_i\cdot\boldsymbol{r}),\label{eq:potential}
\end{equation}
where $\boldsymbol{\delta}=(1,0)$, the reciprocal lattice vector $\boldsymbol{b}_i=4\pi(3a_0)^{-1}(\cos{2\pi i/3},\sin{2\pi i/3})$. 

\begin{figure*}
    \centering
\includegraphics[width=0.9\textwidth]{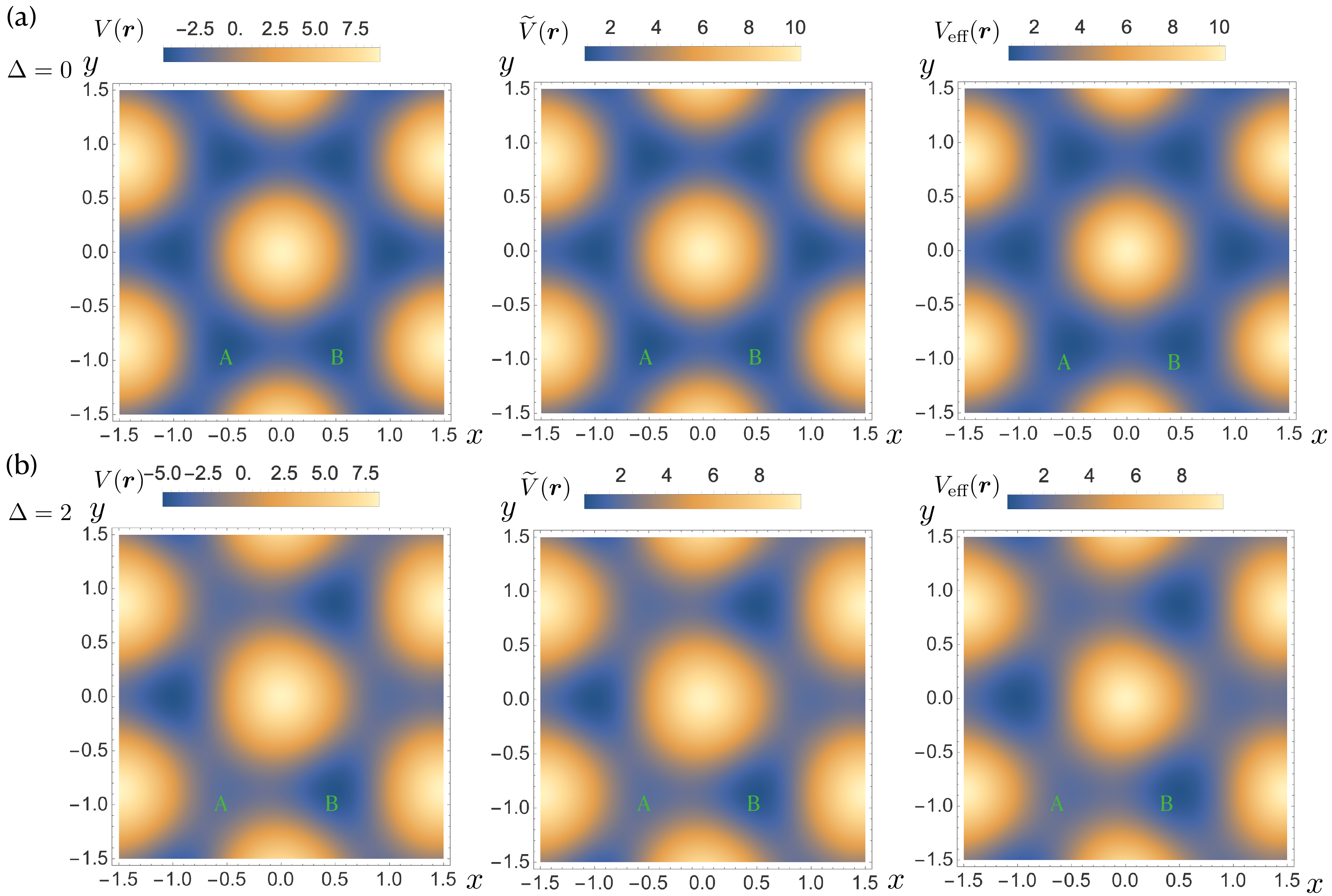}
    \caption{The potential functions of the cavity graphene at the coupling strength $g/\omega_c=1$. $V(\boldsymbol{r})$, $\widetilde{V}(\boldsymbol{r})$ and $V_{\text{eff}}(\boldsymbol{r})$ are defined in Eqs.~(\ref{eq:potential}), (\ref{eq:tildeV}) and (\ref{eq:eff_V}), respectively. We have chosen the material parameters $V_0=m=1.5$, and the cavity parameter $\omega_c=2$.  The corresponding sub-lattice split $\Delta$ in (a) and (b) have been written in the figures. }
    \label{smfig:potentials}
\end{figure*}

 In the effective Hamiltonians Eqs.~\ref{eq:Heff} and (\ref{eq:Heff_A}), we have defined a modified potential function $\widetilde{V}(\boldsymbol{r})$ and an effective potential $ V_{\text{eff}}(\boldsymbol{r})$. We now check whether these two potential functions still hold the honeycomb lattice structure as the original $V(\boldsymbol{r})$. In Figs. ~\ref{smfig:potentials} (a)-(b), we show these potential functions at the coupling strength $g/\omega_c=1$. Throughout our work, we set $a_0=1$, $\hbar=1$, and $e=1$ for convenience. Also, We will fix the material parameters $V_0=m=1.5$, and the cavity parameter $\omega_c=2$ in the later discussion. The sub-lattice onsite energy split $\Delta$ is $0$ or $2$ in (a) or (b), respectively. We see that the modified and effective potential functions still hold the same pattern as the potential of the bare graphene.

\section{Plane wave expansion of the effective Hamiltonian}
For the numerical calculation of the spectrum and the quantum geometric tensor of the cavity materials, we use plane-wave states $|\vec{K}\rangle$ ($\vec{K}=\sum_{i=1,2}K_i\vec{b}_i=(K_x,K_y)$) to expand the periodic Bloch states $|u_{n\vec{k}}\rangle$ ($\vec{k}=(k_x,k_y)$) as 
\begin{align}
|u_{n\vec{k}}\rangle=\sum_{\vec{K}}c_{\vec{K}}|\vec{K}\rangle.
\end{align}
Thus, the effective Hamiltonian at momentum $\vec{k}$ becomes
\begin{align} 
\hat{H}^{\text{eff}}_{\vec{k}}&=\sum_{\vec{K}}\left[\frac{\hbar^2(K_x+k_x)^2}{2m_x}+\frac{\hbar^2(K_y+k_y)^2}{2m_y}\right]|\vec{K}\rangle\langle\vec{K}|\nonumber\\
    &+\sum_{\vec{b}}\widetilde{V}_{\vec{b}}\sum_{\vec{K}}e^{\mathrm{i}\frac{\xi^2\cos(2\theta)}{2}[(\vec{k}+\vec{K})\times\vec{e}_z]\cdot\vec{b}}|\vec{K}+\vec{b}\rangle\langle\vec{K}|.
\end{align}
The corresponding derivative of $\hat{H}^{\text{eff}}_{\vec{k}}$ along $\mu$ direction is
\begin{align}
\partial_{k_\mu}\hat{H}^{\text{eff}}_{\vec{k}}&=\sum_{\vec{K}}\left[\frac{\hbar^2(K_x+k_x)}{m_x}+\frac{\hbar^2(K_y+k_y)}{m_y}\right]|\vec{K}\rangle\langle\vec{K}|\nonumber\\
    &+\sum_{\vec{b}}\widetilde{V}_{\vec{b}}\left(-\mathrm{i}\frac{\xi^2\cos(2\theta)}{2}\epsilon_{\mu\nu}b_\nu\right)\sum_{\vec{K}}e^{\mathrm{i}\frac{\xi^2\cos(2\theta)}{2}[(\vec{k}+\vec{K})\times\vec{e}_z]\cdot\vec{b}}|\vec{K}+\vec{b}\rangle\langle\vec{K}|.
\end{align}

\section{Brief review of the Quantum geometric tensor}
The quantum geometric tensor (QGT), or the Fubini-Study metric, describes the associated geometric and topological properties of the band structures. For the single occupied band, it is defined as
\begin{align}
    Q_{\mu\nu}(\vec{k})=\langle\partial_{k_\mu}u_{n}(\vec{k})|\partial_{k_\nu}u_{n}(\vec{k})\rangle-\mathcal{A}_\mu(\vec{k})\mathcal{A}_\nu(\vec{k}),
\end{align}
where 
\begin{align}
    \mathcal{A}_\mu(\vec{k})=\mathrm{i}\langle u_n(\vec{k})|\partial_{k_\mu}u_n(\vec{k})\rangle.
\end{align}
Since the periodic Bloch wave function can be made orthonormalized as
\begin{align}
    \int_{\text{unit cell}} d^D\vec{r}\,u^\ast_{n\vec{k}'}(\vec{r})u_{n\vec{k}}(\vec{r})=\delta_{\vec{k}'\vec{k}},
\end{align}
we can prove $\mathcal{A}_\mu$ is real. The real part of the quantum geometric tensor, which is also called the quantum metric, can be expressed as
\begin{align}
    g_{\mu\nu}&=\text{Re}[Q_{\mu\nu}]\nonumber\\
    &=-\langle u_{n}(\vec{k})|\partial_{k_\mu}\partial_{k_\nu}u_{n}(\vec{k})\rangle-\mathcal{A}_\mu(\vec{k})\mathcal{A}_\nu(\vec{k}),
\end{align}
which can be derived by calculating the differential change of the transition probability between $|u_n(\vec{k})\rangle$ at $\vec{k}$ and $|u_n(\vec{k})\rangle$ at $\vec{k}+\Delta\vec{k}$:
\begin{align}
  1-|\langle u_n(\vec{k})|u_n(\vec{k}+\Delta\vec{k})\rangle\big|^2=g_{\mu\nu}(\vec{k})dk^\mu dk^\nu.
\end{align}
The imaginary part of the quantum geometric tensor is proportional to the Berry curvature
\begin{align}
    \mathcal{F}_{\mu\nu}(\vec{k})&=-2\,\text{Im}[Q_{\mu\nu}]\nonumber\\
    &=\partial_{k_\mu}\mathcal{A}_\nu-\partial_{k_\nu}\mathcal{A}_\mu.
\end{align}
While the integral of the Berry curvature is a topological invariant, the integral of the quantum metric can be related to the spread of the Wannier states\cite{RMP_Resta1994,Vanderbilt1997,Resta2006,Resta2011}, which is regarded as a geometrical property. Interestingly, there are inequality relations between the real and imaginary part of the quantum geometric tensor as follows~\cite{relation2021},
    \begin{align}
    \text{Tr}(g)&\geq 2\sqrt{\text{det}(g)}\geq|\mathcal{F}_{xy}|.
\end{align}
The first and second inequalities above are caused by the semi-definitive properties of the tensor $g_{\mu\nu}$ and $Q_{\mu\nu}$, respectively. These inequality relations give the lower bound of the integral of the quantum metric~\cite{relation2021}. This lower bound recently has been used to explain the phase stiffness of the twisted bilayer graphene flat band~\cite{topobound2020}. 

Now, we introduce a useful formula to calculate the quantum geometric tensor of a single band (labeled by $n$):
\begin{align}
    Q_{\mu\nu}=\sum_{m}^{m\neq n}\frac{\langle u_{n}(\vec{k})|\partial_{k_\mu}\hat{H}_{\vec{k}}|u_{m}(\vec{k})\rangle\langle u_{m}(\vec{k})|\partial_{k_\nu}\hat{H}_{\vec{k}}|u_{n}(\vec{k})\rangle}{\left(E_{n}(\vec{k})-E_{m}(\vec{k})\right)^2},
\end{align}
where $\hat{H}_{\vec{k}}=\exp(-\mathrm{i}\vec{k}\cdot\vec{r})\hat{H}\exp(\mathrm{i}\vec{k}\cdot\vec{r})$. The power of this formula in the numerical calculation is that the Bloch wave functions do not need to have a smooth gauge at different crystal momentum $\vec{k}$. 

To compare the geometric tensor of the cavity graphene and the Haldane model, we show the tensor for a general two-band model. If we consider a two-band Dirac-type Hamiltonian 
\begin{align}
    \mathcal{H}(\vec{k})=\vec{d}(\vec{k})\cdot\vec{\sigma}.\label{Hb}
\end{align}
Here all components of $\vec{d}$ are real functions. Normally, we call $\vec{d}$ the characteristic vector. First, we show the eigenstates of Eq.~(\ref{Hb}) as follows
\begin{align}
    |\psi_+\rangle&=\left(\begin{array}{c}
    e^{-\mathrm{i}\phi}\cos{\frac{\theta}{2}}\\
    \sin{\frac{\theta}{2}}
    \end{array}\right),\\
    |\psi_-\rangle&=\left(\begin{array}{c}
    e^{-\mathrm{i}\phi}\sin{\frac{\theta}{2}}\\
    -\cos{\frac{\theta}{2}}
    \end{array}\right).\label{psi-}
\end{align}
where the eigenstates $|\psi_\pm\rangle$ have the eigenenergies $E_{\pm}(\vec{k})=\pm|\vec{d}(\vec{k})|$. The characteristic vector $\vec{d}$ is parameterized by the spherical angles $\theta$ and $\phi$. Then the quantum geometric tensor for the lowest band can be expressed as
    \begin{align}
    g_{\mu\nu}&=\frac{1}{4}\frac{\partial\vec{n}}{\partial k_\mu}\cdot\frac{\partial\vec{n}}{\partial k_\nu},\label{d_gij}\\
    \mathcal{F}_{xy}&=\frac{1}{2}(\frac{\partial\vec{n}}{\partial k_x}\times \frac{\partial\vec{n}}{\partial k_y})\cdot\vec{n},\\
    \vec{n}&=\frac{\vec{d}}{|\vec{d}|}.
\end{align}
The imaginary part of $Q_{\mu\nu}$ is related to a topological number $C_1$~\cite{QWZmodel2006}, which is the first Chern number~\cite{TKNN}:
 \begin{align}
 C_1&=\frac{1}{2\pi}\int_{\text{FBZ}}d^2k\,\mathcal{F}_{xy}=\frac{1}{4\pi}\int_{\text{FBZ}}d^2k\left(\frac{\partial\vec{n}}{\partial k_x}\times\frac{\partial \vec{n}}{\partial k_y}\right)\cdot\vec{n}.\label{eq:chern_1}
  \end{align}
The integrand element $(\frac{\partial\vec{n}}{\partial k_x}\times\frac{\partial \vec{n}}{\partial k_y})\cdot\vec{n}$ in Eq.~(\ref{eq:chern_1}) is the differential solid angle on the unit sphere formed by $\vec{n}$, which maps the FBZ (a Torus manifold) to a sphere. Hence, the first Chern number counts how many times $T^2$ wraps around $S^2$ through the mapping $f$. The phases with non-zero $C_1$ are non-trivial topological phases, while $C_1=0$ corresponds to the trivial phase. However, the integral of the quantum metric tensor $g_{\mu\nu}$ is not quantized by a topological invariant, which is thus a geometric quantity.

For a general Dirac Hamiltonian
  \begin{align}
      \mathcal{H}(\mathbf{k})=\sum_{i}f_i(\mathbf{k})\Gamma_i,
  \end{align}
where $f_i(\mathbf{k})$ is a real function and $d_M$-dimensional $\Gamma_i$ matrices satisfy the Clifford's algebra $\{\Gamma_i,\Gamma_j\}$,  $Q_{\mu\nu}$ has the following form~\cite{guangyu2020}
\begin{align}
    Q_{\mu\nu}=\sum_{i,j}\partial_\mu f_i\partial_\nu f_j\frac{d_M(\delta_{ij}-\hat{f}_i\hat{f}_j-i\epsilon_{ijk}\hat{f}_k)}{8f^2}.
\end{align}

\section{Quantum geometry of graphene and Haldane model}

This section will discuss the quantum geometry of the single-layer graphene lattice models. First, we introduce the tight-binding model of graphene with nearest-neighbor hopping $t$ and onsite energy split $\Delta$. The Hamiltonian of the graphene is
\begin{align}
    h=\left(\begin{array}{cc}
\Delta  & t[1+T(\vec{a}_1)+T^\dagger(\vec{a}_2)] \\ 
t[1+T^\dagger(\vec{a}_1)+T(\vec{a}_2)] &-\Delta
\end{array}\right),\label{eq:h_graphene}
\end{align}
where $T(\vec{a}_i)$ is the translation operator for $\vec{a}_i$ displacement, which can be expressed as $\sum_{\vec{R}}|\vec{R}+\vec{a}_i\rangle\langle \vec{R}|$. In momentum space, the Hamiltonian becomes
\begin{align}
     h(\vec{k})=\left(\begin{array}{cc}
\Delta  & t[1+e^{-\mathrm{i}k_1}+e^{\mathrm{i}k_2}] \\ 
t[1+e^{\mathrm{i}k_1}+e^{-\mathrm{i}k_2}] &-\Delta
\end{array}\right),\label{eq:hk_graphene}
\end{align}
where
\begin{align}
    \vec{k}&=\sum_{i}\frac{k_i}{2\pi}\vec{b}_i,\\
     k_i&=\vec{a}_i\cdot\vec{k}.
\end{align}
The corresponding characteristic vector is
\begin{align}
    \vec{d}(\vec{k})&=\left(t(1+\cos{k_1}+\cos{k_2}),t(1+\sin{k_1}-\sin{k_2}),\Delta\right).
\end{align}
There are two important points in the Brillouin zone called $K$ and $K'$ where the two energy bands touch, and the direction of the characteristic vector becomes ill-defined. The position vectors of $K$ and $K'$ are $\vec{K}=2(\vec{b}_1+\vec{b}_2)/3$ and $\vec{K}'=(\vec{b}_1+\vec{b}_2)/3=-\vec{K}+\vec{b}_1+\vec{b}_2$, respectively. Around $K$ and $K'$ points, the Hamiltonian becomes a massive Dirac Hamiltonian:
\begin{align}
   h_K(\vec{\kappa})&= h(\vec{K}+\vec{\kappa})\approx\Delta\sigma_z-\frac{3t}{2}(\kappa_x\sigma_y-\kappa_y\sigma_x),\\
   h_{K'}(\vec{\kappa})&= h(\vec{K}'+\vec{\kappa})\approx\Delta\sigma_z-\frac{3t}{2}(\kappa_x\sigma_y+\kappa_y\sigma_x).
\end{align}
We see that at $K$ and $K'$,  there is an energy gap $2\Delta$. When $\Delta\to0$, these Hamiltonians approximately describe Dirac-Weyl particles. The Berry curvature at $K$ ($\vec{K}$) or $K'$ ($-\vec{K}$)  point is approximated as
\begin{align}
    \mathcal{F}_{xy}(\pm\vec{K}+\vec{\kappa})&=\frac{1}{2}(\frac{\partial\vec{d}}{\partial \kappa_x}\times \frac{\partial\vec{d}}{\partial \kappa_y})\cdot\frac{\vec{d}}{d^3}\approx\pm\frac{\delta}{2(\vec{\kappa}^2+\delta^2)^{3/2}},\label{eq:grapheneFxy}
\end{align}
where $\delta=2\Delta/(3|t|)$. States around $K$ ($\vec{K}$) or $K'$ ($-\vec{K}$)  point have opposite signs of the Berry curvature and will produce inverse Berry phases as follows
\begin{align}
    \int d^2\vec{\kappa}\,\mathcal{F}_{xy}(\pm\vec{K}+\vec{\kappa})=\pm\pi\,\,\text{sign}(\delta),
\end{align}
When summarizing the Berry phases at $K$ and $K'$, one observes zero since the time-reversal symmetry always makes the total Berry phase vanish. 

We use the same way to calculate the quantum metric of Dirac particles around $K$ and $K'$ points. Let us first derive the results around $K$ point,
\begin{align}
    \frac{\partial \vec{n}}{\partial \kappa_\mu}&=-\frac{\epsilon_{\mu\nu}\mathbf{e}_\nu}{\sqrt{\kappa^2+\delta^2}}-\frac{\kappa_\mu \vec{n}}{\kappa^2+\delta^2},\\
    g_{xx}&=\frac{1}{4}\frac{\kappa_y^2+\delta^2}{(\kappa^2+\delta^2)^2},\\
    g_{yy}&=\frac{1}{4}\frac{\kappa_x^2+\delta^2}{(\kappa^2+\delta^2)^2},\\
    g_{xy}&=-\frac{1}{4}\frac{\kappa_x\kappa_y}{(\kappa^2+\delta^2)^2}.
\end{align}
We can see that all of the $g_{\mu\nu}$ components have the order $1/\kappa^2$ as $\kappa\to\infty$. Thus, the integrals of them all have ultraviolet divergence. However, we find that the quantum geometric volume $d^2k\sqrt{\text{det}(g)}$ for the Dirac particles has a convergent value after integrating over the total momentum space since the volume form
\begin{align}
    \sqrt{\text{det}(g(\pm\vec{K}+\vec{\kappa}))}=\frac{1}{2}|\mathcal{F}_{xy}(\pm\vec{K}+\vec{\kappa})|
\end{align} 
is equal to the absolute value of the Berry curvature. Unlike the total Berry phase, which vanishes, the quantum geometric volume 
\begin{align}
 \mathcal{V}= \frac{1}{\pi} \int_{\text{BZ}}d^2\vec{k}  \sqrt{\text{det}(g(\vec{k}))}\label{eq:QV}
\end{align}
for the graphene is approximated as 1.

\begin{figure}
    \centering
    \includegraphics[width=0.7\textwidth]{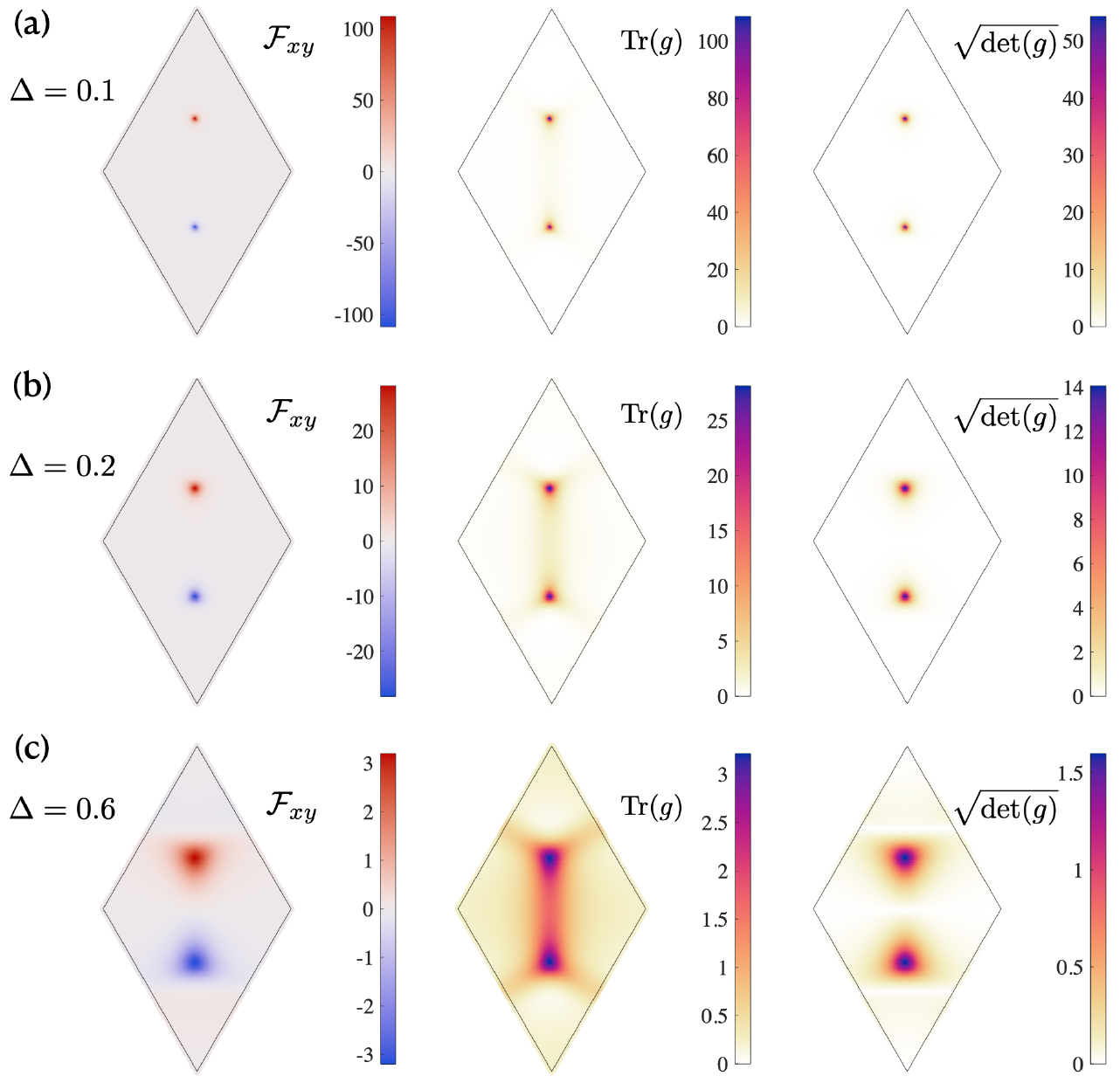}
    \caption{The components of the quantum geometric tensor of the graphene lowest band. The on-site energy split is set as $\Delta=0.1,0.2,0.6$ in (a), (b) and (c), respectively. The nearest-neighboring hopping $t=-1$.}\label{smfig:grapheneQGT}
\end{figure}

So far, we have looked at the quantum geometric phases produced by the particles around $K$ and $K'$ points when $\Delta\to0$. For a finite split energy $\Delta$, we use a numerical calculation to study the quantum geometry. By discretizing the BZ by $200\times200$ lattices, we calculate the QGT of graphene with parameters $t=-1$ and $\Delta=0.1,0.2,0.6$. We show the distributions of the QGT for the lowest band in Figs.~\ref{smfig:grapheneQGT} (a)-(c). We can see that as $\Delta$ increases, which makes the energy band more flat, the quantum geometry becomes more flattened.

\begin{figure}
    \centering
    \includegraphics[width=0.7\textwidth]{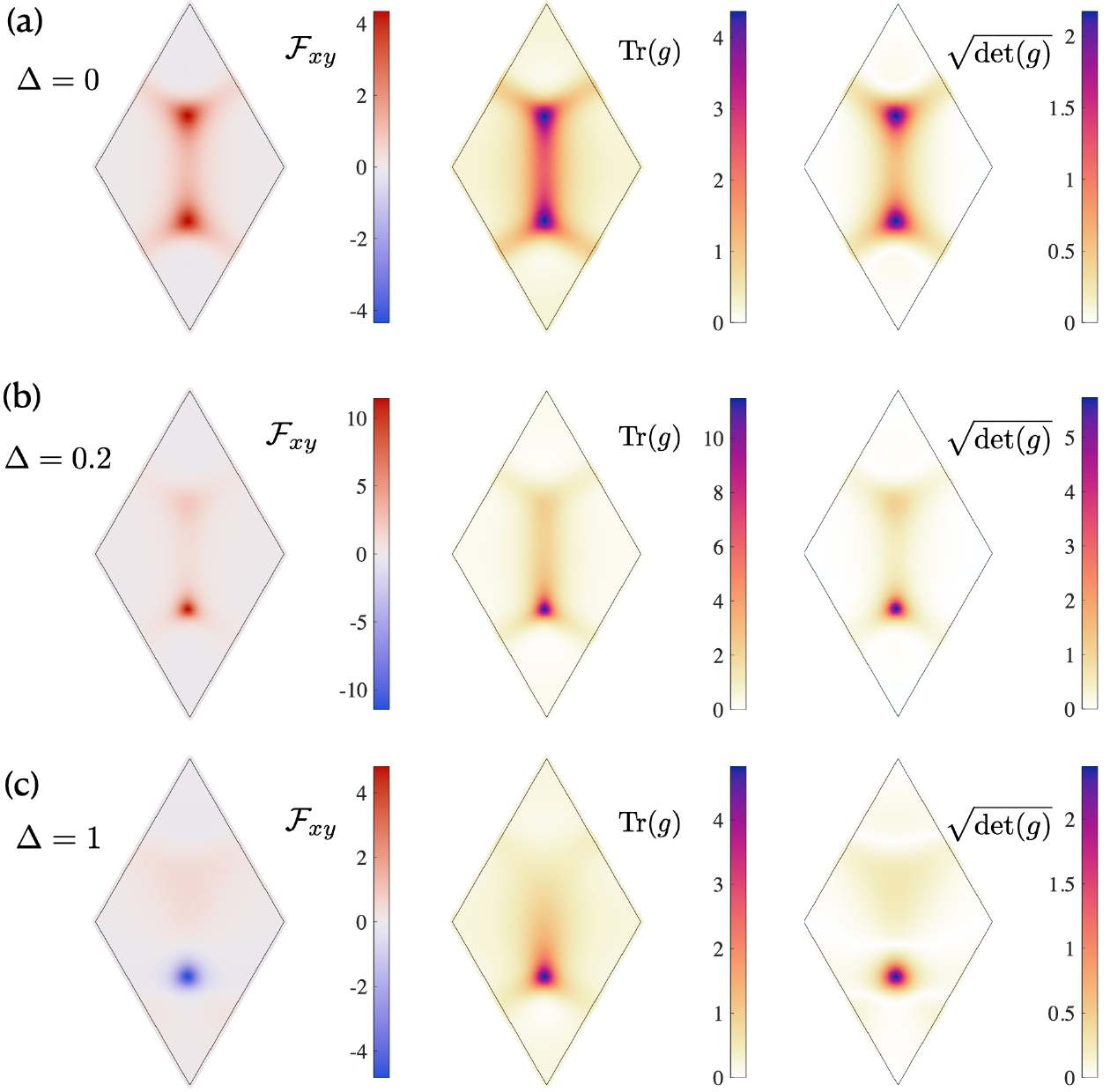}
    \caption{Quantum geometric tensor of the lowest band of the Haldane model described by the Hamiltonian Eq.~(\ref{eq:Haldane}). The fixed parameters are $t=-1$, $u=0.14$, $\varphi=3\pi/4$. We have chosen different onsite energy splits $\Delta$ as 0, 0.2, and 1 in (a)-(c) respectively. }\label{smfig:Haldane_geo}
\end{figure}
Let us review the Haldane model on honeycomb lattices. We can obtain the Haldane model with a non-trivial Chern number by generalizing the graphene model shown in Fig.~\ref{smfig:graphene} with next-nearest hopping carrying magnetic flux $\varphi$. Here, we first write its lattice Hamiltonian as follows
\begin{align}
    H&=h+u\{\left(\begin{array}{cc}{e^{-\mathrm{i}\varphi}}&{}\\ {} & {e^{\mathrm{i}\varphi}}\end{array}\right)[T(\boldsymbol{a}_1)+T(\boldsymbol{a}_2)+T^\dagger(\boldsymbol{a}_1+\boldsymbol{a}_2)]+\text{h.c.}\}.
\end{align}
Here, $h$ is the graphene lattice Hamiltonian Eq.~(\ref{eq:h_graphene}), and we have removed the onsite energy. By using discrete Fourier transformation, we obtain the Bloch Hamiltonian as
\begin{align}
    \mathcal{H}(\boldsymbol{k})&=  h(\boldsymbol{k})+\sum_{i=1}^{3}2u\cos\varphi\cos{(\boldsymbol{k}\cdot\boldsymbol{a}_i)}\,\sigma_0-2u\sin\varphi\sin{(\boldsymbol{k}\cdot\boldsymbol{a}_i)}\,\sigma_z,\label{eq:Haldane}
\end{align}
where $h(\boldsymbol{k})$ is defined in Eq.~(\ref{eq:hk_graphene}) and $\vec{a}_i=\sqrt{3}a_0(\cos(\pi/6+2i\pi/3),\sin(\pi/6+2i\pi/3))$ ($i=1,2$). The characteristic vector is 
\begin{align}
    \vec{d}(\vec{k})&=\left(t(1+\cos{k_1}+\cos{k_2}),t(1+\sin{k_1}-\sin{k_2}),\Delta-2u\sin\varphi\sum_{i=1}^{3}\sin{(\vec{k}\cdot\vec{a}_i)}\right).
\end{align}
Around $K$ and $K'$ points, the Hamiltonian becomes 
\begin{align}
   h_K(\vec{\kappa})&= h(\vec{K}+\vec{\kappa})\approx\Delta_K\sigma_z-\frac{3t}{2}(\kappa_x\sigma_y-\kappa_y\sigma_x)-\frac{3}{2}u\cos\varphi,\\
   h_{K'}(\vec{\kappa})&= h(\vec{K}'+\vec{\kappa})\approx\Delta_{K'}\sigma_z-\frac{3t}{2}(\kappa_x\sigma_y+\kappa_y\sigma_x)-\frac{3}{2}u\cos\varphi.
\end{align}
Here, we define
\begin{align}
\Delta_K&=\Delta+2u\sqrt{3}\sin\varphi=\frac{3|t|}{2}\delta_K,\\
\Delta_{K'}&=\Delta-2u\sqrt{3}\sin\varphi=\frac{3|t|}{2}\delta_{-K}.
\end{align}
By replacing the $\delta$ in Eq.~(\ref{eq:grapheneFxy}) by $\delta_{K}$ or $\delta_{-K}$, we obtain the quantum geometric tensor around $K$ or $K'$ point:
\begin{align}
\mathcal{F}_{xy}^{\pm\vec{K}}(\vec{\kappa})&=\pm\frac{\delta_{\pm K}}{2(\vec{\kappa}^2+\delta_{\pm K}^2)^{3/2}},\label{eq:FxyHaldane}\\
g^{\pm K}_{xx}(\vec{\kappa})&=\frac{1}{4}\frac{\kappa_y^2+\delta_{\pm K}^2}{(\kappa^2+\delta_{\pm K}^2)^2},\\
    g^{\pm K}_{yy}(\vec{\kappa})&=\frac{1}{4}\frac{\kappa_x^2+\delta_{\pm K}^2}{(\kappa^2+\delta_{\pm K}^2)^2},\\
    g^{\pm K}_{xy}(\vec{\kappa})&=-\frac{1}{4}\frac{\kappa_x\kappa_y}{(\kappa^2+\delta_{\pm K}^2)^2}.
\end{align}
and the corresponding total Chern number
\begin{align}
\frac{\text{sign}(\delta_{K})-\text{sign}(\delta_{-K})}{2}=\frac{\text{sign}(\Delta+2u\sqrt{3}\sin\varphi)-\text{sign}(\Delta-2u\sqrt{3}\sin\varphi)}{2}.\label{eq:HaldaneChern}
\end{align}

We now show the quantum geometries of the Haldane model in the Brillouin zone in Fig.~\ref{smfig:Haldane_geo}. The hopping amplitudes used in Fig.~\ref{smfig:Haldane_geo} are chosen as $t=-1$ and $u=0.14$. The corresponding onsite energy split of the sublattices in (a)-(c) is written underneath the label. Unlike the Landau level, which has constant Berry curvature and quantum geometric metric over BZ, the Haldane model has a non-uniform distribution of the geometric tensor in the momentum space~\cite{bandgeometry2014,fracTBG2020,relation2021}. When $\Delta$ crosses $\Delta_c=\sqrt{6}u\approx0.343$, the Chern number becomes trivial. Thus, the Berry curvature around $K'$ point becomes negative when $\Delta=1>\Delta_c$.

In the main text, we have used the emergent gauge field and Peierls substitution to derive the Hamildane model. For the fixed parameters (we mentioned under Eq.~(\ref{eq:potential})) in the main text, we obtain $\varphi=3\pi/4\exp(-\pi^2/27)$ at $g/\omega_c=1$. In principle, increasing the coupling strength can increase the magnetic flux and probably lead to the topological phase transition based on Eq.~(\ref{eq:HaldaneChern})~\cite{HaldaneModel}. However, it does not happen in the numerical results as we see the reentrant trivial phase when $g/\omega_c$ is high in the main text and in Ref.~\cite{Ashida2023}. We argue that this is because Peierl substitution is a qualitative approximation to deal with the classical vector potential in the Hamiltonian, and the accurate phase factor of the hoppings may deviate from the magnetic flux $\varphi$~\cite{Kohn1959}. In the tight-binding Hamiltonian built from the lattice model, $t$, $u$, and $\varphi$ in Eq.~(\ref{eq:Haldane}) are independent parameters. However, they are all related to the more fundamental parameters of the system, especially the coupling strength $g/\omega_c$. 

\section{Chern number calculation through Wilson loop formalism }
\begin{figure}
    \centering
\includegraphics[width=0.6\textwidth]{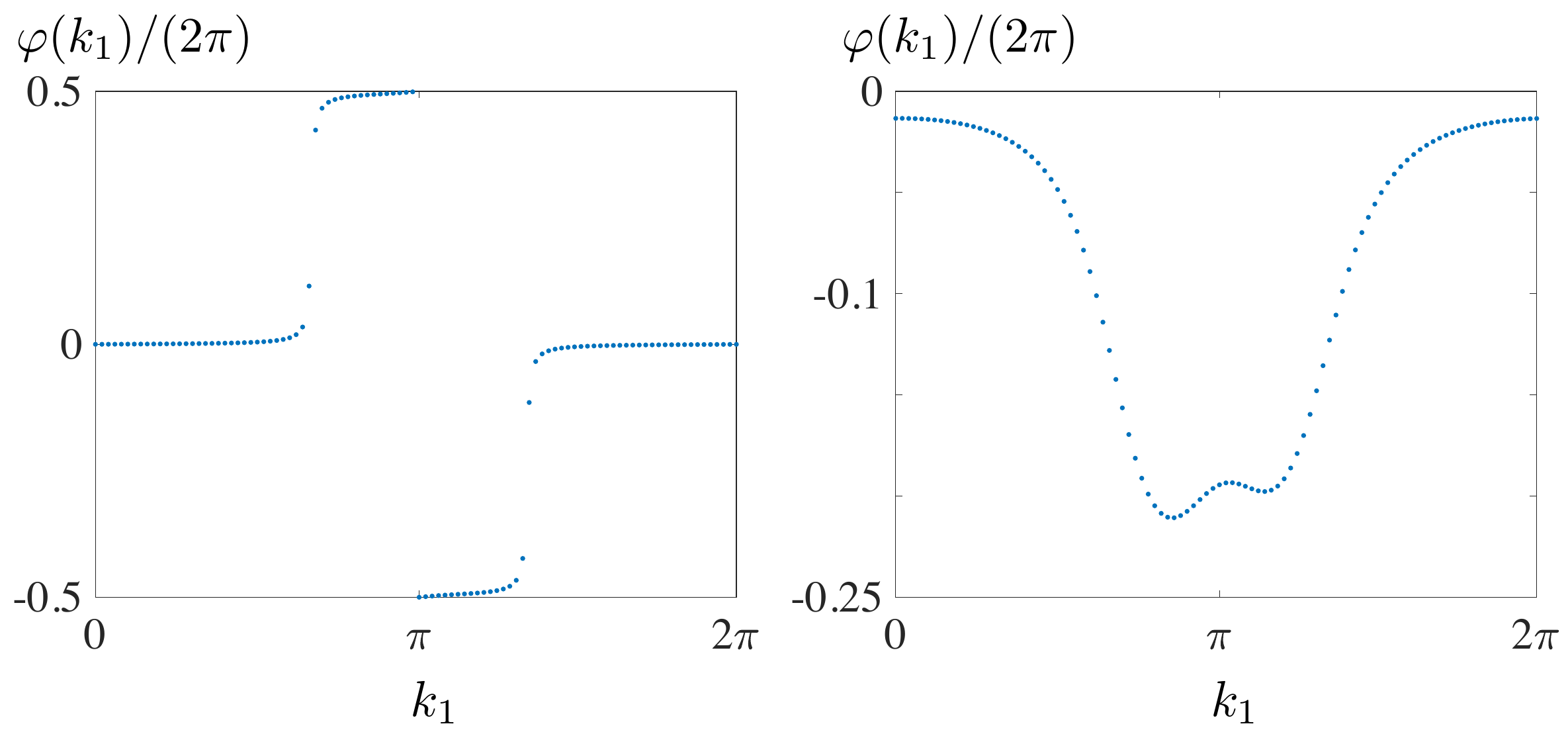}
    \caption{Phase flow of the Wilson loop along $k_1:0\to2\pi$ of the lowest band of the cavity graphene. The cavity parameter used here are $\omega_c=2$, $g/\omega_c=1$, and the material parameters are set as $V_0=m=1.5$ and $\Delta=0$ (left) or $1$ (right). We can see that the smooth change of the phase is $2\pi$ for the left figure and 0 for the right.}\label{smfig:wilsonphase}
\end{figure}

If we numerically integrate the discrete values of the Berry curvatures over the BZ mesh, we will not get an exact integer Chern number. To confirm the non-trivial Chern number in Fig. 2 in the main text, one should resort to the Wilson loop formalism~\cite{RMP_Resta1994,wilsonloop2014}. The (abelian) Wilson loop over the $n$-th band along the loop $k_2:0\to2\pi$ is defined as follow
\begin{align}
    \mathcal{W}_{k_2}(k_1)=\lim_{N_2\to\infty}\prod_{m=1}^{N_2} \langle u_n(k_1,k_2+\frac{2\pi m}{N_2})| u_n(k_1,k_2+\frac{2\pi (m-1)}{N_2})\rangle,
\end{align}
where $k_i$ is defined through Eq.~(\ref{smeq:kv}), and $|u_n(\vec{k})\rangle$ is the Bloch wave function with the gauge $|u_n(\vec{k})\rangle=|u_n(\vec{k}+\vec{K})\rangle$. Then, the Chern number of the $n$-th band is related to the phase flow of the Wilson loop from $k_1$ to $k_1+2\pi$:
\begin{align}
    \mathcal{C}&=\frac{\varphi(k_1+2\pi)-\varphi(k_1)}{2\pi},\\
    \varphi(k_1)&=\text{arg}(\mathcal{W}_{k_2}(k_1)).
\end{align}
Note that $\varphi(k_1)$ should be made continuous. In Fig.~\ref{smfig:wilsonphase}, we show the numerical results of $\varphi(k_1)$ for the lowest band of the cavity graphene model in the domain $k_1\in[0,2\pi]$. The corresponding parameters are chosen as $\omega_c=2$, $g/\omega_c=1$, $V_0=m=1.5$ and $\Delta=0$ (left) or $\Delta=1$ (right). In the numerical calculation, we discretize the loops of $k_1:0\to2\pi$ and $k_2:0\to2\pi$ by $N_1=100$ and $N_2=30$ lattice points. We see that the smooth change of the phase $\varphi$ over the loop $k_1:0\to2\pi$ is $2\pi$ for the left figure, and 0 for the right figure, respectively. This confirms the non-trivial Chern number of the lowest-band Berry curvature in Fig.~2 of the main text. 

\section{Valley energy splitting in a general cavity with hybrid chirality}

\begin{figure}
    \centering
\includegraphics[width=0.6\textwidth]{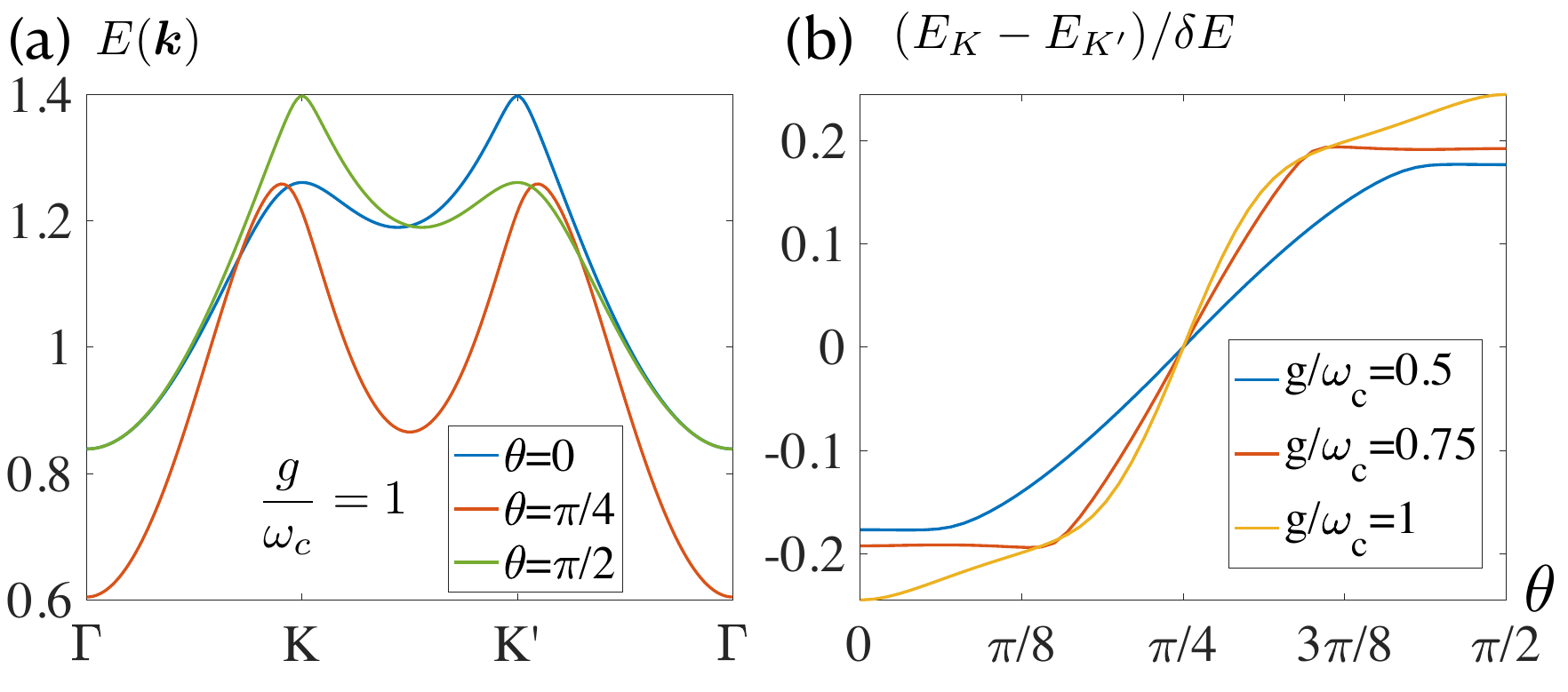}
    \caption{(a) The lowest energy band of the graphene in a cavity with hybrid chirality defined in Eq.~\ref{eq:Atheta}. The onsite energy split here is $\Delta=0.6$. By tuning $\theta$ from 0 to $\pi/2$, the energy difference of the two valleys changes the sign. At $\theta=\pi/4$ when the photons in the cavity become all linearly polarized, the two valleys become degenerate. (b) The valley energy split ratio $(E_K-E_{K'})/\delta E$ as $\theta$ changes. We show the results with different coupling strength factors $g/\omega_c=0.5,0.75,1$.}
    \label{smfig:valley_theta}
\end{figure}

We numerically calculate the lowest energy band of the effective Hamiltonian Eq.~(\ref{eq:Heff}) with different hybrid degree $\theta$. In Fig.~\ref{smfig:valley_theta} (a), we show the energy shift of valleys $K$ and  $K'$ for the system with $\Delta=0.6$. We find that if the cavity photon has net chiral component, the energy shift of valleys becomes different. Apart from that, $K'$ ($K$) valley becomes more energetic in the hybrid-chiral cavity when $0<\theta<\pi/4$ ($\pi/4<\theta<\pi/2$). However, if the photon is linearly polarized at $\theta=\pi/4$, the two valleys are degenerate since the time reversal symmetry holds again.

For more details about how the valley energy split can be changed by the hybrid degree of the photon chirality, we calculate the ratio of the energy difference $E_K-E_{K'}$ to the bandwidth $\delta E$, which is estimated by $\max(E_K,E_{K'})-E_\Gamma$. The calculation is processed with fixed sub-lattice energy split $\Delta=0.6$ at several coupling strength factors $g/\omega_c=0.5,0.75,1$. In Fig.~\ref{smfig:valley_theta}) (b), we see that the split of the two valleys changes the sign as $\theta$ exceeds $\pi/4$. Also, for the chiral cavities, the ratio $(E_K-E_{K'})/\delta E$ is increased if the coupling strength becomes larger. This means that for a $C_2$-symmetry-breaking material in a chiral cavity, the energy imbalance of the two valleys can be increased if one makes the light-matter coupling stronger.

\section{The photon number change in a function of the coupling strength}
\begin{figure*}
    \centering
\includegraphics[width=0.7\textwidth]{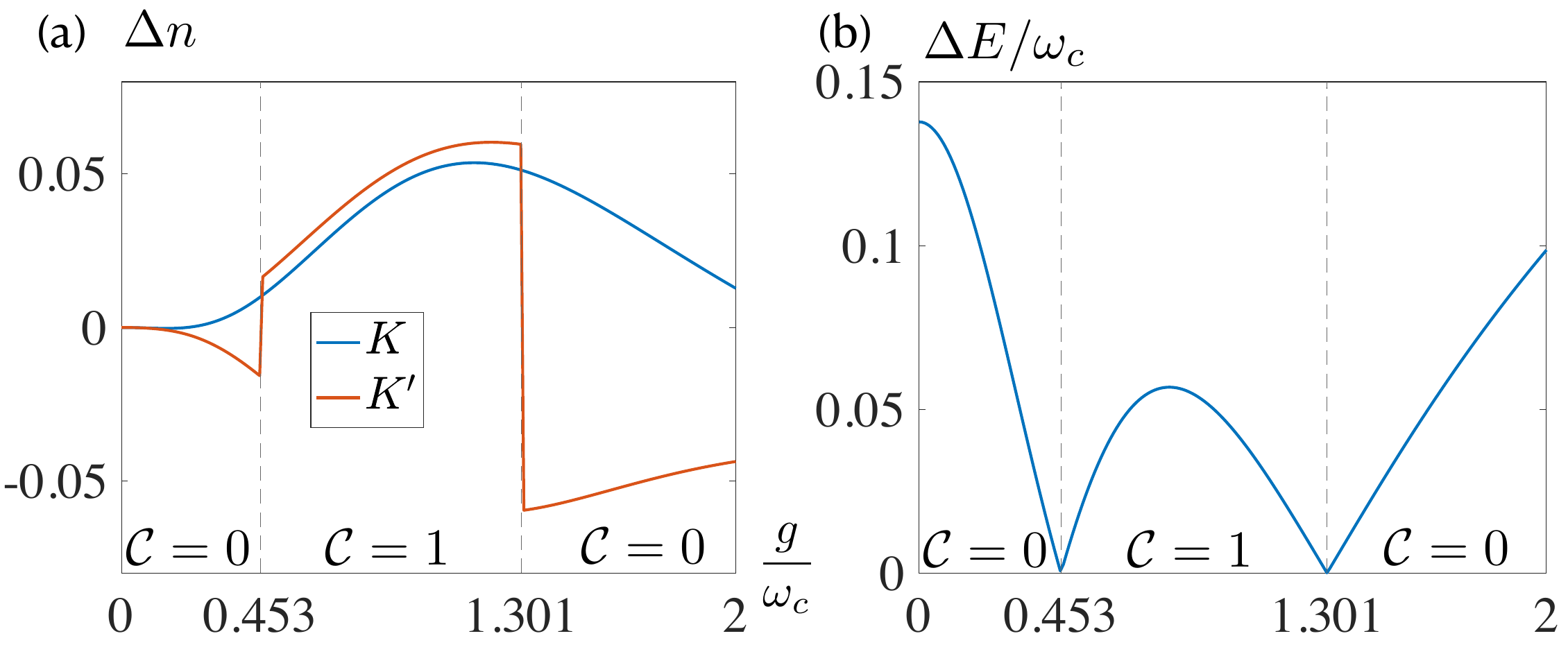}
    \caption{(a): The photon number changes during the interband excitation from the lowest band to the upper band in a function of the coupling strength $g/\omega_c$ ranging from 0 to 2. The sub-lattice split is set to be $\Delta=0.6$. The dashed lines label two topological transition points at $g/\omega_c=0.453$ and $g/\omega_c=1.301$. (b): The energy gap between the lowest energy band and the upper band. We observe that the system is off-resonant, indicated by the condition $\Delta E\ll \hbar\omega_c$, and thus the tunneling from the lower band to the upper band at $K'$ point cost the minimal excitation energy.}
    \label{smfig:photon_gw}
\end{figure*}

In the main text, we have analyzed the photon number of the light-matter interacting bands. Here, we complement the results of the photon number change during the interband excitation in a function of the coupling strength $g/\omega_c$ with a fixed sub-lattice split $\Delta$. In Fig.~\ref{smfig:photon_gw} (a), we present the change of photon numbers $\Delta n_p$ for a light-matter interacting state at $K$ and $K'$ points during the transition from the lowest to the upper band for $\Delta=0.6$ across a range of $g/\omega_c$ from 0 to 2. The associated energy gap within the same range is illustrated in (b), aiding in identifying points of topological phase transition where the gap closes. Through (a), we observe that $\Delta n_p(K)$ consistently remains positive with increasing $g/\omega_c$, while $\Delta n_p(K')$ changes sign as the Chern number becomes zero. The transition in Chern number aligns with the sign change of Berry curvature around $K'$ (refer to Eqs.~(\ref{eq:FxyHaldane}) and (\ref{eq:HaldaneChern})). Consequently, we can conclude that the signs of $\Delta n_p(K)$ and $\Delta n_p(K')$ are governed by the Berry curvature's signs in that region. Note that this conclusion is within the regime where the energy gap is much smaller than the single-photon energy $\hbar\omega_c$ (see (b)).

\bibliography{ref.bib}